\documentclass[allclo,dvips]{FBSart}
\usepackage{amsfonts}
\usepackage{amssymb}
\usepackage{graphicx} 

\newcommand{\sfrac}[2]{\mbox{\footnotesize $\displaystyle \frac{#1}{#2}$}}
 
\newcommand{\lsim}{\mathrel{\rlap{\lower4pt\hbox{\hskip0pt$\sim$}} 
\raise1pt\hbox{$<$}}}           
\newcommand{\gsim}{\mathrel{\rlap{\lower4pt\hbox{\hskip0pt$\sim$}} 
\raise1pt\hbox{$>$}}}           

\newcounter{dumbone}

\usepackage{color}
\definecolor{purple}{rgb}{0.5,0,0.5}
\definecolor{blue}{rgb}{0.0,0,0.9}

\title{Survey of nucleon electromagnetic form factors}

\author{I.\,C.~Clo\"et,\instnr{1,2}
G.~Eichmann,\instnr{1,3} 
B.~El-Bennich,\instnr{1}
T.~Kl\"ahn\instnr{1}
and C.\,D.\ Roberts\instnr{1,4} } 

\instlist{
Physics Division, Argonne National Laboratory, Argonne IL
60439, USA \and
Department of Physics, University of Washington, Seattle WA 98195, USA \and
Institut f\"ur Physik, Karl-Franzens-Universit\"at Graz, A-8010 Graz, Austria \and
School of Physics, The University of New South Wales, Sydney NSW 2052, Australia
}

\runningauthor{I.\,C.~Clo\"et, et al.}
\runningtitle{Nucleon electromagnetic form factors}
\begin{document}

\maketitle 
\begin{abstract}
A dressed-quark core contribution to nucleon electromagnetic form factors is calculated.  It is defined by the solution of a Poincar\'e covariant Faddeev equation in which dressed-quarks provide the elementary degree of freedom and correlations between them are expressed via diquarks.  The nucleon-photon vertex involves a single parameter; i.e., a diquark charge radius.  It is argued to be commensurate with the pion's charge radius.  A comprehensive analysis and explanation of the form factors is built upon this foundation.  A particular feature of the study is a separation of form factor contributions into those from different diagram types and correlation sectors, and subsequently a flavour separation for each of these.  Amongst the extensive body of results that one could highlight are: $r_1^{n,u}>r_1^{n,d}$, owing to the presence of axial-vector quark-quark correlations; and for both the neutron and proton the ratio of Sachs electric and magnetic form factors possesses a zero.
%


\end{abstract}

\section{Introduction}
Owing in part to the relatively simple nature of the virtual photon as a probe, a reliable explanation of electromagnetic form factors provides information on the distribution of a nucleon's characterising properties; e.g., total- and angular-momentum, amongst its QCD constituents.  Since contemporary experiments employ $Q^2>M_N^2$; i.e., momentum transfers in excess of the nucleon's mass, a veracious understanding of the body of extant data requires a Poincar\'e covariant description of the nucleon.  Poincar\'e covariance and the vector exchange nature of QCD guarantee the existence of nonzero quark orbital angular momentum in a hadron's rest-frame bound-state amplitude \cite{Bhagwat:2006xi,Cloet:2007pi}.

In fact the challenge is compounded owing, e.g., to the running of the dressed-quark mass \cite{Lane:1974he,Politzer:1976tv,Roberts:1994dr,bowman,Bhagwat:2003vw,Bhagwat:2006tu}.  This entails that a quantum field theoretic treatment of hadron structure and electromagnetic interactions is generally necessary in order to provide understanding in terms of QCD's genuine elementary degrees of freedom.  The dressed light-quark mass function at infrared momenta is roughly $100$-times larger than the current-quark mass.  This marked enhancement is a corollary of dynamical chiral symmetry breaking (DCSB) and owes primarily to a dense cloud of gluons that clothes a low-momentum quark \cite{Bhagwat:2007vx}.  (The dressing gluons also acquire mass dynamically \cite{Bowman:2007du}.)  It means that the Higgs mechanism is largely irrelevant to the bulk of normal matter in the universe.  Instead the single most important mass generating mechanism for light-quark hadrons is the strong interaction effect of DCSB; e.g., one can identify DCSB as being responsible for 98\% of a proton's mass.  It has long been argued that form factors are a sensitive probe of this effect \cite{Roberts:1994hh}.

Recent years have seen rapid experimental and theoretical progress in the study of nucleon electromagnetic form factors, which is reviewed, e.g., in Refs.\,\cite{Arrington:2006zm,Perdrisat:2006hj}.  Despite this, questions remain unanswered, amongst them:
can one formulate an impulse-like approximation for hadron form factors and, if so, in terms of which degrees of freedom;
is there a valid mapping of form factors into statements about the distribution of charge and magnetisation within a nucleon; 
and what role is played by pseudoscalar mesons in hadron electromagnetic structure and can one describe this in a quantitative, model-independent fashion?
Herein we contribute to the discussion of these issues.

In Sect.\,\ref{nucleonmodel} we recapitulate briefly upon a Poincar\'e covariant Faddeev equation for the nucleon, in which the primary element is the dressed-quark with its strongly momentum dependent mass function.  The Faddeev equation solution defines a nucleon's dressed-quark core.  The study of baryons in this way sits squarely within the ambit of the application of Dyson-Schwinger equations (DSEs) in QCD \cite{Roberts:2007ji}.  Since the DSEs admit a nonperturbative symmetry-preserving truncation scheme \cite{Munczek:1994zz,Bender:1996bb,Bender:2002as,Bhagwat:2004hn}, which e.g. has enabled the proof of numerous exact results for pseudoscalar mesons \cite{Maris:1997hd,Maris:1997tm,Holl:2004fr,Holl:2005vu}, the approach holds particular promise as a means of unifying the treatment of meson and baryon observables that preserves all global and local corollaries of DCSB without fine-tuning \cite{Eichmann:2008ef}.  The coupling of a photon to the nucleon's dressed-quark core is detailed in Sect.\,\ref{sec:MMR}.  

In Sect.\,\ref{sec:interpreting} we discuss the interpretation of form factors and present a perspective on the circumstances under which the three dimensional Fourier  transform of a Breit-frame Sachs form factor can reasonably be understood in terms of a charge or magnetisation density.  

Sections~\ref{FFall}--\ref{FFn} are extensive.  They detail our computed results and the understanding they provide.  All electromagnetic form factors of the proton and neutron are described along with their decomposition into individual flavour, diagram and diquark contributions, the meaning of which will subsequently become apparent.

We consider form factor contributions arising from pseudoscalar meson loops in Sect.\,\ref{sec:chiral} and exemplify the manner in which they add to the dressed-quark core results.  We wrap-up in Sect~\ref{sec:epilogue}.

\section{Nucleon Model}
\label{nucleonmodel}
In quantum field theory a nucleon appears as a pole in a six-point quark Green function.  The pole's residue is proportional to the nucleon's Faddeev amplitude, which is obtained from a Poincar\'e covariant Faddeev equation that adds-up all possible quantum field theoretical exchanges and interactions that can take place between three dressed-quarks. Canonical normalisation of the Faddeev amplitude guarantees unit residue for the $s$-channel nucleon pole in the $J^P= \frac{1}{2}^+$ three-quark vacuum polarisation diagram and entails unit charge for the proton.

\begin{figure}[t]
\leftline{%
\includegraphics[clip,width=0.75\textwidth]{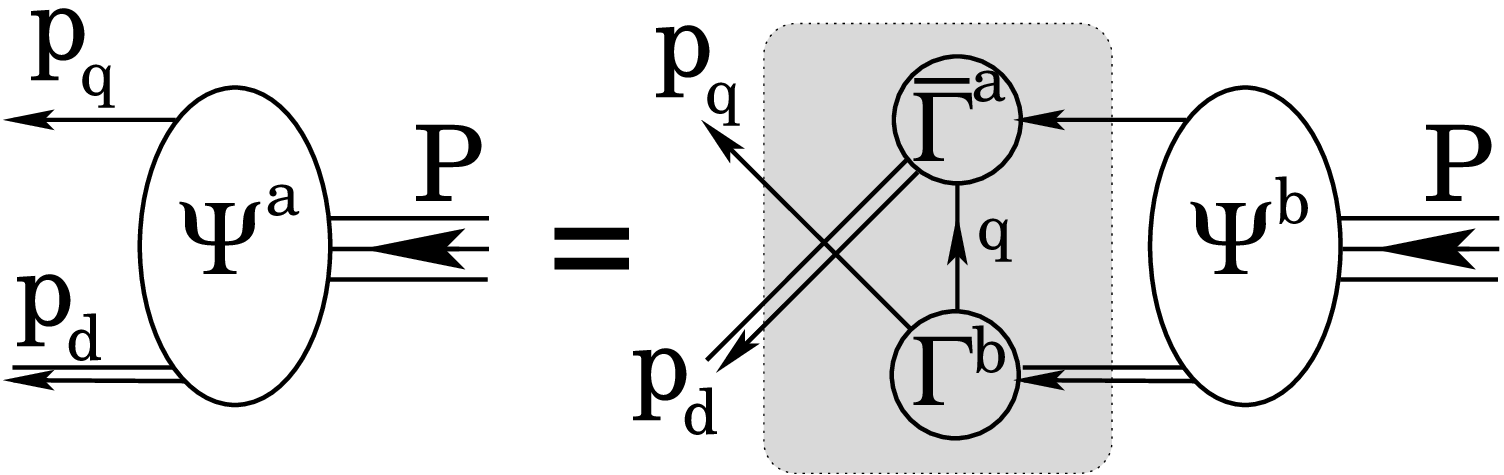}}
\caption{\label{faddeevfigure} Poincar\'e covariant Faddeev equation, Eq.\,(\protect\ref{FEone}), employed herein to calculate nucleon properties.  $\Psi$ in Eq.\,(\protect\ref{PsiNucleon}) is the Faddeev amplitude for a nucleon of total momentum $P= p_q + p_d$.  It expresses the relative momentum correlation between the dressed-quark and -diquarks within the nucleon.  The shaded region demarcates the kernel of the Faddeev equation, Sect.\,\protect\ref{completing}, in which: the \emph{single line} denotes the dressed-quark propagator, Sect.\,\protect\ref{subsubsec:S}; $\Gamma$ is the diquark Bethe-Salpeter-like amplitude, Sect.\,\protect\ref{qqBSA}; and the \emph{double line} is the diquark propagator, Sect.\,\protect\ref{qqprop}.}
\end{figure}

A tractable truncation of the Faddeev equation is based \cite{Cahill:1988dx} on the observation that an interaction which describes mesons also generates diquark correlations in the colour-$\bar 3$ channel \cite{Cahill:1987qr}.  The dominant correlations for ground state octet and decuplet baryons are scalar ($0^+$) and axial-vector ($1^+$) diquarks because, for example, the associated mass-scales are smaller than the baryons' masses \cite{Burden:1996nh,Maris:2002yu}, namely (in GeV)
\begin{equation}
\label{diquarkmass}
m_{[ud]_{0^+}} = 0.7 - 0.8
 \,,\; 
m_{(uu)_{1^+}}=m_{(ud)_{1^+}}=m_{(dd)_{1^+}}=0.9 - 1.0\,.
\end{equation}

The kernel of the Faddeev equation is completed by specifying that the quarks are dressed, with two of the three dressed-quarks correlated always as a colour-$\bar 3$ diquark.  As illustrated in Fig\,\ref{faddeevfigure}, binding is then effected by the iterated exchange of roles between the bystander and diquark-participant quarks.  

The Faddeev equation that we employ is explained in \ref{app:FE}~\textit{Faddeev Equation}.  With all its elements specified, as described therein, the equation can be solved to obtain the nucleon's mass and amplitude.  Owing to Eq.\,(\ref{DQPropConstr}), in this calculation the masses of the scalar and axial-vector diquarks are the only variable parameters.  The axial-vector mass is chosen so as to obtain a desired mass for the $\Delta$,\footnote{This is natural because the spin- and isospin-$3/2$ $\Delta$ contains only an axial-vector diquark.  The relevant Faddeev equation is not different in principle to that for the nucleon.  It is described in Ref.\,\protect\cite{Flambaum:2005kc}.} and the scalar mass is subsequently set by requiring a particular nucleon mass.  

\begin{table}[b]
\begin{center}
\caption{\label{tableNmass} Mass-scale parameters (in GeV) for the scalar and axial-vector diquark correlations, fixed by fitting nucleon and $\Delta$ masses offset to allow for ``pion cloud'' contributions \protect\cite{Hecht:2002ej}.  We also list $\omega_{J^{P}}= \sfrac{1}{\surd 2}m_{J^{P}}$, the width-parameter in the $(qq)_{J^P}$ Bethe-Salpeter amplitude, Eqs.\,(\protect\ref{Gamma0p}) \& (\protect\ref{Gamma1p}):  its inverse is an indication of the diquark's matter radius.  Row~3 illustrates effects of omitting the $1^+$-diquark correlation: the $\Delta$ cannot be formed and $M_N$ is significantly increased.  Evidently, the $1^+$-diquark provides significant attraction in the Faddeev equation's kernel.}
\begin{tabular*}{1.0\textwidth}{
c@{\extracolsep{0ptplus1fil}}c@{\extracolsep{0ptplus1fil}}
c@{\extracolsep{0ptplus1fil}} c@{\extracolsep{0ptplus1fil}}c@{\extracolsep{0ptplus1fil}}c@{\extracolsep{0ptplus1fil}}}
\hline
$M_N$ & $M_{\Delta}$~ & $m_{0^{+}}$ & $m_{1^{+}}$~ &
$\omega_{0^{+}} $ & $\omega_{1^{+}}$ \\
\hline
1.18 & 1.33~ & 0.796 & 0.893 & 0.56=1/(0.35\,{\rm fm}) & 0.63=1/(0.31\,{\rm fm}) \\
1.46 &  & 0.796 &  & 0.56=1/(0.35\,{\rm fm}) &  \\
\hline
\end{tabular*}
\end{center}
\end{table}

We have written here of desired rather than experimental mass values because it is known that the masses of the nucleon and $\Delta$ are materially reduced by pseudoscalar meson loop effects.  This is detailed in Refs.\,\cite{Hecht:2002ej,Young:2002cj}.  Hence, a baryon represented by the Faddeev equation described above must possess a mass that is inflated with respect to experiment so as to allow for an additional attractive contribution from the pseudoscalar mesons.  As in previous work \cite{Flambaum:2005kc,Alkofer:2004yf,Holl:2005zi,Cloet:2008wg} and reported in Table~\ref{tableNmass}, we require $M_N=1.18\,$GeV and $M_\Delta=1.33\,$GeV.  The results and conclusions of our study are essentially unchanged should even larger masses and a smaller splitting $M_\Delta-M_N$ be more realistic, a possibility suggested by Refs.\,\cite{Eichmann:2008ef,AWTErice}.  The relationship between the $\Delta$--$N$ mass splitting and that between the axial-vector and scalar diquark correlations is sketched in Ref.\,\cite{Cloet:2008fw}.

\section{Nucleon Electromagnetic Current}
\label{sec:MMR}
The nucleon's electromagnetic current is
\begin{eqnarray}
\label{Jnucleon}
J_\mu(P^\prime,P) & = & ie\,\bar u(P^\prime)\, \Lambda_\mu(q,P) \,u(P)\,, \\
& = &  i e \,\bar u(P^\prime)\,\left( \gamma_\mu F_1(Q^2) +
\frac{1}{2M}\, \sigma_{\mu\nu}\,Q_\nu\,F_2(Q^2)\right) u(P)\,,
\label{JnucleonB}
\end{eqnarray}
where $P$ ($P^\prime$) is the momentum of the incoming (outgoing) nucleon, $Q= P^\prime - P$, and $F_1$ and $F_2$ are, respectively, the Dirac and Pauli form factors.  They are the primary calculated quantities, from which one obtains the nucleon's electric and magnetic (Sachs) form factors 
\begin{equation}
\label{GEpeq}
G_E(Q^2)  =  F_1(Q^2) - \frac{Q^2}{4 M^2} F_2(Q^2)\,,\;  
G_M(Q^2)  =  F_1(Q^2) + F_2(Q^2)\,.
\end{equation}

Static electromagnetic properties are associated with the behaviour of these form factors in the neighbourhood of $Q^2\simeq 0$. The nucleons' magnetic moments are defined through
\begin{equation}
\label{momdef}
\mu_n = \kappa_n = G_M^n(0)\,, \; \mu_p = 1 + \kappa_p = G_M^p(0)\,,
\end{equation}
where $\kappa_N$, $N=n,p$, are referred to as the anomalous magnetic moments; and the electric and magnetic rms radii via
\begin{eqnarray}
\label{chargeradii}
r_p^2 &:=&  -6 \left. \frac{d}{ds} G_E^p(s) \right|_{s=0} ,\; 
r_n^2 := -6 \left. \frac{d}{ds} \, G_E^n(s) \right|_{s=0} ,
\\
(r_N^\mu)^2 &:=& -6 \left. \frac{d}{ds} \ln G_M^N(s) \right|_{s=0}.
\end{eqnarray}

\begin{figure}[t]
\begin{minipage}[t]{\textwidth}
\begin{minipage}[t]{0.45\textwidth}
\leftline{\includegraphics[width=0.90\textwidth]{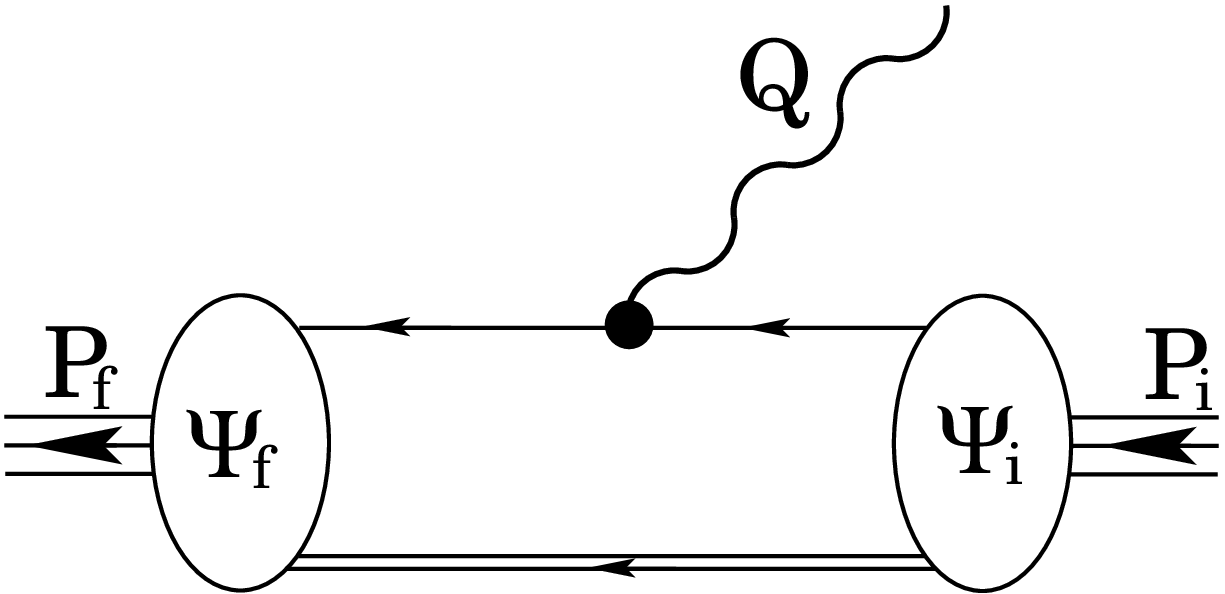}}
\end{minipage}
\begin{minipage}[t]{0.45\textwidth}
\rightline{\includegraphics[width=0.90\textwidth]{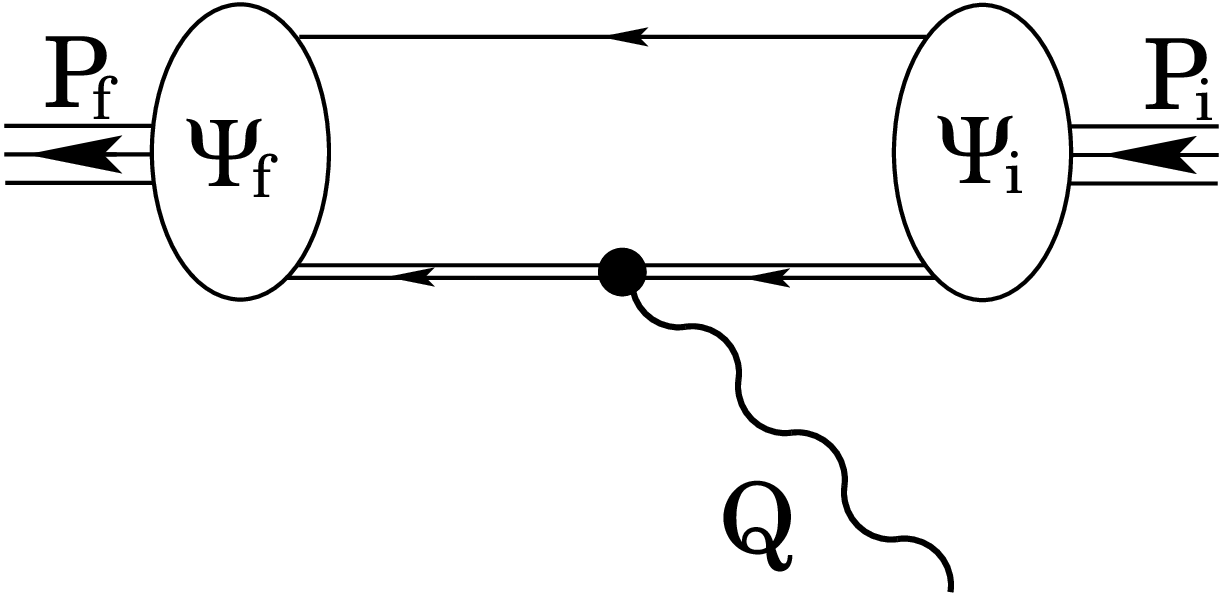}}
\end{minipage}\vspace*{3ex}

\begin{minipage}[t]{0.45\textwidth}
\leftline{\includegraphics[width=0.90\textwidth]{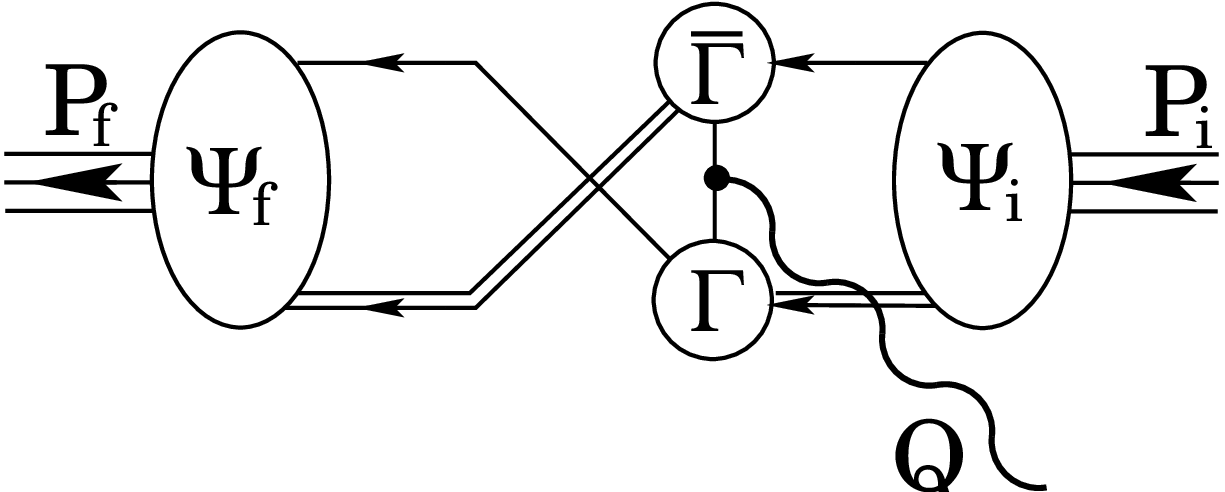}}
\end{minipage}
\begin{minipage}[t]{0.45\textwidth}
\rightline{\includegraphics[width=0.90\textwidth]{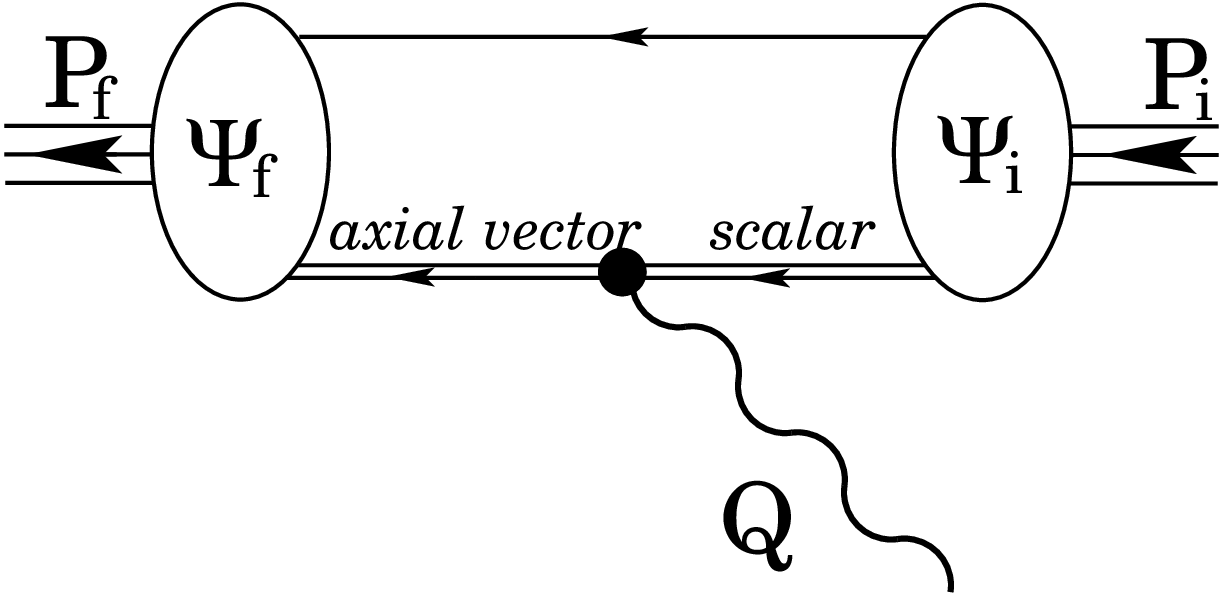}}
\end{minipage}\vspace*{3ex}

\begin{minipage}[t]{0.45\textwidth}
\leftline{\includegraphics[width=0.90\textwidth]{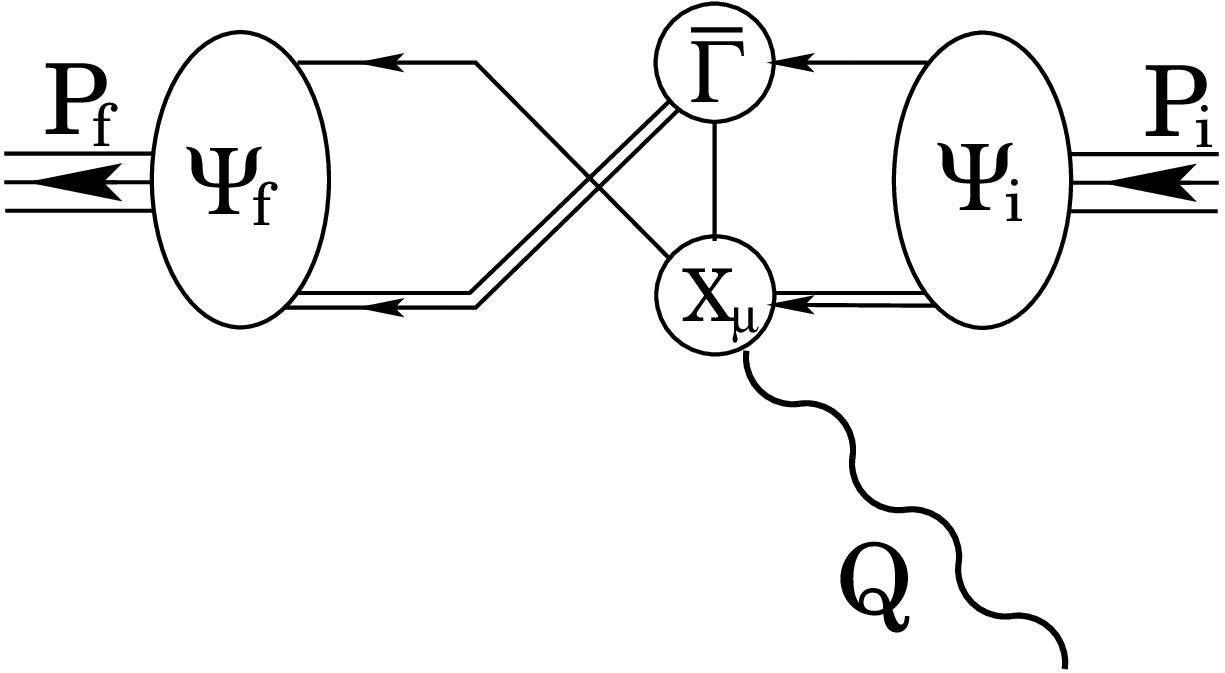}}
\end{minipage}
\begin{minipage}[t]{0.45\textwidth}
\rightline{\includegraphics[width=0.90\textwidth]{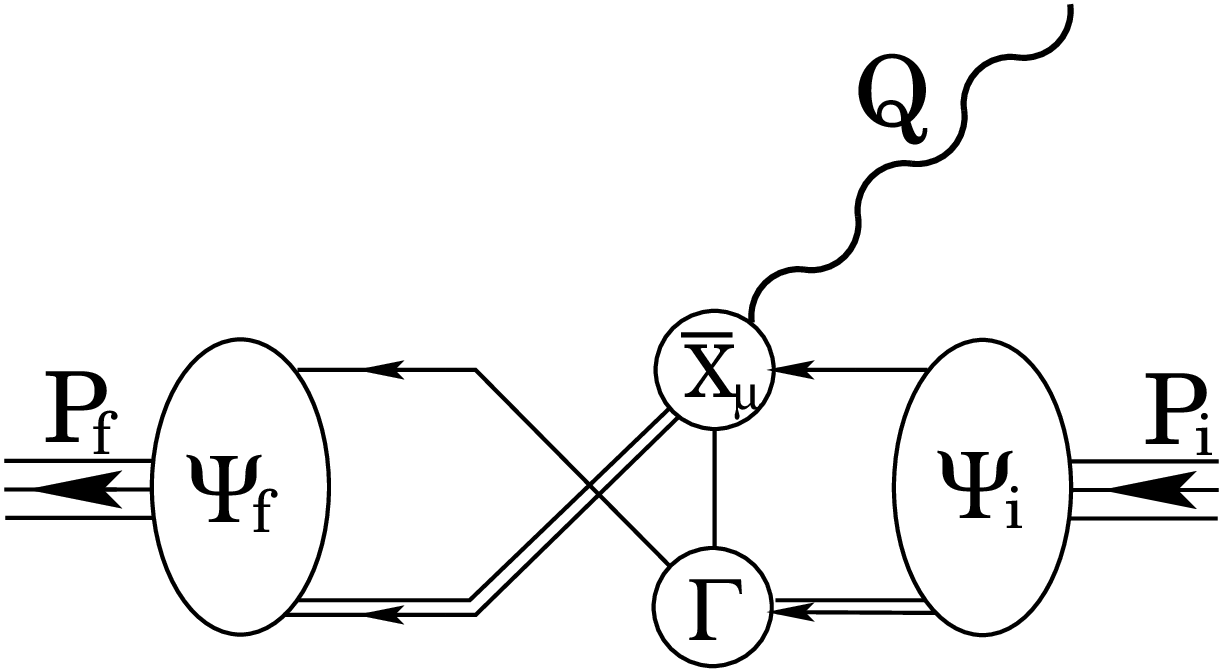}}
\end{minipage}
\end{minipage}
\caption{\label{vertex} Vertex which ensures a conserved current for on-shell nucleons described by the Faddeev amplitudes, $\Psi_{i,f}$, described in Sect.\,\protect\ref{nucleonmodel} and \protect\ref{app:FE}~\emph{Faddeev Equation}.  The single line represents $S(p)$, the dressed-quark propagator, Sec.\,\protect\ref{subsubsec:S}, and the double line, the diquark propagator, Sec.\,\protect\ref{qqprop}; $\Gamma$ is the diquark Bethe-Salpeter amplitude, Sec.\,\protect\ref{qqBSA}; and the remaining vertices are described in \ref{NPVertex} the top-left image is Diagram~1; the top-right, Diagram~2; and so on, with the bottom-right image, Diagram~6.}
\end{figure}

In order to calculate the electromagnetic form factors one must know the manner in which the nucleon described in Sect.\,\ref{nucleonmodel} couples to a photon.  That is derived in Ref.\,\cite{Oettel:1999gc}, illustrated in Fig.\,\ref{vertex} and detailed in \ref{NPVertex}~\emph{Nucleon-Photon Vertex}.  As apparent in that Appendix, the current depends on the electromagnetic properties of the diquark correlations.  

Estimates exist of the size of diquark correlations.  For example, a first Faddeev equation study of nucleon form factors \cite{Bloch:1999ke} found a scalar diquark radius of $r_{[ud]_{0^+}}=0.8\,r_\pi$, where $r_\pi$ is the pion charge radius within the same model.  One obtains a similar result in a DSE calculation \cite{Maris:2004bp} that provides a good description of pseudoscalar and vector meson properties; i.e., 
\begin{equation}
r_{[ud]_{0^+}} \approx 0.7\,{\rm fm}\,,\; r_{(ud)_{1^+}} \sim 0.8\,{\rm fm}\,,
\label{dqradii}
\end{equation}
where the last result is an estimate based on the $\rho$-meson/$\pi$-meson radius-ratio \cite{Maris:2000sk,Bhagwat:2006pu}.  From another perspective, numerical simulations of quenched lattice-regularised QCD suggest a scalar-diquark matter-radius \cite{Alexandrou:2006cq}
\begin{equation}
r^\rho_{[ud]_{0^+}}=1.1\pm 0.2\,{\rm fm}.
\end{equation}

It is thus evident that diquark correlations within a baryon are not pointlike.  Hence, with increasing $Q^2$, interaction diagrams in which the photon resolves a diquark's substructure must be suppressed with respect to contributions from diagrams that describe a photon interacting with a bystander or exchanged quark.  These latter are the only hard interactions with dressed-quarks allowed in a nucleon.  One can therefore improve on Refs.\,\cite{Alkofer:2004yf,Holl:2005zi} by introducing a diquark form factor.  This is expressed in Eqs.\,(\ref{Fqqform}), (\ref{Gamma0plus}) and (\ref{AnsatzF1}).  

We use a one-parameter dipole because the system involves two quarks.  The parameter is a length-scale that characterises the diquark radius.  In the absence of an explicit calculation of the axial-vector diquark's radius, we employ the same value for scalar and axial-vector diquarks.  Owing to differences between the formulation of our nucleon model and the DSE truncation employed in Ref.\,\cite{Maris:2004bp}, the values quoted in Eq.\,(\ref{dqradii}) provide only a loose constraint on this parameter.  It's value does not have a large effect on form factors for $Q^2\lesssim 2\,$GeV$^2$ but does influence their evolution thereafter.  For example, it influences the position of the zero in $G_E^p(Q^2)$: a larger diquark radius shifting the zero further from the origin.  Computations have been analysed with four values: $r_{qq} = 0.0$, $0.4$, $0.8$ and $1.2\,$fm.  Unless otherwise stated, the results reported herein were obtained with 
\begin{equation}
r_{qq}=0.8\,{\rm fm}\,.
\label{rqqvalue}
\end{equation}

\section{On Interpreting Form Factors}
\label{sec:interpreting}
Now that the Faddeev equation and a consistent Ward-Takahashi-identity conserving current are completely defined, the calculation of a nucleons' electromagnetic form factors is a straightforward numerical exercise.  However, in light of Refs.\,\cite{Miller:2007uy,Miller:2008jc} we judge it worthwhile to comment on their putative interpretation in terms of charge and magnetisation densities before presenting our results.

Such an interpretation rests on the existence of a quantitatively reliable expression for the form factors in terms of a current in which the interacting constituents are well-defined and distinct, for then the charge and current carrying quanta are unambiguous.  This is achieved through a current of impulse approximation type, which may include small non-single-particle contributions that arise owing to the Ward-Takahashi identity.  

In QCD the relevant degrees of freedom change as the wavelength of the probe evolves.  This feature is encoded, e.g., in the dressed-quark mass function, which is discussed in connection with Eq.\,(\ref{ZMdef}).  The nature of the mass function is model-independent and one consequence is that to a long wavelength probe a light-quark appears to have a large inertial mass $\sim 350\,$MeV.  

Figure~\ref{vertex} expresses a nucleon current in which the primary degrees of freedom are dressed-quarks.  Along with the Faddeev equation described in App.~A, it is an extension to baryons of the systematic and symmetry preserving rainbow-ladder truncation of QCD's DSEs, which provides a sound description of pseudoscalar and vector mesons and, in particular, a veracious description of the pion as both a Goldstone mode and a bound state of dressed-quarks \cite{Roberts:2007ji}.  It is a valid impulse approximation, which provides a systematically improvable continuum prediction for nucleon form factors.

Subject to this understanding the question of whether a connection exists between the spatial distribution of charge or magnetisation and the three-dimensional Fourier transform of a Sachs form factor involves a consideration of recoil-corrections experienced by dressed-quarks.  The interpretation is appropriate if recoil corrections are small and can be calculated perturbatively.  In that case the relevant expectation values in quantum mechanics are validly approximated by the Fourier transform of the Sachs form factor.

Consider the Breit frame and a photon probe with momentum $Q=(0,0,q,0)$.  In the scattering process this momentum is absorbed by the dressed-quarks within the proton.  It is elastic scattering so all the dressed-quarks must recoil together, which means they can each be considered as absorbing a momentum fraction\footnote{Faddeev and Bethe-Salpeter amplitudes are peaked at zero relative momentum.  Hence, the domain of greatest support in the impulse approximation calculation is that with each quark absorbing $Q/3$.  This is demonstrated explicitly, e.g., in Ref.\,\protect\cite{Roberts:1994hh}.  
}  $Q/3$.  The magnitude of a recoil correction is then measured by the mass-squared scale
\begin{equation}
s_r:=\frac{q^2}{9} \,.
\end{equation}

We will consider that recoil corrections are \emph{small} so long as 
\begin{equation}
\label{recoilsmall}
s_r < \frac{1}{9} M^2(s_r)\,,
\end{equation}
where $M(s)$ is the dressed-quark mass function.  This constraint means 
\begin{equation}
\label{recoilbound}
q \lsim  \, M(\frac{q^2}{9}) \; \Rightarrow \; q \lsim 0.4\,{\rm GeV}\,,
\end{equation}
a value determined from Eqs.\,(\ref{ZMdef}) -- (\ref{tableA}).  This momentum bound corresponds to a length-scale
\begin{equation}
\lambda_{\rm 0.1} = 0.49\,{\rm fm} = 0.57\,r_p\,,
\end{equation}
where $r_p$ is the proton's charge radius.  Hence in the three-dimensional Fourier transform of a Sachs form factor, recoil corrections are on the order of 10\% or less throughout the domain $r \gsim 0.57\,r_p$; namely, over 81\% of the nucleons' volume.

In measuring the total charge one must evaluate
\begin{equation}
Q_N= \lim_{a\to 0} 4 \pi \int_a^\infty dr\, r^2 \rho_N(r)\,.
\end{equation}
It is interesting to reckon the amount of charge that is contained within the domain on which recoil corrections are not negligible.  It is
\begin{equation}
\label{Qex}
Q_N^{0.1} = 4 \pi \int_0^{0.57 r_p} dr\, r^2 \rho_N(r)\,.
\end{equation}
A Gau{\ss}ian charge form factor can be used to obtain an algebraic and hence easily understood estimate; viz., 
\begin{equation}
G_p(q^2) = {\rm e}^{- \frac{1}{6} q^2 r_p^2},
\end{equation}
yields
\begin{equation}
\rho_p(r) = \frac{3 \sqrt{6 \pi}}{4\pi^2 r_p^3}\, {\rm e}^{- 3 r^2/[2 r_p^2]}\,,
\end{equation}
from which follows
\begin{equation}
Q_p^{\rm 0.1} = 0.19\,.
\end{equation}
It is apparent that this region contains only 19\% of the proton's charge.  Expressed another way, the domain on which recoil corrections can be neglected contains 81\% of the proton's charge.  (For the neutron's charge form factor the illustration can be made using a difference of two Gau{\ss}ians, each of which may be said to represent either the $u$- or $d$-quark contribution to the form factor.)  If instead of Eq.\,(\ref{recoilsmall}) one were to consider recoil corrections as small for $s_r < M^2(s_r)/6$, then the upper bound in Eq.\,(\ref{Qex}) is $0.48r_p$ and the region contains only 12\% of the proton's charge.

On the other hand, recoil corrections are certainly large and essentially nonperturbative for
\begin{equation}
s_r \gtrsim M^2(s_r)\; \Rightarrow \; q \gtrsim 1\,{\rm GeV}\,,
\end{equation}
a momentum boundary which corresponds to lengths
\begin{equation}
\lambda_{\rm 1.0} \lesssim 0.2\,{\rm fm} = 0.23\,r_p\,.
\end{equation}
On this domain no quantum mechanical connection can be made between three-dimensional Fourier transforms of Sachs form factors and the density distribution of distinct charge and current carriers.  It corresponds to 1.2\% of the nucleon's volume and contains just 1.6\% of the proton's charge.

This analysis elucidates the circumstances under which the three-dimensional Fourier transform of a Breit-frame Sachs form factor can be viewed as providing a useful, qualitatively and semi-quantitatively reliable description of the configuration space distribution of a nucleon's charge or magnetisation over dressed-quarks.  Dressed-quarks are an emergent feature of QCD.  The requisite conditions pertain within 81\% -- 99\% of a nucleon's volume.  Moreover, notwithstanding any caveats, Poincar\'e invariant form factors are always a gauge of a hadron's structure because they are a measurable and physical manifestation of the nature of the hadron's constituents and the dynamics that binds them together.  

\section{Calculated Form Factors}
\label{FFall}
In the following two sections we present and discuss the results that our model of the dressed-quark core produces for nucleon form factors.  Notably, we made significant modifications to the computer codes used to obtain the results in Ref.\,\cite{Alkofer:2004yf}.  In addition to that described in App.~D, which defines a convergent continuation of the Faddeev amplitude into the Breit frame, we succeeded in reducing execution times by an order of magnitude.   These two improvements enabled us to use a desk-top computer and obtain, within hours, numerically accurate results for the form factors on the domain $Q^2 \in [0,12]\,$GeV$^2$.

In order to explain our results we must introduce our notation.  The Pauli, Dirac and Sachs form factors are all represented by their usual symbols.  Hence, the notation can be introduced by a single example.  We choose the proton's Dirac form factor, $F_1^p$, and list the definitions in \ref{FFN}~\emph{Form Factor Notation}.

It is also worth noting here that our analysis assumes $m_u=m_d$.  Hence the only difference between the $u$- and $d$-quarks is their electric charge.  Our equations, computer codes and results therefore exhibit the following charge symmetry relations:
\begin{equation}
\label{CSrelations}
e_d F_i^{p,u} = e_u F_i^{n,d}\,, \; 
e_u F_i^{p,d} = e_d F_i^{n,u}\,, \; i=1,2.
\end{equation}

\section{Proton Form Factors}
\label{FFp}
\subsection{Dirac proton}
\begin{figure}[t]
\begin{minipage}[t]{\textwidth}
\begin{minipage}[t]{0.49\textwidth}
\leftline{\includegraphics[width=0.99\textwidth,height=14em]{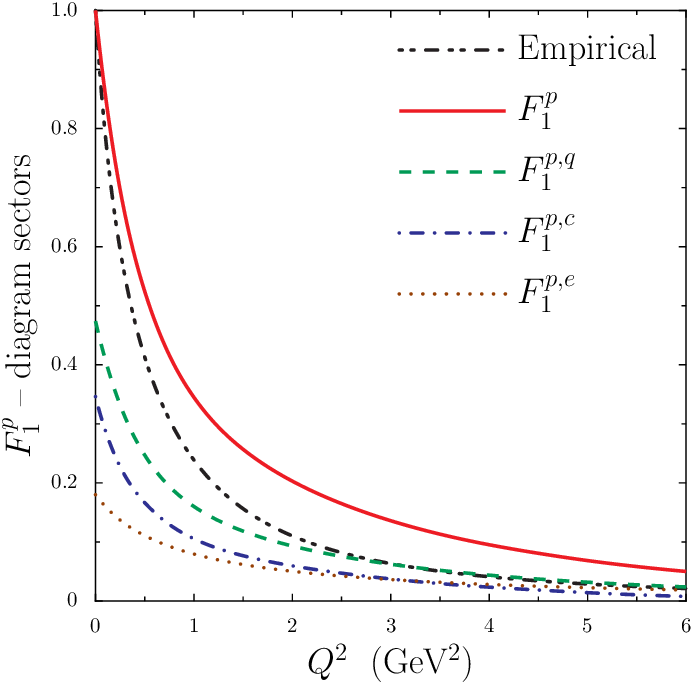}}
\end{minipage}
\begin{minipage}[t]{0.49\textwidth}
\rightline{\includegraphics[width=0.99\textwidth,height=14em]{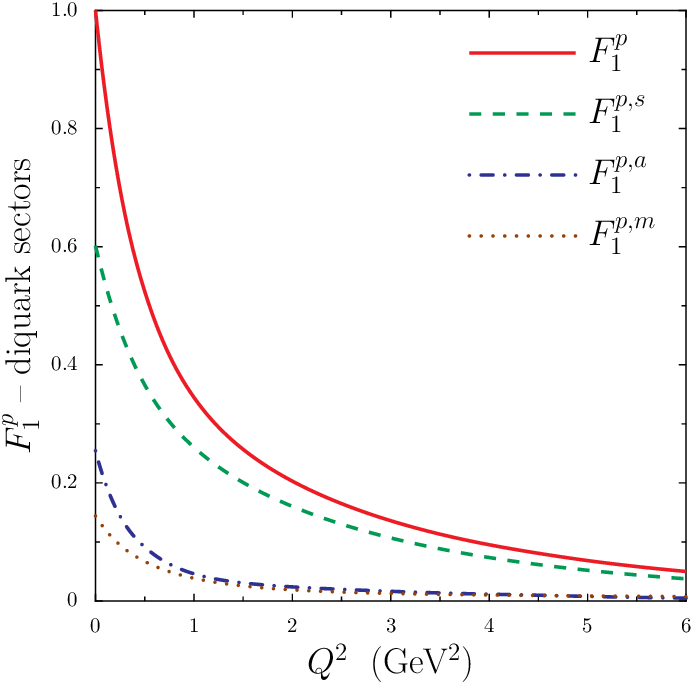}}
\end{minipage}\vspace*{2ex}

\begin{minipage}[t]{0.49\textwidth}
\leftline{\includegraphics[width=0.99\textwidth,height=14em]{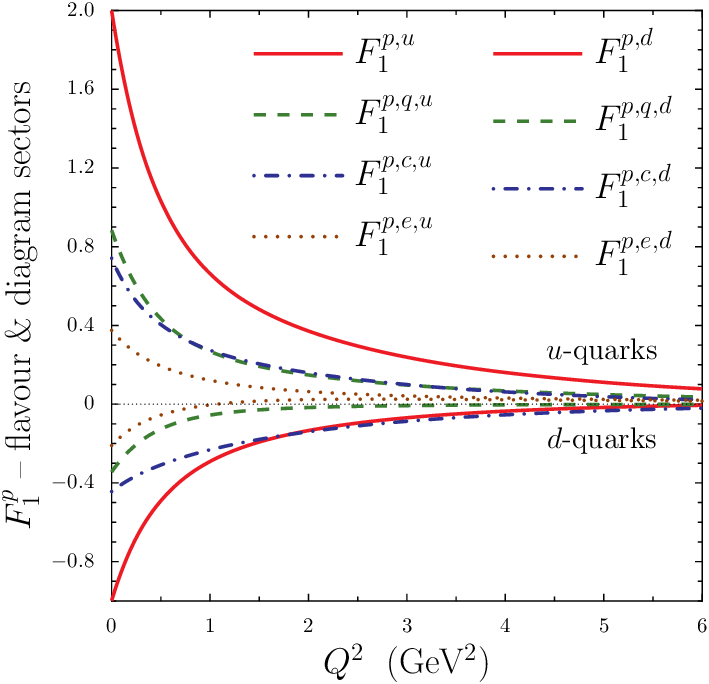}}
\end{minipage}
\begin{minipage}[t]{0.49\textwidth}
\rightline{\includegraphics[width=0.99\textwidth,height=14em]{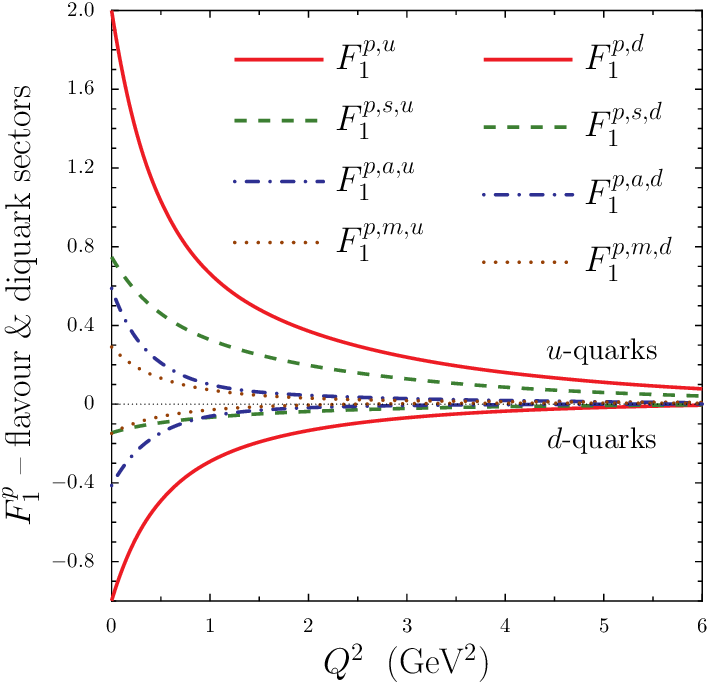}}
\end{minipage}
\end{minipage}

\caption{\label{fig:protonF1} Proton's Dirac form factor. \emph{Upper left} -- Full result and decomposition according to diagram classes; \emph{lower left} -- flavour breakdown of these contributions, expressed in units of the magnitude of the relevant quark's charge; viz., $|e_u|=\frac{2}{3}$ or $|e_d|=\frac{1}{3}$.  \emph{Upper right} -- Full result and decomposition according to diagram diquark content; \emph{lower right} -- flavour breakdown of these contributions.  A parametrisation of experimental data \protect\cite{Kelly:2004hm} is also presented in the \emph{upper left} panel.  A full explanation of the notation is provided in App.\,E.  
} 
\end{figure}

In Fig.\,\ref{fig:protonF1} we depict the proton's Dirac form factor and a breakdown into contributions from various subclasses of diagrams.  The figures deserve careful study.  

The \emph{upper left} panel shows the $Q^2$-evolution of the quark, diquark and exchange (or \emph{two-body}) contributions to the form factor.  Their $Q^2=0$ values measure, respectively, the probability that the photon interacts with a bystander quark or a diquark correlation, or acts in association with diquark breakup: 
\begin{equation}
{\rm quark-}P_1^{p,q}=0.47 \;:\; {\rm diquark-}P_1^{p,c}=0.35 \;:\; {\rm exchange-}P_1^{p,e}=0.18.
\label{probdef}
\end{equation}
These and analogous probabilities are collected in Table~\ref{tableprob}.  For $F_1^p$ the diquark and exchange contributions switch in importance at $Q^2 \sim 3\,$GeV$^2$.  Moreover, while the net result is always positive, the diquark contribution becomes negative at $Q^2\sim 9\,$GeV$^2$.  This panel here, and in kindred figures to follow, also displays a parametrisation of experimental results \cite{Kelly:2004hm} for illustrative comparison with our computation.  The manner by which that comparison should be understood is canvassed in Sect.\,\ref{sec:chiral}.

\begin{table}[t]
\begin{center}
\caption{\label{tableprob} Probabilities associated with the $F_1^p$ form factors evaluated at $Q^2=0$, defined in App.\,E.}
\begin{tabular*}{1.0\textwidth}{
c@{\extracolsep{0ptplus1fil}}
c@{\extracolsep{0ptplus1fil}}c|@{\extracolsep{0ptplus1fil}}
c@{\extracolsep{0ptplus1fil}}c@{\extracolsep{0ptplus1fil}}
c@{\extracolsep{0ptplus1fil}}
}
\hline
$P_1^{p,q}$ & $P_1^{p,c}$ & $P_1^{p,e}$ & $P_1^{p,s}$ & $P_1^{p,a}$ & $P_1^{p,m}$ \\
0.474 & 0.346 & 0.180 & 0.602 & 0.254 & 0.144 \\\hline
$P_1^{p,q,u}$ & $P_1^{p,c,u}$ & $P_1^{p,e,u}$ & $P_1^{p,s,u}$ & $P_1^{p,a,u}$ & $P_1^{p,m,u}$\\
0.441 & 0.371 & 0.188 & 0.561 & 0.294 & 0.145 \\\hline
$P_1^{p,q,d}$ & $P_1^{p,c,d}$ & $P_1^{p,e,d}$ & $P_1^{p,s,d}$ & $P_1^{p,a,d}$ & $P_1^{p,m,d}$\\
0.345 & 0.444 & 0.210 & 0.437 & 0.414 & 0.149\\\hline
\end{tabular*}
\end{center}
\end{table}

A radius can be associated with each of the form factors.  We exemplify its definition via $F_1^{p,q}$; viz., 
\begin{equation}
\label{radiusexample}
(r_1^{p,q})^2 := - \frac{6}{F_1^{p,q}(0)} \left.\frac{d}{dQ^2} F_1^{p,q}(Q^2)\right|_{Q^2=0},
\end{equation}
and remark that
\begin{equation}
(r_1^p)^2 = P_1^{p,q} (r_1^{p,q})^2 + P_1^{p,c} (r_1^{p,c})^2 + P_1^{p,e} (r_1^{p,e})^2. 
\end{equation}
The calculated Dirac radii are reported in Table~\ref{radii}.  Their values emphasise that so far as the Dirac form factor is concerned, the diquark component of the nucleon is softest.

\begin{table}[b]
\begin{center}
\caption{\label{radii} Radii associated with $F_1^p$, defined by analogy with Eq.\,(\protect\ref{radiusexample}).  All entries in fm.}
\begin{tabular*}{1.0\textwidth}{
c|@{\extracolsep{0ptplus1fil}}c@{\extracolsep{0ptplus1fil}}
c@{\extracolsep{0ptplus1fil}}c|@{\extracolsep{0ptplus1fil}}
c@{\extracolsep{0ptplus1fil}}c@{\extracolsep{0ptplus1fil}}
c@{\extracolsep{0ptplus1fil}}
}
\hline
$r_1^p$ & $r_1^{p,q}$ & $r_1^{p,c}$ & $r_1^{p,e}$ & $r_1^{p,s}$ & $r_1^{p,a}$ & $r_1^{p,m}$ \\
0.615 & 0.598 & 0.673 & 0.537 & 0.526 & 0.766 & 0.623\\\hline
$r_1^{p,u}$ & $r_1^{p,q,u}$ & $r_1^{p,c,u}$ & $r_1^{p,e,u}$ & $r_1^{p,s,u}$ & $r_1^{p,a,u}$ & $r_1^{p,m,u}$\\
0.617 & 0.620 & 0.615 & 0.614 & 0.520 & 0.749 & 0.656 \\\hline
$r_1^{p,d}$ & $r_1^{p,q,d}$ & $r_1^{p,c,d}$ & $r_1^{p,e,d}$ & $r_1^{p,s,d}$ & $r_1^{p,a,d}$ & $r_1^{p,m,d}$\\
0.624 & 0.696 & 0.454 & 0.745 & 0.494 & 0.715 & 0.665\\\hline
\end{tabular*}
\end{center}
\end{table}

The \emph{lower left} panel provides a flavour decomposition of the quark, diquark and exchange contributions to the form factor.  While the other two $u$-quark components are  positive definite, $F_1^{p,c,u}$ changes sign at $Q^2\sim 9\,$GeV$^2$.  
Up quarks are doubly represented in the proton and from Table~\ref{tableprob} it is evident that they are almost equally likely to be struck by a photon whether a bystander or a diquark participant.  This explains the near equality of the radii associated with each term in the subclass of these form factor contributions in which a $u$-quark is struck.
The same is not true for the $d$-quark, for which the probabilities show that it is more likely to be struck while a diquark participant.  This signals that the $d$-quark is less free to move throughout the proton's volume and hence explains the small value of $r_1^{p,c,d}$.

The \emph{upper right} panel of Fig.\,\ref{fig:protonF1} shows the $Q^2$-evolution of the contributions to $F_1^p$ that involve a scalar diquark, an axial-vector diquark, or one of each.  It is clear from Table~\ref{tableprob} that the scalar diquark component of the proton is dominant.  All contributions are positive definite, and the relative strength of the axial-vector and mixed contributions switches at $Q^2\sim 5\,$GeV$^2$.  From Table~\ref{radii} one reads that the softest contribution to the proton's Dirac form factor is provided by diagrams involving an axial-vector diquark.  One can picture this as stemming from the axial-vector correlation being more massive than the scalar and hence a bystander quark of any flavour ranges further from a collective centre-of-mass.

The \emph{lower right} panel provides a flavour decomposition of the diquark contributions just discussed.  All $u$-quark components are positive definite.  For the singly-represented $d$-quark, however, each of the form factors changes sign: $F_1^{p,s,d}$ becomes positive at $Q^2\sim 8\,$GeV$^2$; $F_1^{p,a,d}$ at $Q^2\sim 5\,$GeV$^2$; and $F_1^{p,m,d}$ at $Q^2\sim 3\,$GeV$^2$.  Axial-vector contributions to the Dirac form factor are the softest in each flavour sector.

Evident in Table~\ref{radii} is a notable feature of our calculation; viz., 
\begin{equation}
\label{r1dr1u}
r_1^{p,d}>r_1^{p,u}.
\end{equation}
Owing to charge symmetry this entails 
\begin{equation}
\label{r1nund}
r_1^{n,u}>r_1^{n,d},
\end{equation}
a result also obtained and explained in Ref.\,\cite{Eichmann:2008ef}.  Equation~(\ref{r1dr1u}) follows from the presence of axial-vector diquark correlations in the nucleon.  One reads from Table~\ref{tableprob} that the proton's singly represented $d$-quark is more likely to be struck in association with an axial-vector diquark correlation than with a scalar, and form factor contributions involving an axial-vector diquark are soft.  On the other hand, the doubly-represented $u$-quark is predominantly linked with harder scalar-diquark contributions.

\subsection{Pauli proton}
\begin{figure}[t]
\begin{minipage}[t]{\textwidth}
\begin{minipage}[t]{0.49\textwidth}
\leftline{\includegraphics[width=0.99\textwidth]{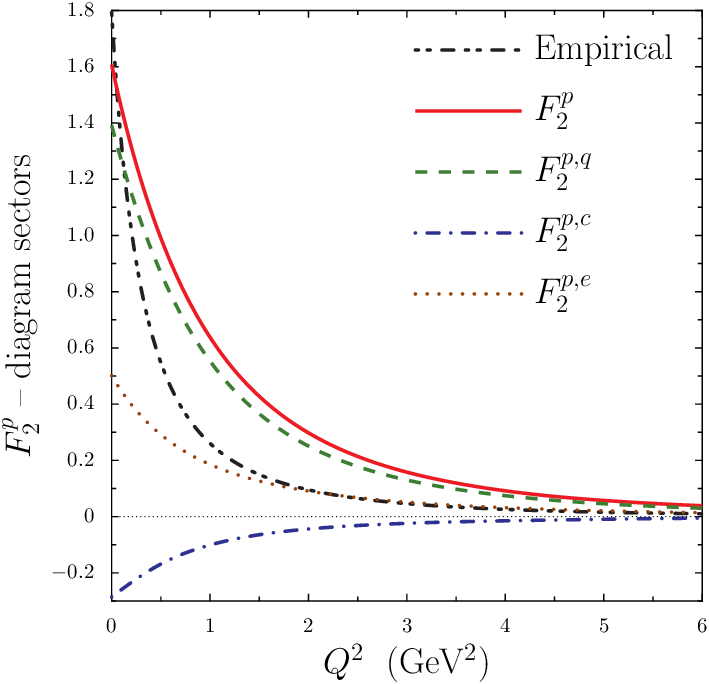}}
\end{minipage}
\begin{minipage}[t]{0.49\textwidth}
\rightline{\includegraphics[width=0.99\textwidth]{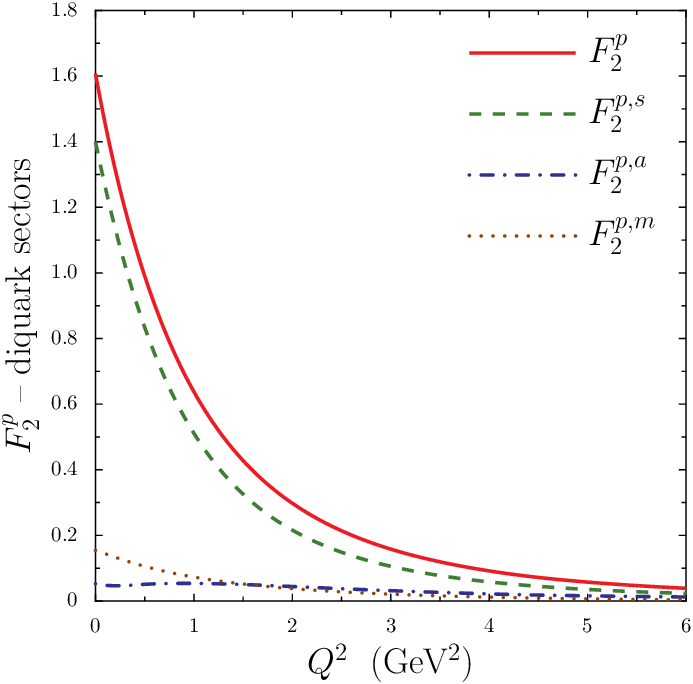}}
\end{minipage}
\end{minipage}

\caption{\label{fig:protonF2} Proton's Pauli form factor. \emph{Left panel} -- Full result and decomposition according to diagram classes; \emph{right panel} -- full result and decomposition according to diagram diquark content.  Form factors are expressed in magnetons defined by the calculated nucleon mass, $M_N$ in Table~\protect\ref{tableNmass}.  A parametrisation of experimental data \protect\cite{Kelly:2004hm} is also presented in the \emph{left panel}.  A full explanation of the notation is provided in App.\,E.  
} 
\end{figure}

\begin{figure}[t]
\begin{minipage}{\textwidth}
\begin{minipage}[t]{0.49\textwidth}
\leftline{\includegraphics[width=0.99\textwidth]{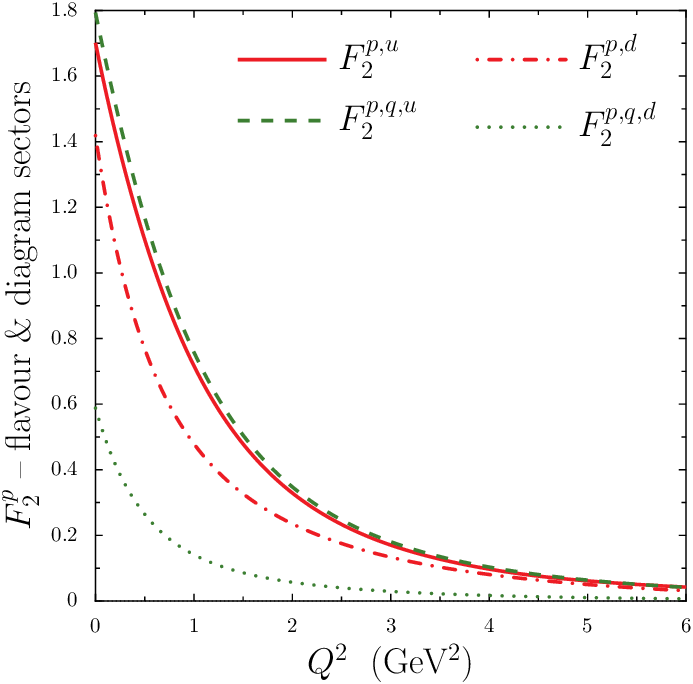}}
\end{minipage}
\begin{minipage}[t]{0.49\textwidth}
\rightline{\includegraphics[width=0.99\textwidth]{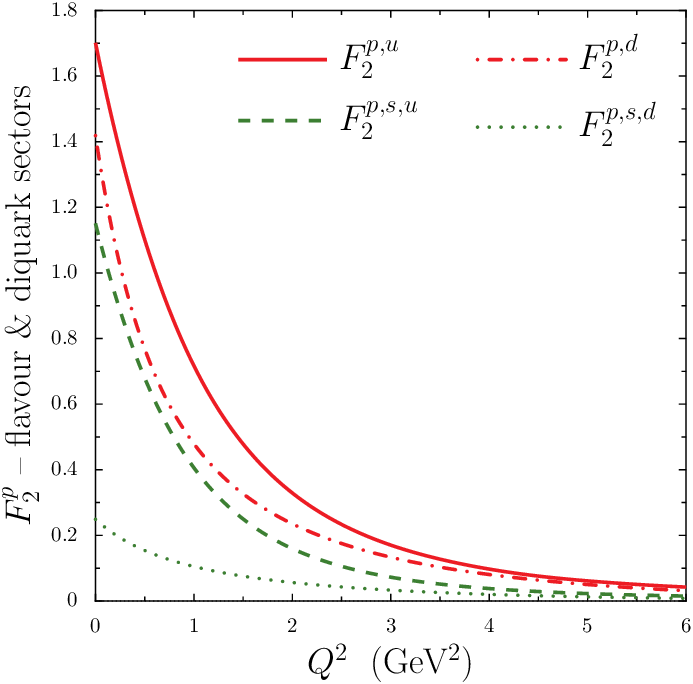}}
\end{minipage}\vspace*{2ex}

\begin{minipage}[t]{0.49\textwidth}
\leftline{\includegraphics[width=0.99\textwidth]{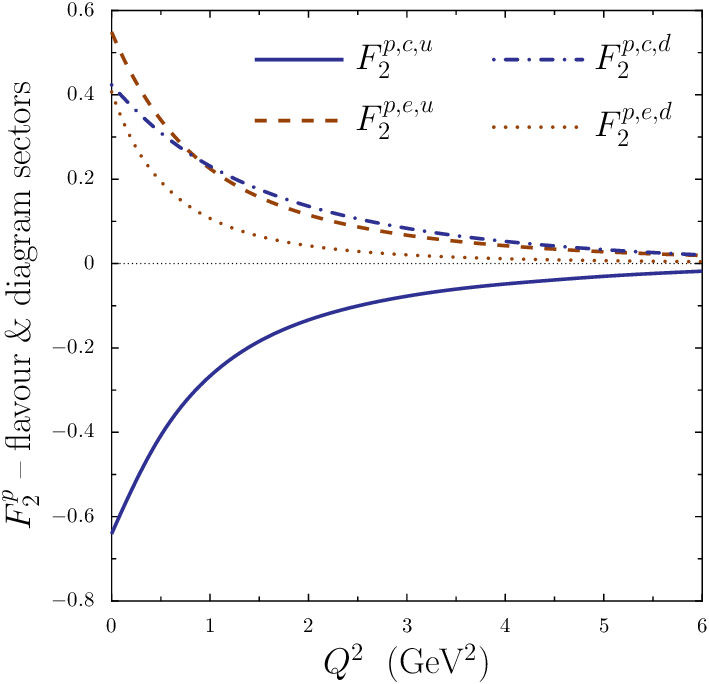}}
\end{minipage}
\begin{minipage}[t]{0.49\textwidth}
\rightline{\includegraphics[width=0.99\textwidth]{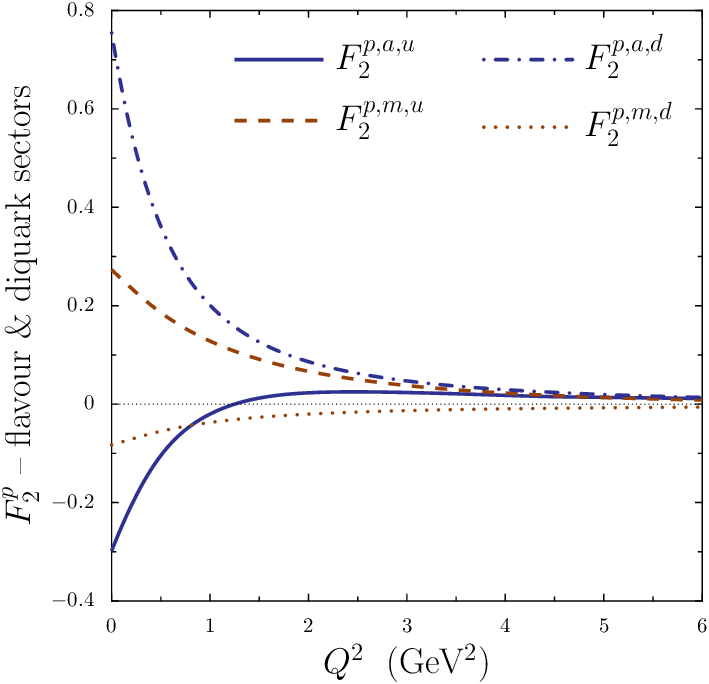}}
\end{minipage}
\end{minipage}

\caption{\label{fig:protonF2A} Proton's Pauli form factor: \emph{left panels} -- flavour breakdown of left panel in Fig.\,\ref{fig:protonF2}; and \emph{right panels} -- flavour breakdown of right panel in Fig.\,\ref{fig:protonF2}.  A full explanation of the notation is provided in App.\,E.  
} 
\end{figure}

In Figs.\,\ref{fig:protonF2} and \ref{fig:protonF2A} we depict the proton's Pauli form factor and a breakdown into contributions from various subclasses of diagrams.  

\begin{table}[b]
\begin{center}
\caption{\label{tablepmag} Flavour and diagram breakdown of contributions to the proton's anomalous magnetic moment; viz., the $F_2^p$ form factors evaluated at $Q^2=0$, measured in magnetons defined by the calculated nucleon mass, $M_N$.}
\begin{tabular*}{1.0\textwidth}{
c|@{\extracolsep{0ptplus1fil}}
c@{\extracolsep{0ptplus1fil}}
c@{\extracolsep{0ptplus1fil}}c|@{\extracolsep{0ptplus1fil}}
c@{\extracolsep{0ptplus1fil}}c@{\extracolsep{0ptplus1fil}}
c@{\extracolsep{0ptplus1fil}}
}
\hline
$\kappa_p$ & $\kappa_p^{q}$ & $\kappa_p^{c}$ & $\kappa_p^{e}$ 
            & $\kappa_p^{s}$ & $\kappa_p^{a}$ & $\kappa_p^{m}$ \\
1.674 & 1.445 & -0.297 &  0.526 & 1.460 & 0.0556 & 0.159~  \\\hline
$\kappa_p^u$ & $\kappa_p^{q,u}$ & $\kappa_p^{c,u}$ & $\kappa_p^{e,u}$ & $\kappa_p^{s,u}$ & $\kappa_p^{a,u}$ & $\kappa_p^{m,u}$\\
1.174 & 1.235 & -0.441 & 0.381 &  1.199 & -0.211~~ & 0.187~  \\\hline
$\kappa_p^d$ & $\kappa_p^{q,d}$ & $\kappa_p^{c,d}$ & $\kappa_p^{e,d}$ & $\kappa_p^{s,d}$ & $\kappa_p^{a,d}$ & $\kappa_p^{m,d}$\\
0.500 & 0.210 & ~0.145 & 0.145 & 0.260 & 0.268~  & -0.0284  \\\hline
\end{tabular*}
\end{center}
\end{table}

The \emph{left panel} of Fig.\,\ref{fig:protonF2} shows the $Q^2$-evolution of the quark, diquark and exchange contributions to the form factor.  Listed in Table~\ref{tablepmag}, their $Q^2=0$ values measure, respectively, the contribution to the proton's anomalous magnetic moment from the photon interacting with a bystander quark, a diquark or in association with diquark breakup.  The net contribution from Diagrams~2 and 4 in Fig.\,\ref{vertex} is negative.  This remains the case until $Q^2\sim 9\,$GeV$^2$, at which point the net diquark contribution changes sign, as was also the case in $F_1^p$.  The Pauli radii are listed in Table~\ref{radiiF2}, from which it is evident that Diagrams~3, 5 and 6 in Fig.\,\ref{vertex} provide the softest contribution.

\begin{table}[b]
\begin{center}
\caption{\label{radiiF2} Radii associated with $F_2^p$, defined by analogy with Eq.\,(\protect\ref{radiusexample}).  All entries in fm.}
\begin{tabular*}{1.0\textwidth}{
c|@{\extracolsep{0ptplus1fil}}c@{\extracolsep{0ptplus1fil}}
c@{\extracolsep{0ptplus1fil}}c|@{\extracolsep{0ptplus1fil}}
c@{\extracolsep{0ptplus1fil}}c@{\extracolsep{0ptplus1fil}}
c@{\extracolsep{0ptplus1fil}}
}
\hline
$r_2^p$ & $r_2^{p,q}$ & $r_2^{p,c}$ & $r_2^{p,e}$ 
    & $r_2^{p,s}$ & $r_2^{p,a}$ & $r_2^{p,m}$ \\
0.490 & 0.469 & 0.456 & 0.528 & 0.493 & 0.596 & 0.399 \\\hline
$r_2^{p,u}$ & $r_2^{p,q,u}$ & $r_2^{p,c,u}$ & $r_2^{p,e,u}$ & $r_2^{p,s,u}$ & $r_2^{p,a,u}$ & $r_2^{p,m,u}$\\
0.449 & 0.432 & 0.434 & 0.485 & 0.489 & 0.573 & 0.399 \\\hline
$r_2^{p,d}$ & $r_2^{p,q,d}$ & $r_2^{p,c,d}$ & $r_2^{p,e,d}$ & $r_2^{p,s,d}$ & $r_2^{p,a,d}$ & $r_2^{p,m,d}$\\
0.577 & 0.644 & 0.378 & 0.628 & 0.511 & 0.616 & 0.394\\\hline
\end{tabular*}
\end{center}
\end{table}

The \emph{left panels} in Fig.\,\ref{fig:protonF2A} provide a flavour decomposition of the quark, diquark and exchange contributions to the proton's Pauli form factor.  We remark that $F_2^{p,q,u}$ is positive definite whereas $F_2^{p,c,u}$ changes sign at $Q^2\gtrsim 10\,$GeV$^2$ and $F_2^{p,e,u}$ at $Q^2\gtrsim 17\,$GeV$^2$.  (The latter should be interpreted qualitatively because our calculations are not truly reliable beyond $12\,$GeV$^2$.)
It is evident upon comparison between Tables~\ref{radii} and \ref{radiiF2} that the pattern exhibited by the Pauli radii is kindred to that of the Dirac radii, with the origin alike.

The \emph{right panel} of Fig.\,\ref{fig:protonF2} shows the $Q^2$-evolution of the contributions to $F_2^p$ that involve a scalar diquark, an axial-vector diquark, or one of each.  It is apparent from the figure and Table~\ref{tablepmag} that diagrams involving the scalar correlation are dominant on a material $Q^2$ domain.  These contributions to a nucleon's Faddeev amplitude have the simplest rest-frame spin--angular-momentum structure \cite{Cloet:2007pi,Oettel:1998bk}.
We find that the scalar and axial-vector contributions are positive definite whereas the mixed contribution changes sign at $Q^2\gtrsim 8\,$GeV$^2$.  The latter provides a larger contribution to the proton's magnetic moment than the axial-vector diagram.  
One reads from Table~\ref{radiiF2} that the softest contribution to the proton's Pauli form factor is provided by the axial-vector diquark diagrams.  This was also the case for the Dirac form factor.  However, in contrast to $F_1^p$, the mixed contribution to $F_2^p$ is hardest, a result which owes primarily to Diagram~4 and the simple \emph{Ansatz} we have made for the interaction therein; viz., Eq.\,(\ref{calTvalue}).

The \emph{right panels} of Fig.\,\ref{fig:protonF2A} provide a flavour decomposition of the diquark contributions just discussed.  It is curious that $\kappa_p^{a,u}<0$, a feature which highlights the presence and role of correlations in the nucleon's Faddeev amplitude.  The associated form factor becomes positive at $Q^2\approx 1.5\,$GeV$^2$.  The contribution with the simplest structure, $F_2^{p,s,u}$, is positive definite whereas $F_2^{p,m,u}$ becomes negative at $Q^2\gtrsim 10\,$GeV$^2$.  
In association with the proton's $d$-quark, the axial-vector diagrams make a positive definite contribution, the scalar diquark form factor becomes negative at $Q^2\approx 12\,$GeV$^2$ and the mixed contribution is negative definite but small.

It is apparent from Table~\ref{radiiF2} that 
\begin{equation}
\label{r2dr2u}
r_2^{p,d}>r_2^{p,u},
\end{equation}
which entails $r_2^{n,u}>r_2^{n,d}$.  These orderings are the same as those exhibited by the Dirac radii, Eq.\,(\ref{r1dr1u}), but the separation in magnitudes is larger.  The presence of axial-vector diquark correlations again plays a large role in producing these results.  We note in addition that $r_2^{p,u}<r_1^{p,u}$ and $r_2^{p,d}<r_1^{p,d}$, with the greater reduction in $r_2^{p,u}$.  Indeed, it is almost uniformly true that the quark-core Pauli form factors are harder than their Dirac counterparts.  The reduction is marked for $r_2^{p,a,u}$ and the only exception to the rule is $r_2^{p,s,d}$.  

\subsection{Pauli--Dirac proton ratio}
\begin{figure}[t]
\leftline{%
\includegraphics[clip,width=0.75\textwidth]{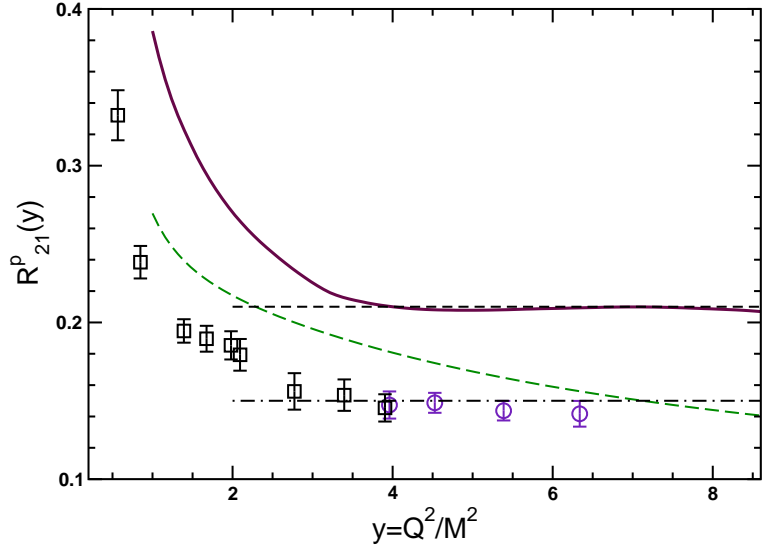}}
\caption{\label{figF2F1} \emph{Solid curve} -- Calculated dimensionless ratio in Eq.\,(\protect\ref{jiratio}) with $\hat \Lambda = 0.44$ and $M_N$ in Table~\protect\ref{tableNmass}. \emph{Short-dashed line} -- constant at $0.21$.  \emph{Long dashed curve} -- Ratio evaluated with the experimental nucleon mass using the parametrisations in Ref.\,\protect\cite{Kelly:2004hm}.  
\emph{Boxes} -- Ratio evaluated with data from Ref.\,\protect\cite{jones}; and 
\emph{Circles} -- from Ref.\,\protect\cite{gayou}.  \emph{Dash-dot line} -- constant at $0.15$.
}
\end{figure}

In Fig.\,\ref{figF2F1} we plot a weighted ratio of Pauli to Dirac form factors; viz., 
\begin{equation}
\label{jiratio}
R_{21}^p(\hat Q^2):=\frac{\hat Q^2}{(\ln [\hat Q^2/\hat \Lambda^2])^2}\frac{F_2^p(\hat Q^2)}{F_1^p(\hat Q^2)}\,,\;
\hat Q^2 = \frac{Q^2}{M_N^2}\,,\;\hat \Lambda^2 = \frac{\Lambda^2}{M_N^2}.
\end{equation}
A perturbative analysis that considers effects arising from both the proton's leading- and subleading-twist light-cone wave functions, the latter of which represents quarks with one unit of orbital angular momentum, suggests that this ratio should be constant for $Q^2\gg \Lambda^2$, where $\Lambda$ is a mass-scale that is said to correspond to an upper-bound on the domain of 
soft momenta \cite{belitsky}.  

We analysed our calculated result in this context and found that with $\hat \Lambda = 0.44$ this weighted ratio is a constant$\,=0.21$ on $\hat Q^2 \geq 4.3$; by which we mean that the rms relative error with respect to the straight-line fit is $0.34$\% with an associated standard deviation of $0.24$\%.  These numbers increase as the minimum value of $\hat Q^2$ included in the fit is decreased and, moreover, the value of $\hat \Lambda$ comes to depend on this minimum value.

In the figure we also plot the ratio in Eq.\,(\ref{jiratio}) as evaluated from extant experimental data, available on the domain $\hat Q^2 \in [3.9,6.3]$.  Using $\hat \Lambda = 0.44$, the result is described by a constant$\,=0.15$ with rms relative error $1.5$\% and an associated standard deviation of $0.98$\%.  It is evident in the figure that on the domain for which the ratio is well described by a constant, our model produces a result that lies above the experimental data.  This is because thereupon our calculation underestimates experimental results for $F_1^p$ by $\sim 15$\% and overestimates those for $F_2^p$ by a similar amount.  (See Sect.\,\ref{sec:chiral} for details.)

It is curious that what might appear to be a low mass-scale, $\Lambda = 0.44 M_N$, should serve to produce a constant value for the ratio in Eq.\,(\ref{jiratio}) \cite{Alkofer:2004yf}.  In seeking to understand the origin of this scale we analysed the pointwise behaviour of our calculated Faddeev amplitude.  The dominant functions are $s_1$, $A_3$, $A_5$, which was to be expected given the associated Dirac structures (see Eqs.\,(\ref{Sexp}) -- (\ref{Afunctions}).)  A Gau{\ss}ian can be fitted to the leading Chebyshev moment of each of these functions.  That procedure yields the following widths (in units of $M_N$): 
\begin{equation}
\omega_{s_1^1} = 0.48 \,,\;
\omega_{A_3^1} = 0.47 \,,\;
\omega_{A_5^1} = 0.46 \,.
\label{FAwidths}
\end{equation}
The similarity between these widths and $\Lambda$ is notable.  It highlights the point that while $\Lambda$ \emph{per se} is not an elemental input to our calculation, such a mass-scale can arise dynamically as a derivative quantity which may be expressed in the relative-momentum support of the Faddeev amplitude.  A challenge now is to determine whether an algebraic relationship exists between $\Lambda$ in Eq.\,(\ref{jiratio}) and the widths characterising the Faddeev amplitude.  

\subsection{Sachs proton electric}
\begin{figure}[t]
\begin{minipage}[t]{\textwidth}
\begin{minipage}[t]{0.49\textwidth}
\leftline{\includegraphics[width=0.99\textwidth]{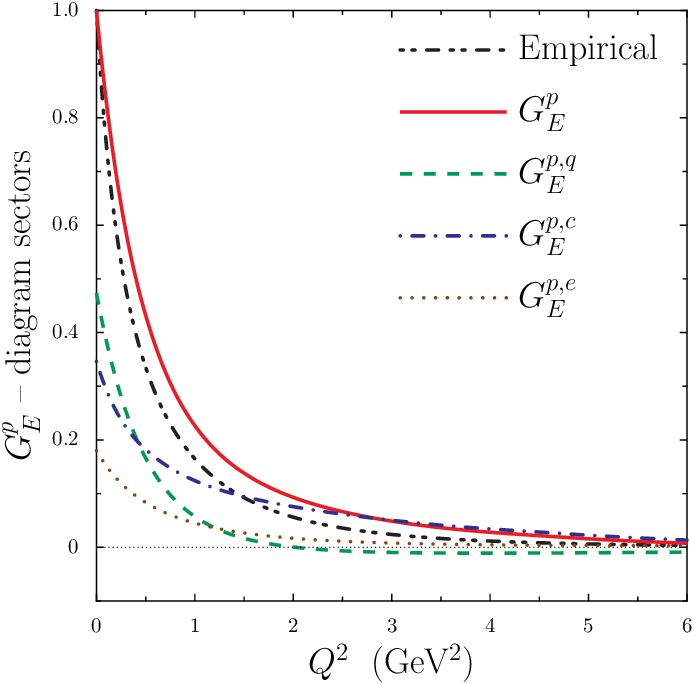}}
\end{minipage}
\begin{minipage}[t]{0.49\textwidth}
\rightline{\includegraphics[width=0.99\textwidth]{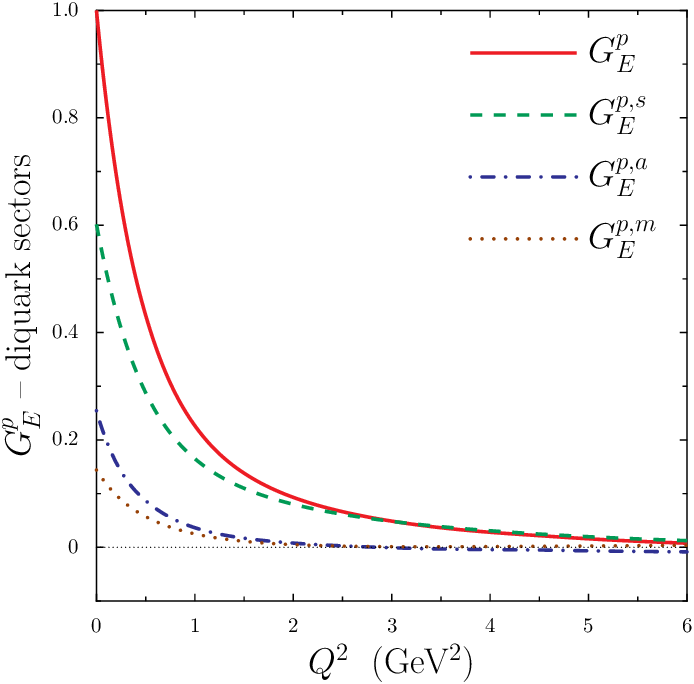}}
\end{minipage}\vspace*{2ex}

\begin{minipage}[t]{0.49\textwidth}
\leftline{\includegraphics[width=0.99\textwidth]{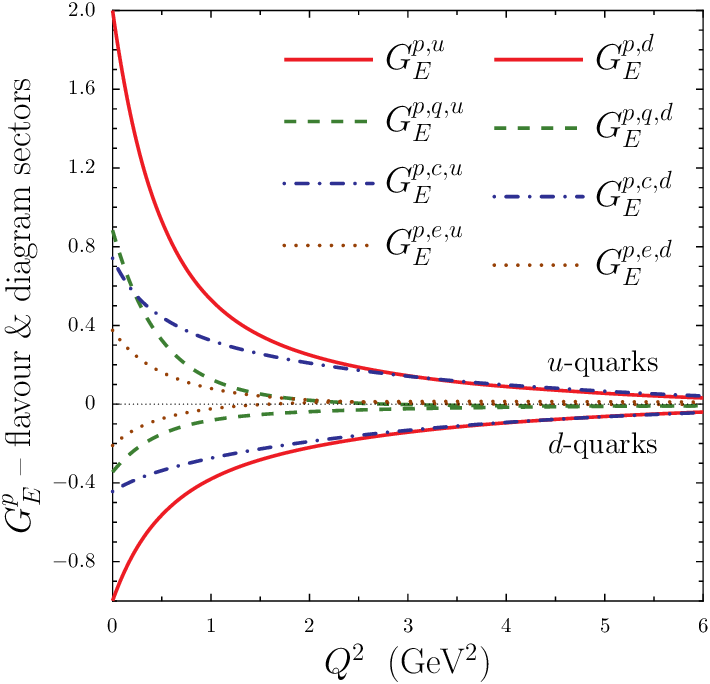}}
\end{minipage}
\begin{minipage}[t]{0.49\textwidth}
\rightline{\includegraphics[width=0.99\textwidth]{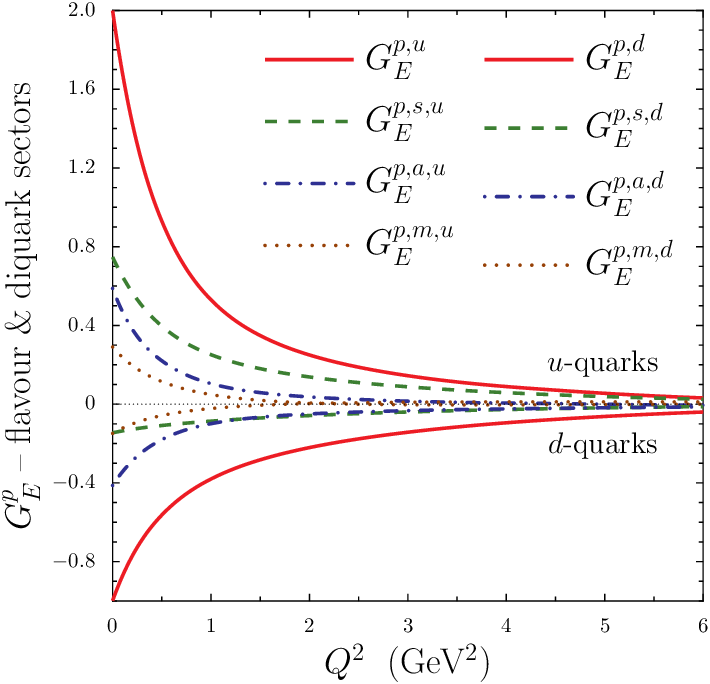}}
\end{minipage}
\end{minipage}

\caption{\label{fig:protonGE} Proton's Sachs electric form factor. \emph{Upper left} -- Full result and decomposition according to diagram classes; \emph{lower left} -- flavour breakdown of these contributions, expressed in units of the magnitude of the relevant quark's charge.  \emph{Upper right} -- Full result and decomposition according to diagram diquark content; \emph{lower right} -- flavour breakdown of these contributions.  A parametrisation of experimental data \protect\cite{Kelly:2004hm} is also presented in the \emph{upper left} panel.  A full explanation of the notation is provided in App.\,E.
} 
\end{figure}

In Fig.\,\ref{fig:protonGE} we present the proton's Sachs electric form factor and a separation into contributions from various subclasses of diagrams.  While in principle, given Eqs.\,(\ref{GEpeq}), these panels contain no information that cannot be constructed from material already presented, they are nevertheless practically useful and informative.

The \emph{upper left} panel shows the $Q^2$-evolution of the quark, diquark and exchange (or \emph{two-body}) contributions to $G_E^p(Q^2)$.  Their $Q^2=0$ values have precisely the same value and interpretation as those associated with the Dirac form factor, which are presented in Table~\ref{tableprob}.  It is notable that the quark contribution; namely, Diagram~1 in Fig.\,\ref{vertex}, possesses a zero at $Q^2\approx 3.0\,$GeV$^2$.  It is only because the diquark contribution remains positive until $Q^2\approx 9.0\,$GeV$^2$ and the exchange contribution is positive definite that the complete result for $G_E^p(Q^2)$ does not exhibit a zero until $Q^2\approx 8.0\,$GeV$^2$.\footnote{A zero in $G_E^p(Q^2)$ was seen in the light-front constituent-quark model of Ref.\,\protect\cite{Frank:1995pv}.  In Ref.\,\protect\cite{Bloch:2003vn} it was shown to be a property of the scalar-diquark Faddeev model of Ref.\,\protect\cite{Bloch:1999ke} but its appearance and location were argued to be dependent on dynamics, consistent with Refs.\,\protect\cite{JuliaDiaz:2003gq,Melde:2007zz} and the present study.}  We list the Sachs radii in Table~\ref{radiiGE}.  In comparison with the Dirac radii in Table~\ref{radii}, they are relatively uniform owing to Foldy-term contributions.

\begin{table}[t]
\begin{center}
\caption{\label{radiiGE} Radii associated with $G_E^p$, defined by analogy with Eq.\,(\protect\ref{radiusexample}).  NB.\ The value in this table yields $M_N r_E^p = 4.01$ cf.\ experiment \cite{Yao:2006px} $M_N r_E^p = 4.18$.  Tabulated entries in fm.}
\begin{tabular*}{1.0\textwidth}{
c|@{\extracolsep{0ptplus1fil}}c@{\extracolsep{0ptplus1fil}}
c@{\extracolsep{0ptplus1fil}}c|@{\extracolsep{0ptplus1fil}}
c@{\extracolsep{0ptplus1fil}}c@{\extracolsep{0ptplus1fil}}
c@{\extracolsep{0ptplus1fil}}
}
\hline
$r_E^p$ & $r_E^{p,q}$ & $r_E^{p,c}$ & $r_E^{p,e}$ & $r_E^{p,s}$ & $r_E^{p,a}$ & $r_E^{p,m}$ \\
0.666 & 0.681 & 0.645 & 0.639 & 0.613 & 0.767 & 0.681 \\\hline
$r_E^{p,u}$ & $r_E^{p,q,u}$ & $r_E^{p,c,u}$ & $r_E^{p,e,u}$ & $r_E^{p,s,u}$ & $r_E^{p,a,u}$ & $r_E^{p,m,u}$\\
0.645 & 0.675 & 0.583 & 0.660 & 0.581 & 0.733 & 0.681 \\\hline
$r_E^{p,d}$ & $r_E^{p,q,d}$ & $r_E^{p,c,d}$ & $r_E^{p,e,d}$ & $r_E^{p,s,d}$ & $r_E^{p,a,d}$ & $r_E^{p,m,d}$\\
0.573 & 0.644 & 0.405 & 0.706 & 0.410 & 0.663 & 0.679 \\\hline
\end{tabular*}
\end{center}
\end{table}

The \emph{lower left} panel provides a flavour decomposition of the quark, diquark and exchange contributions to the form factor.  Once more their $Q^2=0$ values have precisely the same value and interpretation as those associated with the Dirac form factor.  $G_E^{p,u}$ has a zero at $Q^2\approx 9.0\,$GeV$^2$ and no contribution to $G_E^{p,u}$ is positive definite: $G_E^{p,q,u}$ possesses a zero at $Q^2\approx 3.0\,$GeV$^2$; $G_E^{p,c,u}$ at $Q^2\approx 10.0\,$GeV$^2$; and $G_E^{p,e,u}$ is negative on the domain $4.0\lsim Q^2\lsim 7.0$GeV$^2$.  On the other hand, $G_E^{p,d}$ has a zero at $Q^2\approx 10.0\,$GeV$^2$ but $G_E^{p,q,d}$ is negative definite.  $G_E^{p,c,d}$ possesses a zero at $Q^2\approx 11.0\,$GeV$^2$ and $G_E^{p,e,d}$ at $Q^2\approx 2.0\,$GeV$^2$.
We list the $u$- and $d$-quark Sachs radii in Table~\ref{radiiGE}.  Their values are readily computed and understood from Tables~\ref{radii} and \ref{tablepmag}.

The \emph{upper right} panel of Fig.\,\ref{fig:protonGE} shows the $Q^2$-evolution of the contributions to $G_E^p$ that involve a scalar diquark, an axial-vector diquark, or one of each.  $G_E^{p,s}$ has a zero at $Q^2\approx 10.0\,$GeV$^2$ and $G_E^{p,a}$ at $Q^2\approx 3.0\,$GeV$^2$, whereas $G_E^{p,m}$ is positive definite.  The \emph{lower right} panel provides a flavour decomposition of the diquark contributions.  $G_E^{p,s,u}$ exhibits a zero at $Q^2\approx 10.0\,$GeV$^2$, $G_E^{p,a,u}$ at $Q^2\approx 5.0\,$GeV$^2$ and $G_E^{p,m,u}$ is negative on the domain $2.0\lsim Q^2\lsim 6.0$GeV$^2$.  On the other hand, $G_E^{p,s,d}$ passes through zero at $Q^2\approx 11.0\,$GeV$^2$ and $G_E^{p,m,d}$ at $Q^2\approx 2.0\,$GeV$^2$ but $G_E^{p,a,d}$ is negative definite.  The associated Sachs radii are listed in Table~\ref{radiiGE}.

\begin{figure}[t]
\begin{minipage}[t]{\textwidth}
\begin{minipage}[t]{0.49\textwidth}
\leftline{\includegraphics[width=0.99\textwidth]{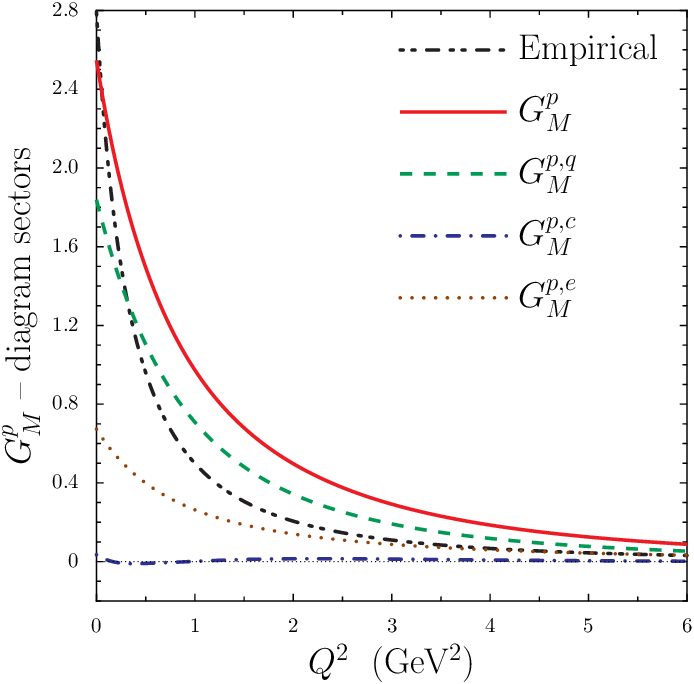}}
\end{minipage}
\begin{minipage}[t]{0.49\textwidth}
\rightline{\includegraphics[width=0.99\textwidth]{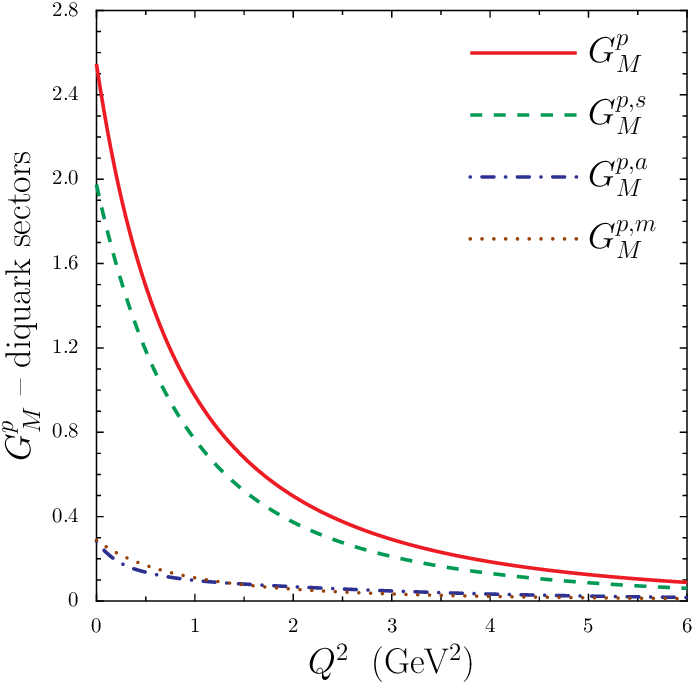}}
\end{minipage}
\end{minipage}

\caption{\label{fig:protonGM} Proton's Sachs magnetic form factor. \emph{Left panel} -- Full result and decomposition according to diagram classes; and \emph{right panel} -- full result and decomposition according to diagram diquark content.  Form factors are expressed in magnetons defined by the calculated nucleon mass, $M_N$ in Table~\protect\ref{tableNmass}.  A parametrisation of experimental data \protect\cite{Kelly:2004hm} is also presented in the \emph{upper left panel}.  A full explanation of the notation is provided in App.\,E.
} 
\end{figure}

\begin{figure}[t]
\begin{minipage}[t]{\textwidth}
\begin{minipage}[t]{0.49\textwidth}
\leftline{\includegraphics[width=0.99\textwidth]{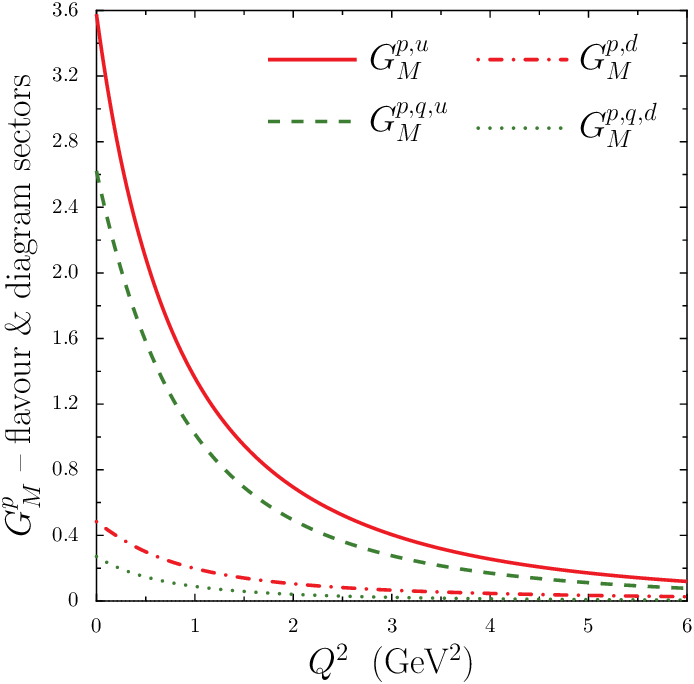}}
\end{minipage}
\begin{minipage}[t]{0.49\textwidth}
\rightline{\includegraphics[width=0.99\textwidth]{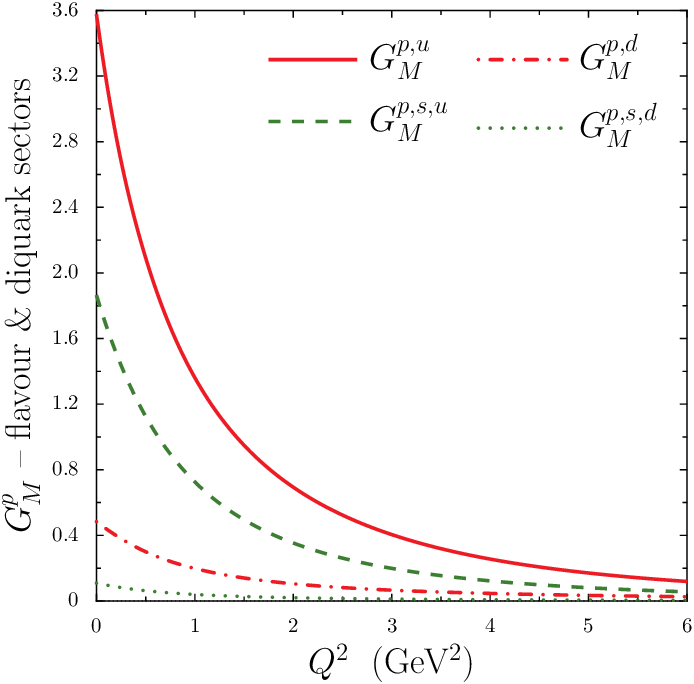}}
\end{minipage}\vspace*{2ex}

\begin{minipage}[t]{0.49\textwidth}
\leftline{\includegraphics[width=0.99\textwidth]{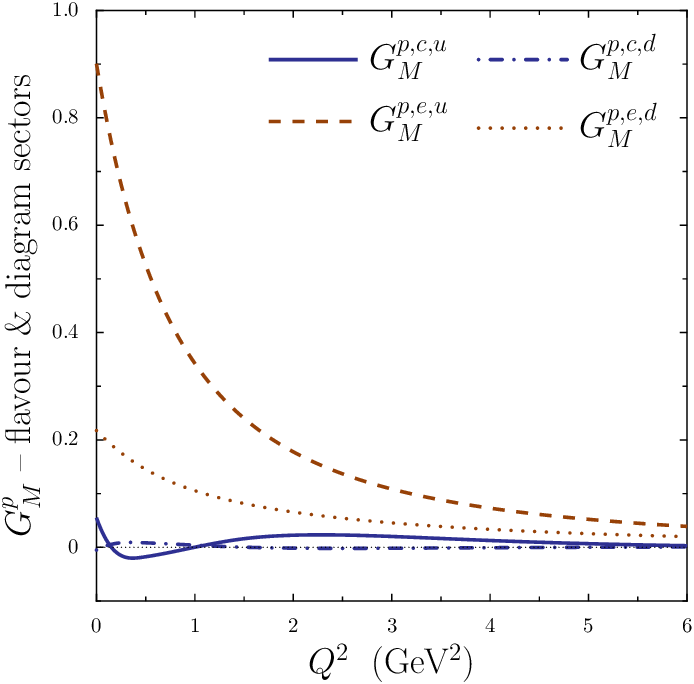}}
\end{minipage}
\begin{minipage}[t]{0.49\textwidth}
\rightline{\includegraphics[width=0.99\textwidth]{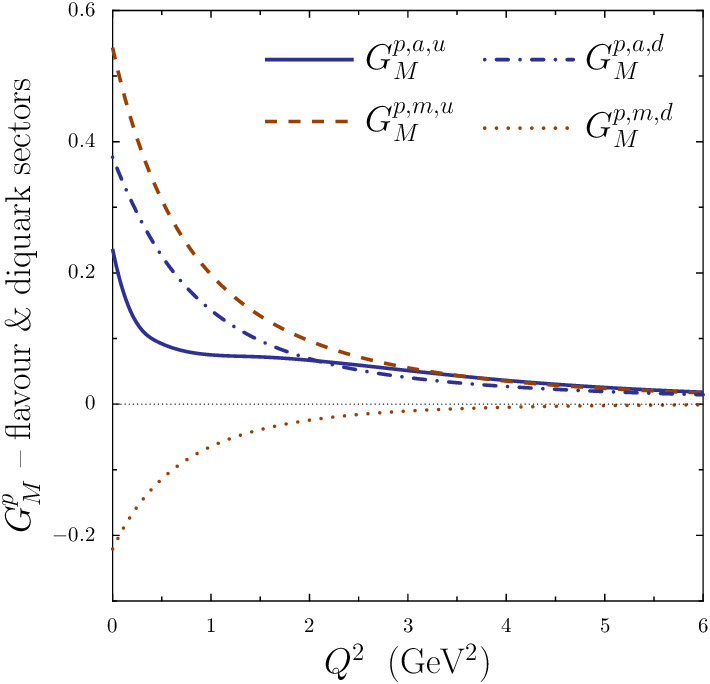}}
\end{minipage}
\end{minipage}

\caption{\label{fig:protonGMA} Proton's Sachs magnetic form factor. \emph{Left panels} -- Flavour breakdown of contributions in \emph{left panel} of Fig.\,\protect\ref{fig:protonGM}.  \emph{Right panels} -- Flavour breakdown of contributions in \emph{right panel} of Fig.\,\protect\ref{fig:protonGM}.  A full explanation of the notation is provided in App.\,E.
} 
\end{figure}

\subsection{Sachs proton magnetic}
In Figs.\,\ref{fig:protonGM} and \ref{fig:protonGMA} we depict the proton's Sachs magnetic form factor and a separation into contributions from various subclasses of diagrams.  Again, while in principle these panels only contain information that can be constructed from material already presented, they are nonetheless practically useful and informative.  

\begin{table}[b]
\begin{center}
\caption{\label{tableGMp} Flavour and diagram decomposition of contributions to the proton's magnetic moment; viz., the $G_M^p$ form factors evaluated at $Q^2=0$, measured in magnetons defined by the calculated nucleon mass, $M_N$.  Experimentally \cite{Yao:2006px}, $\mu_p = 2.79$.}
\begin{tabular*}{1.0\textwidth}{
c|@{\extracolsep{0ptplus1fil}}
c@{\extracolsep{0ptplus1fil}}
c@{\extracolsep{0ptplus1fil}}c|@{\extracolsep{0ptplus1fil}}
c@{\extracolsep{0ptplus1fil}}c@{\extracolsep{0ptplus1fil}}
c@{\extracolsep{0ptplus1fil}}
}
\hline
$\mu_p$ & $\mu_p^{q}$ & $\mu_p^{c}$ & $\mu_p^{e}$ & $\mu_p^{s}$ & $\mu_p^{a}$ & $\mu_p^{m}$ \\
2.674 & 1.919 & \,0.0495 & 0.706~ & 2.061 & 0.311 & 0.303~  \\\hline
$\mu_p^u$ & $\mu_p^{q,u}$ & $\mu_p^{c,u}$ & $\mu_p^{e,u}$ & $\mu_p^{s,u}$ & $\mu_p^{a,u}$ & $\mu_p^{m,u}$\\
2.507 & 1.824 & \,0.0527 & 0.631~ &  1.947 & 0.181 & 0.381~  \\\hline
$\mu_p^d$ & $\mu_p^{q,d}$ & $\mu_p^{c,d}$ & $\mu_p^{e,d}$ & $\mu_p^{s,d}$ & $\mu_p^{a,d}$ & $\mu_p^{m,d}$\\
0.168 & 0.210 & -0.00322 & 0.0751 & 0.115 & 0.131  & -0.0779  \\\hline
\end{tabular*}
\end{center}
\end{table}

The \emph{left panel} of Fig.\,\ref{fig:protonGM} shows the $Q^2$-evolution of the quark, diquark and exchange contributions to the form factor.  
$G_M^p$, $G_M^{p,q}$ and $G_M^{p,e}$ are positive definite and monotonically decreasing.  On the other hand, the net contribution from Diagrams~2 and 4 in Fig.\,\ref{vertex}; namely, $G_M^{p,c}$, is uniformly small, negative in the vicinity of $Q^2\sim 0.5\,$GeV$^2$ and again for $Q^2 \gsim 8\,$GeV$^2$.
The pattern is qualitatively similar in the flavour breakdown of these form factors, depicted in the \emph{left panels} of Fig.\,\ref{fig:protonGMA}.

The \emph{right panel} of Fig.\,\ref{fig:protonGM} exhibits the $Q^2$-evolution of the contributions to $G_M^p$ that involve a scalar diquark, an axial-vector diquark, or one of each.  All contributions are positive definite, diagrams involving only a scalar diquark are dominant and contributions involving at least one axial-vector diquark are uniformly of comparable magnitude.  The flavour breakdown is contained in the \emph{right panels} of Fig.\,\ref{fig:protonGMA}: all contributions are positive definite except $G_M^{p,m,d}$, which is uniformly small but becomes positive at $Q^2 \approx 9.0\,$GeV$^2$ and remains so until $Q^2\approx 17\,$GeV$^2$.  (NB.\ The latter should be interpreted qualitatively because the feature appears at a larger value of $Q^2$ than we consider our computation reliable.)

In Table~\ref{tableGMp} we list the $Q^2=0$ values of all the form factors, which measure, respectively, the contributions to the proton's magnetic moment.  These values can be obtained from $\mu_p^\alpha = F_1^\alpha(0)\,P_1^\alpha + \kappa_p^\alpha$, where, e.g., $\alpha = (p,q),(p,c)$, etc.
The magnetic radii are listed in Table~\ref{radiiGM}.  The general pattern has electric radii exceeding magnetic radii.  The few exceptions are easily explained.  For example, $r_E^{p,a,u}<r_M^{p,a,u}$: this is primarily because $F_2^{p,a,u}(Q^2)<0$ and of significant magnitude in the neighbourhood of $Q^2=0$.  As already noted, it is curious that this contribution to the proton's anomalous magnetic moment is negative.

\begin{figure}[t]
\leftline{\includegraphics[width=0.80\textwidth]{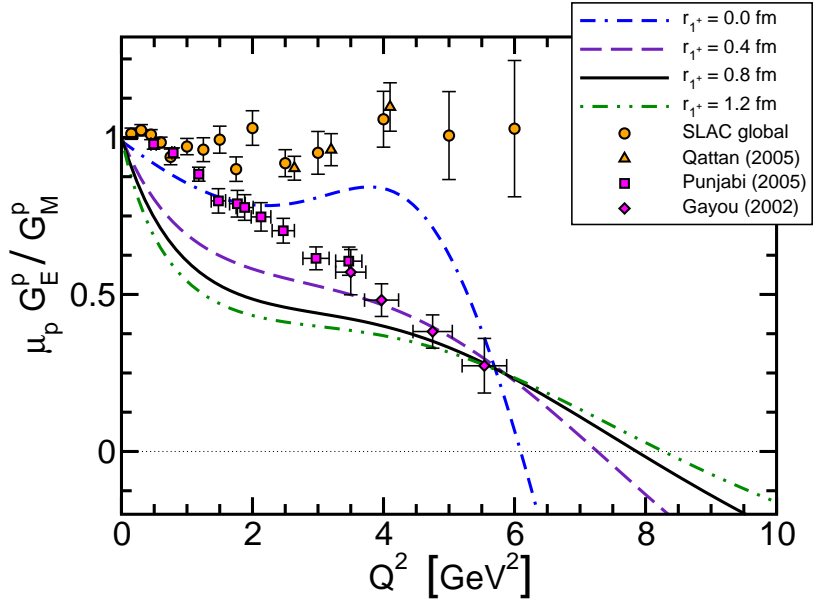}}
\caption{\label{fig:GEGM} Result for the normalised ratio of Sachs electric and magnetic form factors computed with four different diquark radii, $r_{1^+}$.  
Data: diamonds -- \cite{gayou}; squares -- \cite{Punjabi:2005wq}; triangles -- \cite{Qattan:2004ht}; and circles \cite{Walker:1993vj}.
} 
\end{figure}

\begin{table}[b]
\begin{center}
\caption{\label{radiiGM} Radii associated with $G_M^p$, defined by analogy with Eq.\,(\protect\ref{radiusexample}).  NB.\ The value in this table yields $M_N r_M^p = 3.23$ cf.\ experiment \cite{Mergell:1995bf} $M_N r_M^p = 3.99$.  Tabulated entries in fm.}
\begin{tabular*}{1.0\textwidth}{
c|@{\extracolsep{0ptplus1fil}}c@{\extracolsep{0ptplus1fil}}
c@{\extracolsep{0ptplus1fil}}c|@{\extracolsep{0ptplus1fil}}
c@{\extracolsep{0ptplus1fil}}c@{\extracolsep{0ptplus1fil}}
c@{\extracolsep{0ptplus1fil}}
}
\hline
$r_M^p$ & $r_M^{p,q}$ & $r_M^{p,c}$ & $r_M^{p,e}$ & $r_M^{p,s}$ & $r_M^{p,a}$ & $r_M^{p,m}$ \\
0.540 & 0.504 & 1.385 & 0.531 & 0.503 & 0.760 & 0.534 \\\hline
$r_M^{p,u}$ & $r_M^{p,q,u}$ & $r_M^{p,c,u}$ & $r_M^{p,e,u}$ & $r_M^{p,s,u}$ & $r_M^{p,a,u}$ & $r_M^{p,m,u}$\\
0.544 & 0.500 & 1.424 & 0.539 & 0.502 & 0.936 & 0.544 \\\hline
$r_M^{p,d}$ & $r_M^{p,q,d}$ & $r_M^{p,c,d}$ & $r_M^{p,e,d}$ & $r_M^{p,s,d}$ & $r_M^{p,a,d}$ & $r_M^{p,m,d}$\\
0.470 & 0.571  & 1.749 & 0.455 & 0.531 & 0.486 & 0.580 \\\hline
\end{tabular*}
\end{center}
\end{table}

\subsection{Sachs electric--magnetic proton ratio}
We plot $\mu_p \, G_E^p(Q^2)/G_M^p(Q^2)$ in Fig.\,\ref{fig:GEGM} in comparison with contemporary data.  A sensitivity to the proton's electromagnetic current is evident, here expressed via the diquarks' radius.  Irrespective of that radius, however, the proton's electric form factor possesses a zero and the magnetic form factor is positive definite.  On $Q^2\lsim 3\,$GeV$^2$ our result lies below experiment.  As discussed in Sect.\,\ref{sec:chiral}, this can likely be attributed to our omission of so-called pseudoscalar-meson-cloud contributions.  

\section{Neutron Form Factors}
\label{FFn}
\subsection{Dirac neutron}
\begin{figure}[t]
\begin{minipage}[t]{\textwidth}
\begin{minipage}[t]{0.49\textwidth}
\leftline{\includegraphics[width=0.99\textwidth]{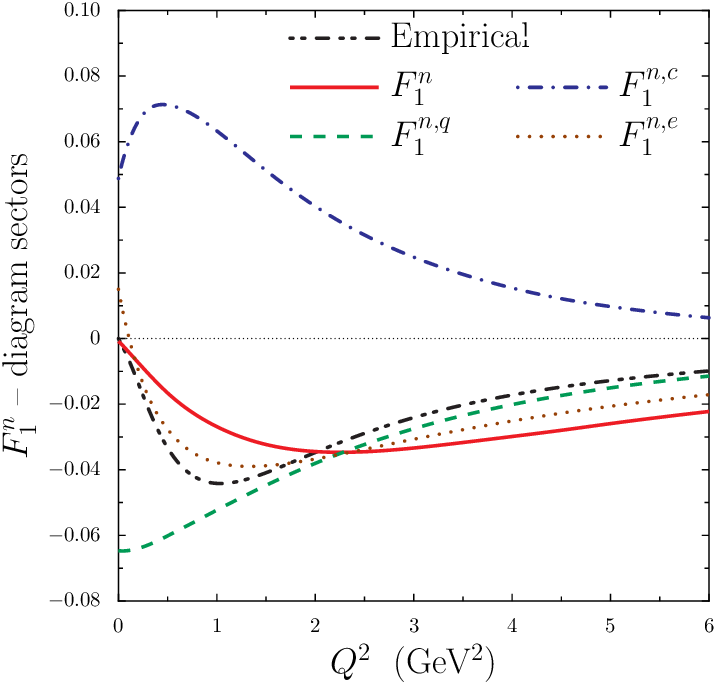}}
\end{minipage}
\begin{minipage}[t]{0.49\textwidth}
\rightline{\includegraphics[width=0.99\textwidth]{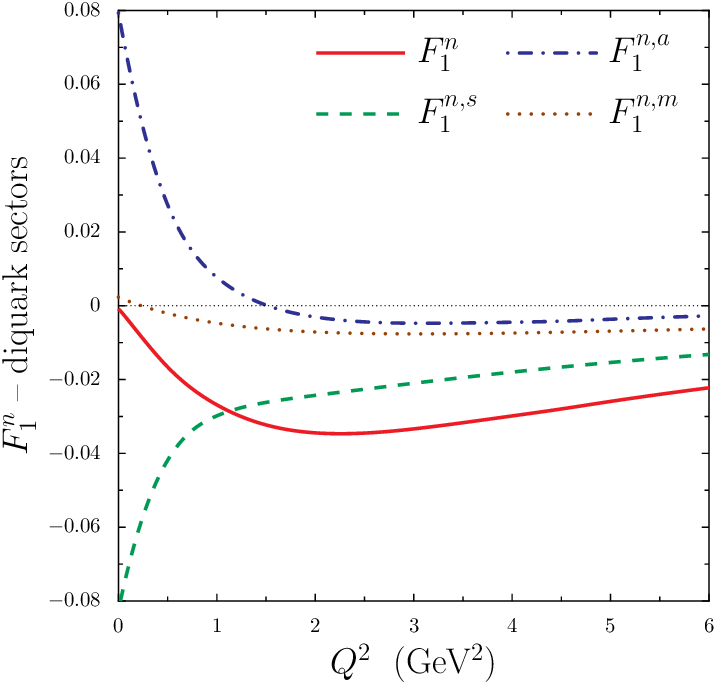}}
\end{minipage}
\end{minipage}

\caption{\label{fig:neutronF1} Neutron's Dirac form factor. \emph{Left panel} -- Full result and decomposition according to diagram classes; \emph{Right panel} -- Full result and decomposition according to diagram diquark content.  A parametrisation of experimental data \protect\cite{Kelly:2004hm} is also presented in the \emph{left} panel.  A full explanation of the notation is provided in App.\,E.
} 
\end{figure}

In Fig.\,\ref{fig:neutronF1} we depict the neutron's Dirac form factor and a decomposition into contributions from various subclasses of diagrams.  Owing to charge symmetry, Eqs.\,(\ref{CSrelations}), it is unnecessary to present a flavour breakdown.  For example, with the normalisation used in our figures, the curve that would be denoted by $F_1^{n,u}(Q^2)$ is simply negative-$F_1^{p,d}(Q^2)$ drawn from Fig.\,\ref{fig:protonF1}.

In addition to that of $F_1^n$ itself, the \emph{left panel} panel depicts the $Q^2$-evolution of the quark, diquark and exchange contributions to this form factor.  $F_1^n$ and $F_1^{n,q}$ are negative definite, and $F_1^{n,e}$ is only positive for $Q^2\lsim 0.5\,$GeV$^2$.  On the other hand, the diquark contribution; viz., $F_1^{n,d}$, is positive until $Q^2\approx 12\,$GeV$^2$.  
The \emph{right panel} renders the $Q^2$-dependence of contributions from diagrams containing a scalar diquark, an axial-vector diquark or one of each.  $F_1^{n,s}$ is negative definite and $F_1^{n,a}$ is negative for $Q^2\gsim 2\,$GeV$^2$.  $F_1^{n,m}$ is small at $Q^2=0$ (only 3\% of the other two form factors) and negative for $Q^2\gsim 0.1\,$GeV$^2$.
These features reflect:
the dominant role played in the Faddeev amplitude by the positively-charged $[ud]$ scalar diquark;
the fact that the $u$-quark is singly-represented and only a bystander in combination with an axial-vector diquark;
and the softness of the diquark correlations, which ensures that only a bystander quark can participate in the scattering process at large-$Q^2$.

We list computed Dirac radii connected with the neutron in Table~\ref{radiiF1n}.  Two entries are imaginary because the associated form factors have an inflexion point away from $Q^2=0$.  We do not currently attribute any real significance to this local feature, which for the neutron is particularly sensitive to details of the \emph{Ansatz} employed for Diagrams~5 and 6 in Fig.\,\ref{vertex}; namely, the two-body piece of the current which is not yet well constrained.

\begin{table}[b]
\begin{center}
\caption{\label{radiiF1n} Radii associated with $F_1^n$, defined by analogy with Eq.\,(\protect\ref{radiusexample}) except when the form factor vanishes at $Q^2=0$, in which case $r^2= -6 F^\prime(Q^2=0)$.  An imaginary result signifies a negative rms radius.  This convention enables a straightforward comparison between the length-scale associated with different radii.  All entries in fm.}
\begin{tabular*}{1.0\textwidth}{
c|@{\extracolsep{0ptplus1fil}}c@{\extracolsep{0ptplus1fil}}
c@{\extracolsep{0ptplus1fil}}c|@{\extracolsep{0ptplus1fil}}
c@{\extracolsep{0ptplus1fil}}c@{\extracolsep{0ptplus1fil}}
c@{\extracolsep{0ptplus1fil}}
}
\hline
$r_1^n$ & $r_1^{n,q}$ & $r_1^{n,c}$ & $r_1^{n,e}$ & $r_1^{n,s}$ & $r_1^{n,a}$ & $r_1^{n,m}$ \\
0.102 & 0.112\,i & 0.812\,i & 1.577 & 0.595 & 0.642 & 1.056 \\\hline
\end{tabular*}
\end{center}
\end{table}

\subsection{Pauli neutron}
\begin{figure}[t]
\begin{minipage}[t]{\textwidth}
\begin{minipage}[t]{0.49\textwidth}
\leftline{\includegraphics[width=0.99\textwidth]{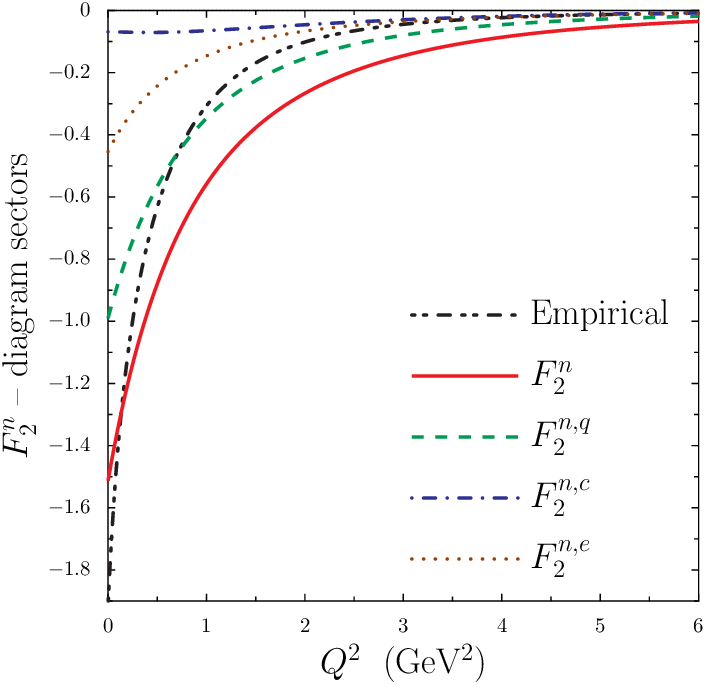}}
\end{minipage}
\begin{minipage}[t]{0.49\textwidth}
\rightline{\includegraphics[width=0.99\textwidth]{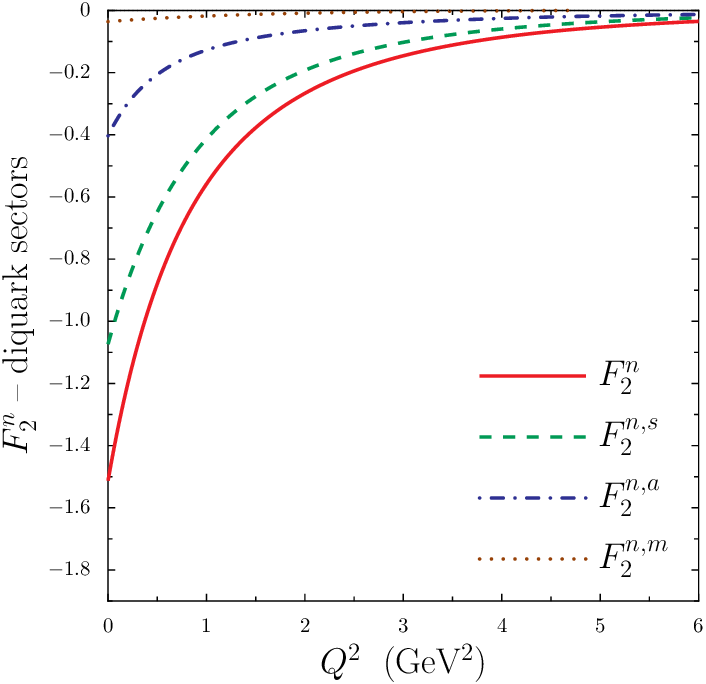}}
\end{minipage}
\end{minipage}

\caption{\label{fig:neutronF2} Neutron's Pauli form factor. \emph{Left panel} -- Full result and decomposition according to diagram classes; \emph{Right panel} -- Full result and decomposition according to diagram diquark content.  A parametrisation of experimental data \protect\cite{Kelly:2004hm} is also presented in the \emph{left} panel.  A full explanation of the notation is provided in App.\,E.
} 
\end{figure}

In Fig.\,\ref{fig:neutronF2} we depict the neutron's Pauli form factor and a decomposition into contributions from various subclasses of diagrams.  Once more, owing to charge symmetry, Eqs.\,(\ref{CSrelations}), it is unnecessary to present a flavour breakdown.  For example, with the normalisation used in our figures, the curve that would be denoted by $F_2^{n,u}(Q^2)$ is simply negative-$F_2^{pd}(Q^2)$ drawn from Fig.\,\ref{fig:protonF2}.

The \emph{left panel} panel depicts the $Q^2$-evolution of $F_2^n$ itself, and that of the quark, diquark and exchange contributions to this form factor.  $F_2^n$, $F_2^{n,q}$ and $F_2^{n,e}$ are negative definite on the domain within which we consider our calculations accurate, and $F_2^{n,c}$ is negative until $Q^2 \approx 12\,$GeV$^2$.  
The \emph{right panel} portrays the $Q^2$-dependence of contributions from diagrams containing a scalar diquark, an axial-vector diquark or one of each.  $F_2^{n,s}$ and $F_2^{n,a}$ are negative definite, and $F_2^{n,m}$ is negative for $Q^2\lsim 5\,$GeV$^2$ and always small in magnitude.
These features are consistent with those of the Dirac form factor.

We list computed anomalous magnetic moments and Pauli radii connected with the neutron in Table~\ref{radiiF2n}.  The small value of $\kappa_n^d$ may be understood via a cancellation between $d(ud)_{1^+}$ and $u(dd)_{1^+}$ contributions.  Along with the small value of $\kappa_{\cal T}$, Eq.\,(\ref{kappaTfit}), this explains the size of $\kappa_n^m$.  With the exception of the uniformly small $F_2^{n,c}$, the Pauli radii follow the same pattern as those of the proton.  

\begin{table}[b]
\begin{center}
\caption{\label{radiiF2n}
\emph{Upper rows}: Diagram decomposition of contributions to the neutron's anomalous magnetic moment; viz., the $F_2^n$ form factors evaluated at $Q^2=0$, measured in magnetons defined by the calculated nucleon mass, $M_N$.  Experimentally \cite{Yao:2006px}, $\mu_n = -1.91$.
\emph{Lower rows}: Radii associated with $F_2^n$, defined by analogy with Eq.\,(\protect\ref{radiusexample}). These entries in fm.  An imaginary result signifies a negative rms radius. 
}
\begin{tabular*}{1.0\textwidth}{
c|@{\extracolsep{0ptplus1fil}}c@{\extracolsep{0ptplus1fil}}
c@{\extracolsep{0ptplus1fil}}c|@{\extracolsep{0ptplus1fil}}
c@{\extracolsep{0ptplus1fil}}c@{\extracolsep{0ptplus1fil}}
c@{\extracolsep{0ptplus1fil}}
}
\hline
$\kappa_n$ & $\kappa_n^{q}$ & $\kappa_n^{c}$ & $\kappa_n^{e}$ & $\kappa_n^{s}$ & $\kappa_n^{a}$ & $\kappa_n^{m}$ \\
-1.588 & -1.038 & -0.0686 & -0.481 & -1.120 & -0.430 & -0.0368 \\\hline
$r_2^n$ & $r_2^{n,q}$ & $r_2^{n,c}$ & $r_2^{n,e}$ & $r_2^{n,s}$ & $r_2^{n,a}$ & $r_2^{n,m}$ \\
\,0.533 & \,0.529 & \,0.120\,i & \,0.576 & \,0.500 & \,0.621 & \,0.405 \\\hline
\end{tabular*}
\end{center}
\end{table}

\begin{figure}[t]
\leftline{%
\includegraphics[clip,width=0.75\textwidth]{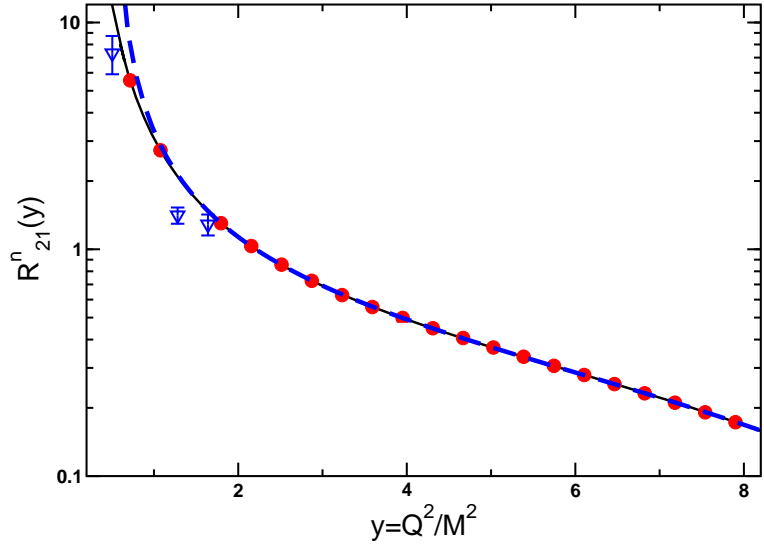}}
\caption{\label{figF2F1n} \emph{Solid circles} and \emph{solid curve} -- Dimensionless ratio in Eq.\,(\protect\ref{jiratio}) calculated for the neutron, with $\hat \Lambda = 0.44$ and $M_N$ in Table~\protect\ref{tableNmass}.  \emph{Dashed curve} -- Right-hand-side of Eq.\,(\ref{F2F1pade}).  Experimental results: 
\emph{down triangles} -- Ref.\,\cite{Madey:2003av}.}
\end{figure}

\subsection{Neutron Pauli--Dirac neutron ratio}
In Fig.\,\ref{figF2F1n} we plot the weighted ratio of Pauli to Dirac form factors in Eq.\,(\ref{jiratio}) for the neutron.  This ratio is constant for the proton, Fig.\,\ref{figF2F1}, however, that is not the case for the neutron.  Moreover, with our calculated neutron form factors there is no value of $\hat\Lambda$ for which this ratio assumes a constant value.  

The apparent cause of this behaviour is a zero in $F_2^n(Q^2)$ at $Q^2\approx 18\,$GeV$^2$.  This point lies beyond the upper bound of the domain within which we consider our computation reliable.  On the other hand, its presence does influence the evolution of the ratio.  This can be seen by analysing the ratio using Pad\'e approximants on subdomains of $Q^2\in [4,12]\,$GeV$^2$, which consistently yields a best fit that possesses a zero at $Q^2\approx 18\,$GeV$^2$; e.g., 
\begin{equation}
\label{F2F1pade}
R_{21}^n(\hat Q^2):= \frac{\hat Q^2}{(\ln \hat Q^2/\hat \Lambda^2)^2}\, \frac{F_2^n(\hat Q^2)}{F_1^n(\hat Q^2)} = \frac{2.85 + 0.274 \,\hat Q^2 - 0.0409 \,\hat Q^4}
{-1 + 1.93 \, \hat Q^2}\,.
\end{equation}
It seems therefore that the zero is not simply the result of inaccurate numerical analysis but either a property of the model itself or an artefact of the numerical method; namely, the Chebyshev expansion described in \ref{CHexpand}~\emph{Chebyshev expansion}.  We are working to resolve this issue.

\subsection{Sachs neutron electric}
\begin{figure}[t]
\begin{minipage}[t]{\textwidth}
\begin{minipage}[t]{0.49\textwidth}
\leftline{\includegraphics[width=0.99\textwidth]{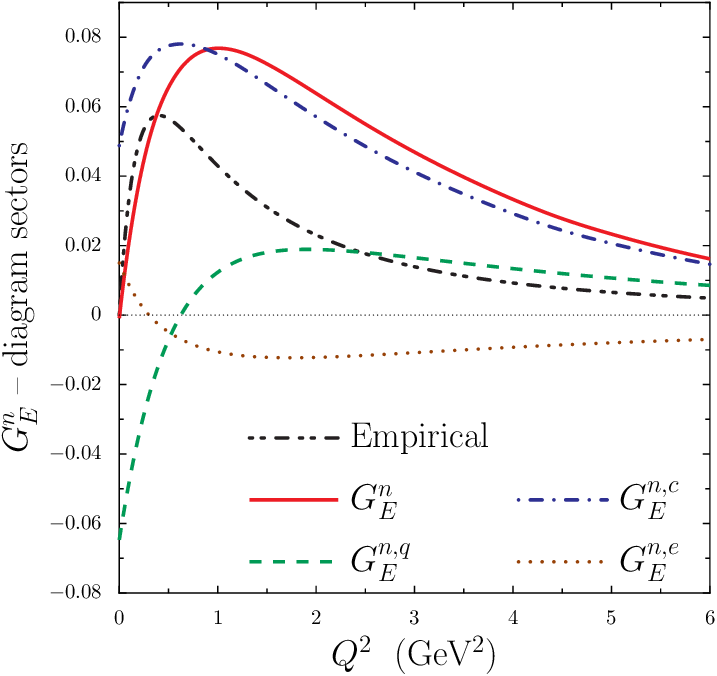}}
\end{minipage}
\begin{minipage}[t]{0.49\textwidth}
\rightline{\includegraphics[width=0.99\textwidth]{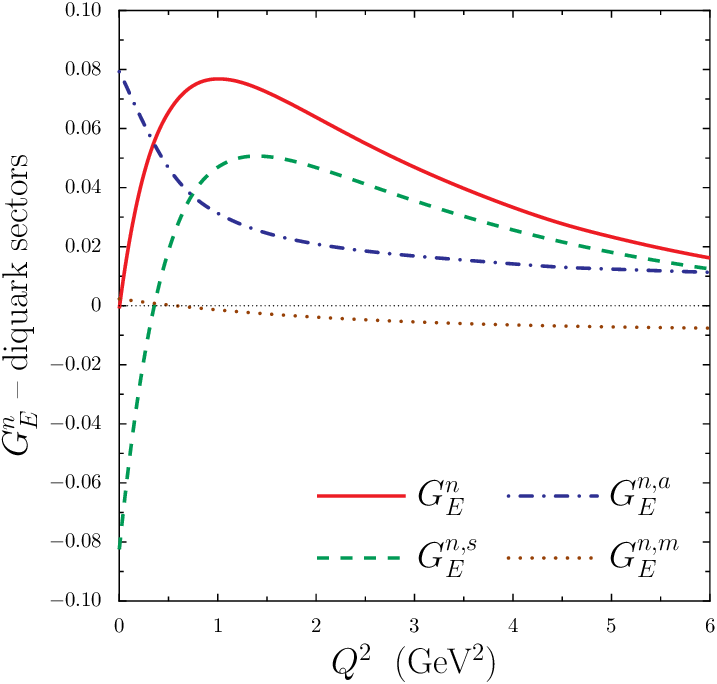}}
\end{minipage}
\end{minipage}

\caption{\label{fig:neutronGE} Neutron's Sachs electric form factor. \emph{Left panel} -- Full result and decomposition according to diagram classes.  \emph{Right panel} -- Full result and decomposition according to diagram diquark content.  A parametrisation of experimental data \protect\cite{Kelly:2004hm} is also presented in the \emph{left} panel.  A complete explanation of the notation is provided in App.\,E.
} 
\end{figure}

In Fig.\,\ref{fig:neutronGE} we present the neutron's Sachs electric form factor and a separation into contributions from various subclasses of diagrams.  Once more, owing to charge symmetry, Eqs.\,(\ref{CSrelations}), it is unnecessary to present a flavour breakdown.  For example, with the normalisation used in our figures, the curve that would be denoted by $G_E^{n,d}(Q^2)$ is simply negative-$G_E^{p,u}(Q^2)$ drawn from Fig.\,\ref{fig:protonGE}.

In addition to that of $G_E^n$ itself, the \emph{left panel} panel depicts the $Q^2$-evolution of the quark, diquark and exchange contributions to this form factor.  Each exhibits a zero, with that for the net result lying at $Q^2\approx 11\,$GeV$^2$.
In the \emph{right panel} we plot the $Q^2$-dependence of contributions from diagrams containing a scalar diquark, an axial-vector diquark or one of each.  $G_E^{n,s}$ is positive on the domain $Q^2\in [0.1,11]\,$GeV$^2$ and $G_E^{n,m}$ is negative for $Q^2\gsim 1\,$GeV$^2$, whereas $G_E^{n,a}$ is positive definite.

These features are again consistent with intuition.  For example, the behaviour of $G_E^{n,q}$.  It is negative at small-$Q^2$ because the scalar diquark component of the Faddeev amplitude is dominant and that is paired with a $d$-quark bystander in the neutron.  This dressed-quark is responsible for the preponderance of negative charge at long range.  $G_E^{n,q}$ is positive at large $Q^2$ because $F_2^n$ dominates on that domain, which focuses attention on the axial-vector diquark component of the Faddeev amplitude.  The positively charged $u$-quark is most likely the bystander quark in these circumstances.  

Another interesting illustrative case is provided by $G_E^{n,a}$, which is positive definite.  As already noted, the $u$-quark is the most probable bystander in the neutron's axial-vector diquark configuration and this explains the preponderance of positive charge at small $Q^2$.  This plus the fact that the current's only hard component is that involving a bystander quark also explains the positive charge at large $Q^2$.  The form factor remains positive in the intermediate region because the term which could interfere; viz., $d [ud]_{0^+}$, involves individual charges with smaller magnitude.

We list computed Dirac radii connected with the neutron in Table~\ref{radiiF1n}.  Two entries are imaginary because the associated form factors have an inflexion point away from $Q^2=0$.  As just explained, the origin of such behaviour lies in interference, mediated by the current, between components in the incoming and outgoing neutrons' Faddeev amplitudes. 

\begin{table}[b]
\begin{center}
\caption{\label{radiiGEn} \emph{Upper rows} -- Radii associated with $G_E^n$, defined by analogy with Eq.\,(\protect\ref{radiusexample}) except when the form factor vanishes at $Q^2=0$, in which case $r^2= -6 F^\prime(Q^2=0)$.  An imaginary result signifies a negative rms radius.  
NB.\ The value in this table yields $M_N^2 (r_n^E)^2 = -(1.36)^2$ cf.\ experiment \cite{Yao:2006px} $M_n^2 (r_n^E)^2= - (1.62)^2$.  \emph{Lower rows} -- Radii associated with $G_M^n$, defined by analogy with Eq.\,(\protect\ref{radiusexample}): $M_N r_n^M = 3.17$ cf.\ experiment \protect\cite{Mergell:1995bf} $M_n r_n^M= 4.24$.
Tabulated entries in fm.}
\begin{tabular*}{1.0\textwidth}{
c|@{\extracolsep{0ptplus1fil}}c@{\extracolsep{0ptplus1fil}}
c@{\extracolsep{0ptplus1fil}}c|@{\extracolsep{0ptplus1fil}}
c@{\extracolsep{0ptplus1fil}}c@{\extracolsep{0ptplus1fil}}
c@{\extracolsep{0ptplus1fil}}
}
\hline
$r_E^n$ & $r_E^{n,q}$ & $r_E^{n,c}$ & $r_E^{n,e}$ & $r_E^{n,s}$ & $r_E^{n,a}$ & $r_E^{n,m}$ \\
0.227\,i & 0.812 & 0.847\,i & 1.069 & 0.961 & 0.430 & 0.674 \\\hline
$r_M^n$ & $r_M^{n,q}$ & $r_M^{n,c}$ & $r_M^{n,e}$ & $r_M^{n,s}$ & $r_M^{n,a}$ & $r_M^{n,m}$ \\
0.529 & 0.513 & 1.254 & 0.514 & 0.507 & 0.614 & 0.316 \\\hline
\end{tabular*}
\end{center}
\end{table}

\subsection{Sachs neutron magnetic}
\begin{figure}[t]
\begin{minipage}[t]{\textwidth}
\begin{minipage}[t]{0.49\textwidth}
\leftline{\includegraphics[width=0.99\textwidth]{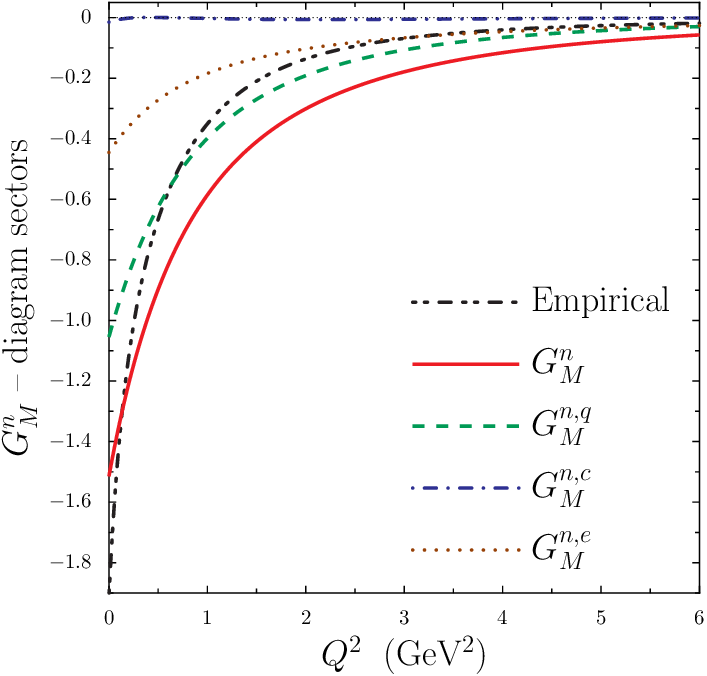}}
\end{minipage}
\begin{minipage}[t]{0.49\textwidth}
\rightline{\includegraphics[width=0.99\textwidth]{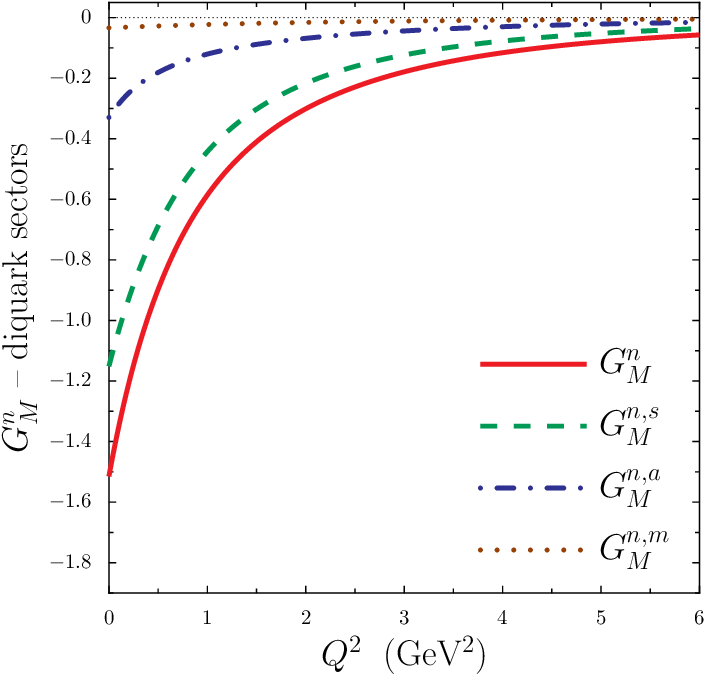}}
\end{minipage}
\end{minipage}

\caption{\label{fig:neutronGM} Neutron's Sachs magnetic form factor. \emph{Left panel} -- Full result and decomposition according to diagram classes.  \emph{Right panel} -- Full result and decomposition according to diagram diquark content.  A parametrisation of experimental data \protect\cite{Kelly:2004hm} is also presented in the \emph{left} panel.  A complete explanation of the notation is provided in App.\,E.
} 
\end{figure}

In Fig.\,\ref{fig:neutronGM} we present the neutron's Sachs magnetic form factor and a decomposition into contributions from various subclasses of diagrams.  Again, owing to charge symmetry, Eqs.\,(\ref{CSrelations}), it is unnecessary to present a flavour breakdown.  For example, with the normalisation used in our figures, the curve that would be denoted by $G_M^{n,u}(Q^2)$ is simply negative-$G_M^{p,d}(Q^2)$ drawn from Fig.\,\ref{fig:protonGM}.

In the \emph{left panel} panel we draw the $Q^2$-evolution of $G_M^n$ itself, and that of the quark, diquark and exchange contributions to this form factor.  $G_M^n$, $G_M^{n,q}$ and $G_M^{n,e}$ are negative definite.   On the other hand, $G_M^{n,c}$ is uniformly small, owing to cancellations between $F_1^n$ and $F_2^n$.  It begins negative, is positive in the vicinity of $Q^2 = 0.5\,$GeV$^2$ and again for $Q^2\gsim 10\,$GeV$^2$.
The \emph{right panel} portrays the $Q^2$-dependence of contributions from diagrams containing a scalar diquark, an axial-vector diquark or one of each.  All are negative definite. 

We list the computed magnetic radii connected with the neutron in Table~\ref{radiiGEn}.  The magnetic moments are the same as the anomalous moments in Table~\ref{radiiF2n}.  With the exception of $G_M^{n,m}$, which at small $Q^2$ is roughly a factor of five smaller than $G_M^{p,m}$, the neutron radii follow the same pattern as those of the proton.  

%
%

\subsection{Sachs electric--magnetic neutron ratio}
\begin{figure}[t]
\leftline{\includegraphics[width=0.80\textwidth]{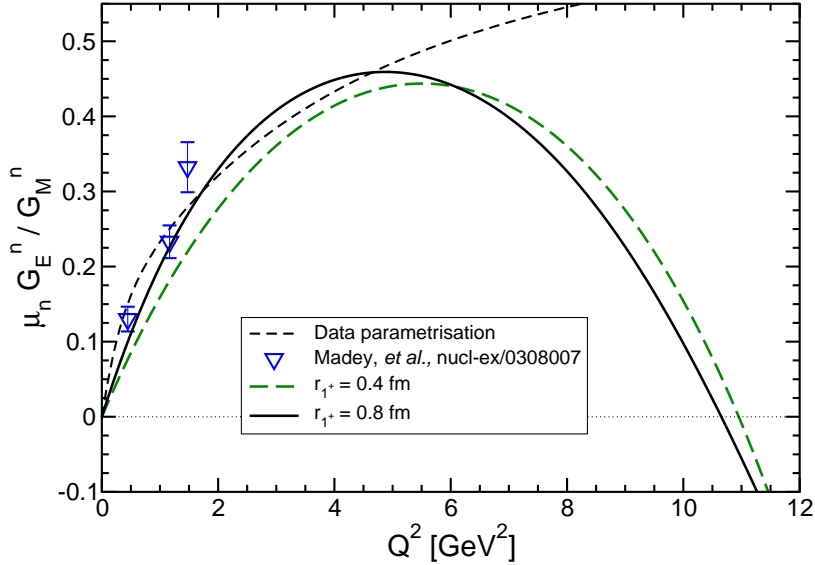}}
\caption{\label{fig:GEGMn} Result for the normalised ratio of Sachs electric and magnetic form factors for the neutron computed with two different diquark radii.  
\emph{Short-dashed curve}: parametrisation of Ref.\,\protect\cite{Kelly:2004hm}.  \emph{Down triangles}: data from Ref.\,\protect\cite{Madey:2003av}.
} 
\end{figure}

We plot $\mu_n G_E^n(Q^2)/G_M^n(Q^2)$ in Fig.\,\ref{fig:GEGMn}.  The figure illustrates a quantitative sensitivity of our results to the neutron's electromagnetic current, here expressed via the diquarks' radius.  Notwithstanding this, the qualitative features are robust, with $G_E^n(Q^2)$ possessing a zero at $Q^2\gsim 10\,$GeV$^2$.  In contrast to the behaviour in Fig.\,\ref{fig:GEGM}, here the zero moves to smaller $Q^2$ with increasing diquark radius.  The effect of our omission of meson cloud contributions is again evident at small $Q^2$.  

\section{Chiral Corrections}
\label{sec:chiral}
The framework we have described hitherto defines a dressed-quark core contribution to the nucleons' electromagnetic form factors.  As with the mass \cite{Hecht:2002ej,Young:2002cj}, the nucleons' magnetic moments, and charge and magnetic radii receive material contributions from the so-called pseudoscalar meson cloud \cite{radiiCh,young}.  There are two types of contribution: regularisation-scheme-dependent terms, which are analytic functions of $m$ in the neighbourhood of vanishing current-quark mass, $m = 0$; and nonanalytic scheme-independent terms.  

For magnetic moments and radii the leading-order scheme-independent contributions are \cite{kubis}
\begin{eqnarray}
\label{mpnpion}
(\mu_{n/p})_{NA}^{\rm 1-loop} &\stackrel{m_\pi \simeq 0}{=}& \pm \, \frac{g_A^2\, M_N}{4\pi^2 f_\pi^2}\, m_\pi\,,\\
\label{rpnpion}
\langle r_{n/p}^2\rangle^{1-loop}_{NA} &\stackrel{m_\pi \simeq 0}{=}& \pm\,\frac{1+5 g_A^2}{32 \pi^2 f_\pi^2} \,\ln (\frac{m_\pi^2}{M_N^2}) \,, \\
\label{rpnmpion}
\langle (r_{N}^\mu)^2\rangle^{1-loop}_{NA} &\stackrel{m_\pi \simeq 0}{=}& -\,\frac{1+5 g_A^2}{32 \pi^2 f_\pi^2} \,\ln (\frac{m_\pi^2}{M_N^2})+ \frac{g_A^2\, M_N}{16 \pi f_\pi^2 \mu_v} \frac{1}{m_\pi} \,,
\end{eqnarray}
where, experimentally, $g_A=1.26$, $f_\pi=0.0924$\,GeV\,$=1/(2.13 \,{\rm fm})$ and $\mu_v=\mu_p-\mu_n=4.7$.  These terms reduce the magnitude of both neutron and proton magnetic moments, and increase the magnitudes of the radii.

Whilst these scheme-independent terms are important, at physical values of the pseudoscalar meson masses they do not usually provide the dominant contribution to observables.  That arises from the regularisation-parameter-dependent terms, as is apparent for baryon masses in Ref.\,\cite{Hecht:2002ej} and for the pion charge radius in Ref.\,\cite{Alkofer:1993gu}.  This is particularly significant for the neutron's charge radius \cite{Alkofer:2004yf} and for the magnetic moments, in which connection the regularisation-scheme-dependent terms provide a nonzero contribution in the chiral limit and have the net effect of \emph{increasing} $|\mu_{N}|$.

Owing to the importance of the chiral loops' regularisation-parameter-dependent parts we estimate the corrections using modified formulae, which incorporate a single parameter that mimics the effect of regularising the integrals; namely \cite{Alkofer:2004yf,Cloet:2008wg,ashley}, 
\begin{eqnarray}
\label{ImproveAH}
(\mu_{n/p})^{{\rm 1-loop}^R} &= &\left( \mu_{n/p}^{\pi 0} \; \pm \; \frac{g_A^2\, M_N}{4\pi^2 f_\pi^2}\, m_\pi \right) \frac{2}{\pi}\arctan(\frac{\lambda^3}{m_\pi^3})\,,\\
\label{rpnpionR}
\langle r_{n/p}^2\rangle^{1-loop^R} &=& \pm\,\frac{1+5 g_A^2}{32 \pi^2 f_\pi^2} \,\ln (\frac{m_\pi^2}{m_\pi^2+\lambda^2}) \,, \\
\nonumber \langle (r_{N}^\mu)^2\rangle^{1-loop^R} &=& -\,\frac{1+5 g_A^2}{32 \pi^2 f_\pi^2} \,\ln (\frac{m_\pi^2}{m_\pi^2+\lambda^2}) 
\\
& & 
+ \frac{g_A^2\, M_N}{16 \pi f_\pi^2 \mu_v} \frac{1}{m_\pi} \,
\frac{2}{\pi}\arctan(\frac{\lambda}{m_\pi})\,.
\label{rpnmpionR}
\end{eqnarray}
wherein $\mu_{n/p}^{\pi 0}$ are the chiral limit values of the meson loop contributions and $\lambda = 0.3\,$GeV$\,=1/[0.66\, {\rm fm}]$ is a regularisation mass-scale.  NB.\ As required physically, the loop contributions vanish when the meson mass is much larger than the regularisation scale: very massive states must decouple from low-energy phenomena.  

In Table~\ref{physicalresults} we exemplify the effect of the corrections in Eqs.\,(\ref{ImproveAH})--(\ref{rpnmpionR}) to nucleon static properties.  The quark-core values are collected from Tables~\ref{radiiGE}--\ref{radiiGM} and \ref{radiiF2n}--\ref{radiiGEn} herein.  The sensitivity of the neutron's charge radius is apparent.  In relation to the magnetic moments, a recent estimate from numerical simulations of lattice-regularised QCD \cite{Wang:2007iw} gives the following chiral-loop contributions to the nucleons' magnetic moments at the physical pion mass:
$\mu_{n}^{\pi} = -0.40$, $\mu_{p}^{\pi} = 0.24$,
which are obtained with 
$\mu_{n}^{\pi 0} = -1.05$, $\mu_{p}^{\pi 0} = 0.88$ 
in Eq.\,(\ref{ImproveAH}).  These results in conjunction with the experimental values point to quark-core magnetic moments of $\mu_n^{q(qq)}=-1.51$, $\mu_p^{q(qq)}=2.55$, which compare well with our computed moments.\footnote{The magnetic moment values in Row~2 of the Table differ slightly ($<8$\%) in magnitude from those reported in Ref.\,\protect\cite{Cloet:2008wg} because an extrapolation is necessary to obtain $G_M(0)$ and herein we've used a $[0,2]$ Pad\'e as opposed to a simple quadratic. Were this significant, it could be corrected by a minor ($\sim 10$\%) adjustment of $\mu_{1^+}$ and $\kappa_{\cal T}$.}

\begin{table}[t]
\begin{center}
\caption{\label{physicalresults} Quark-core and pseudoscalar meson loop [Eqs.\,(\protect\ref{ImproveAH})--(\protect\ref{rpnmpionR})] contributions to the moments and radii, calculated at the physical current-quark mass, Eq.\,(\protect\ref{mcq}).  The radii are listed in fm$^2$. Experimental values are quoted from Ref.\,\protect\cite{Yao:2006px}, where available, and otherwise from Ref.\,\protect\cite{Mergell:1995bf}.}
\begin{tabular*}{1.0\textwidth}
{c@{\extracolsep{0ptplus1fil}}c@{\extracolsep{0ptplus1fil}}c@{\extracolsep{0ptplus1fil}}
c@{\extracolsep{0ptplus1fil}}c@{\extracolsep{0ptplus1fil}}
c@{\extracolsep{0ptplus1fil}}c@{\extracolsep{0ptplus1fil}}}\hline
\rule{0em}{3ex} & $\mu_n$ & $\mu_p$ & $\langle r_n^2 \rangle$ & $\langle r_p^2 \rangle$ & $\langle (r_n^\mu)^2\rangle $ &  $\langle (r_p^\mu)^2\rangle $\\\hline
\rule{0em}{3ex} $q(qq)$ & -1.59 & 2.67 & $-(0.23)^2$ & $(0.67)^2$ & $(0.53)^2$ & $(0.54)^2$  \\
 $\pi$-loop & -0.40 & 0.24 & $-(0.47)^2$ & $(0.47)^2$ & $(0.61)^2$ & $(0.61)^2$ \\\hline
\rule{0em}{3ex} total & -1.99 & 2.91 & $-(0.52)^2$& $(0.82)^2$ & $(0.81)^2$ & $(0.81)^2$ \\\hline
\rule{0em}{3ex} experiment & -1.91 & 2.79& $-(0.34)^2$ & $(0.88)^2$ & $(0.89)^2$ & $(0.84)^2$ \\\hline
\end{tabular*}
\end{center}
\end{table}

\begin{figure}[t]
\leftline{\includegraphics[width=0.80\textwidth]{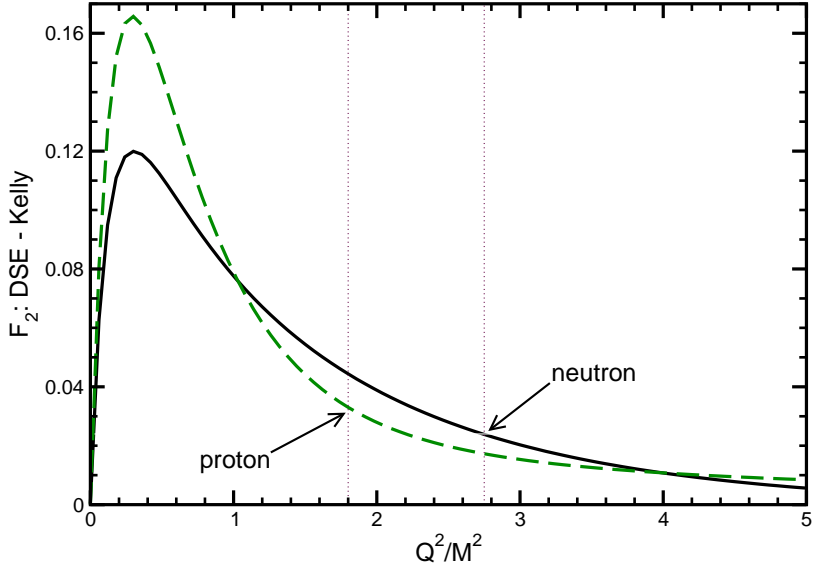}}
\caption{\label{DSEcfKelly} Difference between our calculated Pauli form factor and the parametrisation of experimental data in Ref.\,\protect\cite{Kelly:2004hm}, each normalised by the appropriate anomalous magnetic moment at $Q^2=0$: \emph{dashed curve} -- proton; \emph{solid curve} -- neutron.  The $Q^2$ for which the difference reaches 20\% of its peak value is indicated in each case by a vertical dotted line.
} 
\end{figure}

It is plain in Table~\ref{physicalresults} that pseudoscalar meson loops alter the proton's magnetic radius more than its electric radius.  Indeed, without fine tuning, these two initially rather different radii are brought into agreement.  As observed in Ref.\,\cite{Alkofer:2004yf}, this is important in relation to Fig.\,\ref{fig:GEGM} because it explains why the quark core result disagrees with data at small momentum transfers.  Namely, in the neighbourhood of $Q^2=0$ one has
\begin{equation}
\mu_p\,\frac{ G_E^p(Q^2)}{G_M^p(Q^2)} = 1 - \frac{Q^2}{6} \,\left[ (r_p)^2 - (r_p^\mu)^2 \right]\,,
\end{equation}
and so with $r_p> r_p^\mu$, as is the case for the quark core contribution, the ratio falls immediately with increasing $Q^2$.  This is the behaviour in Fig.\,\ref{fig:GEGM}. However, experimentally, and with addition of a pseudoscalar meson cloud to our quark core, $r_p= r_p^\mu$.  Therefore the complete ratio varies little on $0<Q^2< 0.6\,$GeV$^2$.  

The analysis in this section is rudimentary.  Nonetheless it illustrates that the dressed-quark core defined by our Faddeev equation is uniformly compatible with augmentation by a sensibly regulated pseudoscalar meson cloud.  We emphasise that by construction our Faddeev equation explicitly excludes all diagrams that can be associated with that cloud and so a question of overcounting cannot arise.

It is nevertheless reasonable to inquire into the domain of momentum transfer upon which pseudoscalar meson loops can contribute materially to nucleon form factors.  Regarding this it is relevant to observe that in a meson-nucleon coupled-channel analysis of the $\gamma N \to \Delta$ transition form factors the cloud contributes 50\% of the M1 form factor's magnitude at $Q^2=0$ but is insignificant by $Q^2 \approx 2 M_N^2$ \cite{JuliaDiaz:2006xt}.  We address this question via Fig.\,\ref{DSEcfKelly}, which compares our computed dressed-quark core Pauli form factors with a contemporary parametrisation of experimental data \cite{Kelly:2004hm}.  The differences depicted are consistent with loop corrections providing a necessary quantitative contribution that is important until $Q^2 \approx 2\,$--$\,3 M_N^2$.  An analogous figure for the Dirac form factors presents a comparable picture, although the differences are an order of magnitude smaller and have longer tails.  

\section{Epilogue}
\label{sec:epilogue}
We described a calculation of a dressed-quark core contribution to nucleon electromagnetic form factors.  This core is defined by the solution of a Poincar\'e covariant Faddeev equation in which dressed-quarks provide the elementary degree of freedom and quark-quark correlations are formed therefrom.  The two parameters in the Faddeev equation are diquark masses.  They are set by fitting to required nucleon and $\Delta$ masses.  We allowed one parameter in the nucleon-photon vertex; viz., the diquark charge radius.  Contemporary continuum calculations and comparison with extant data indicate that this radius should be commensurate with the pion's charge radius.  From this foundation we provided a comprehensive analysis and explanation of the form factors.

A feature of our study is the separation of form factor contributions into those from different diagram types and correlation sectors, and subsequently a flavour separation for each of these.  In this way we obtained, for example, Table~\ref{tableprob}, which shows amongst other things that the probability of the photon striking a bystander quark in the proton is 47\%.  It also enables us to determine, Eq.\,(\ref{r1nund}), that $r_1^{n,u}>r_1^{n,d}$; i.e., that the neutron's $u$-quark Dirac radius is greater than that of the $d$-quark, and explain the result in terms of the presence of axial-vector diquark correlations.  The dressed-quark magnetic radii have the same ordering.

From our extensive body of results we will here highlight just a few more.  For the proton a weighted ratio of Pauli to Dirac form factors is constant on a domain that begins at $Q^2/M_N^2 \approx 4$, Fig.\,\ref{figF2F1}.  We correlated this behaviour with the momentum space width of the dominant elements in the proton's Faddeev amplitude, Eq.\,(\ref{FAwidths}).  On the other hand, the same ratio for the neutron is not constant on any domain accessible in our calculation, Fig.\,\ref{figF2F1n}.  In addition, the ratio of Sachs electric and magnetic form factors for the proton exhibits a zero, Fig.\,\ref{fig:GEGM}.  Its position depends on correlations in the Faddeev amplitude and details of the nucleon-photon current.  Our current best estimate for the zero's location is $Q^2\approx 8\,$GeV$^2$.  A similar ratio for the neutron passes through zero at $Q^2\approx 11\,$GeV$^2$, Fig.\,\ref{fig:GEGMn}.

We have defined the nucleon's dressed-quark core via a Poincar\'e covariant Faddeev equation and have seen that pseudoscalar meson loops can be added in a sensible fashion.  The framework is successful and instructive, and unifies a phenomenological description of mesons with that of baryons.  Yet it is simple enough to allow access to numerous form factors and large values of momentum transfer.  Importantly, our approach enables one to chart the interplay between the firmly established and material momentum-dependent dressing of QCD's elementary excitations and the observable properties of hadrons.  In the near term it will be applied to nucleon excited states and transition form factors so as to elucidate their dependence on these fundamental features of QCD.  A medium term goal is to extend Ref.\,\cite{Eichmann:2008ef} and provide a simultaneous description of meson and baryon observables using a single interaction in a truncation of QCD's Dyson-Schwinger equations that can systematically be improved.  

\begin{acknowledge}
We are grateful to J.~Arrington and R.\,D.~Young for profitable discussions. 
This work was supported by: the Department of Energy, Office of Nuclear Physics, 
contract nos.\ DE-FG03-97ER4014 and DE-AC02-06CH11357; 
the Austrian Science Fund \emph{FWF} under grant no.\ W1203;
the Gordon Godfrey Fund of the School of Physics at the University of New South Wales; 
and benefited from the facilities of ANL's Computing Resource Center.
\end{acknowledge}
\medskip

\appendix

\section{Faddeev Equation}
\setcounter{section}{1}
\setcounter{dumbone}{\arabic{section}}
\label{app:FE}
\subsection{General structure}
The nucleon is represented by a Faddeev amplitude
\begin{equation} 
\label{PsiNucleon}
\Psi = \Psi_1 + \Psi_2 + \Psi_3  \,, 
\end{equation} 
where the subscript identifies the bystander quark and, e.g., $\Psi_{1,2}$ are obtained from $\Psi_3$ by a cyclic permutation of all the quark labels.  We employ the simplest realistic representation of $\Psi$.  The spin- and isospin-$1/2$ nucleon is a sum of scalar and axial-vector diquark correlations:
\begin{equation} 
\label{Psi} \Psi_3(p_i,\alpha_i,\tau_i) = {\cal N}_3^{0^+} + {\cal N}_3^{1^+}, 
\end{equation} 
with $(p_i,\alpha_i,\tau_i)$ the momentum, spin and isospin labels of the 
quarks constituting the bound state, and $P=p_1+p_2+p_3$ the system's total momentum.

The scalar diquark piece in Eq.\,(\ref{Psi}) is 
\begin{eqnarray} 
{\cal N}_3^{0^+}(p_i,\alpha_i,\tau_i)&=& [\Gamma^{0^+}(\sfrac{1}{2}p_{[12]};K)]_{\alpha_1 
\alpha_2}^{\tau_1 \tau_2}\, \Delta^{0^+}(K) \,[{\cal S}(\ell;P) u(P)]_{\alpha_3}^{\tau_3}\,,%
\label{calS} 
\end{eqnarray} 
where: the spinor satisfies (\protect\ref{App:EM}~\emph{Euclidean Conventions})
\begin{equation}
(i\gamma\cdot P + M)\, u(P) =0= \bar u(P)\, (i\gamma\cdot P + M)\,,
\end{equation}
with $M$ the mass obtained by solving the Faddeev equation, and it is also a
spinor in isospin space with $\varphi_+= {\rm col}(1,0)$ for the proton and
$\varphi_-= {\rm col}(0,1)$ for the neutron; $K= p_1+p_2=: p_{\{12\}}$,
$p_{[12]}= p_1 - p_2$, $\ell := (-p_{\{12\}} + 2 p_3)/3$; $\Delta^{0^+}$
is a pseudoparticle propagator for the scalar diquark formed from quarks $1$
and $2$, and $\Gamma^{0^+}\!$ is a Bethe-Salpeter-like amplitude describing
their relative momentum correlation; and ${\cal S}$, a $4\times 4$ Dirac
matrix, describes the relative quark-diquark momentum correlation.  (${\cal
S}$, $\Gamma^{0^+}$ and $\Delta^{0^+}$ are discussed in Sect.\,\ref{completing}.)  The colour antisymmetry of $\Psi_3$ is implicit in $\Gamma^{J^P}\!\!$, with the 
Levi-Civita tensor, $\epsilon_{c_1 c_2 c_3}$, expressed via the antisymmetric 
Gell-Mann matrices; viz., defining 
\begin{equation} 
\{H^1=i\lambda^7,H^2=-i\lambda^5,H^3=i\lambda^2\}\,, 
\end{equation} 
then $\epsilon_{c_1 c_2 c_3}= (H^{c_3})_{c_1 c_2}$.  [See 
Eqs.\,(\ref{Gamma0p}), (\ref{Gamma1p}).]

The axial-vector component in Eq.\,(\ref{Psi}) is
\begin{eqnarray} 
{\cal N}^{1^+}(p_i,\alpha_i,\tau_i) & =&  [{\tt t}^i\,\Gamma_\mu^{1^+}(\sfrac{1}{2}p_{[12]};K)]_{\alpha_1 
\alpha_2}^{\tau_1 \tau_2}\,\Delta_{\mu\nu}^{1^+}(K)\, 
[{\cal A}^{i}_\nu(\ell;P) u(P)]_{\alpha_3}^{\tau_3}\,,
\label{calA} 
\end{eqnarray} 
where the symmetric isospin-triplet matrices are 
\begin{equation} 
{\tt t}^+ = \frac{1}{\surd 2}(\tau^0+\tau^3) \,,\; 
{\tt t}^0 = \tau^1\,,\; 
{\tt t}^- = \frac{1}{\surd 2}(\tau^0-\tau^3)\,, 
\end{equation} 
and the other elements in Eq.\,(\ref{calA}) are straightforward generalisations of those in Eq.\,(\ref{calS}). 

The general forms of the matrices ${\cal S}(\ell;P)$ and ${\cal A}^i_\nu(\ell;P)$, which describe the momentum space correlation between the quark and diquark in the nucleon are described in Refs.\,\cite{Cloet:2007pi,Oettel:1998bk}.  The requirement that ${\cal S}(\ell;P)$ represent a positive energy nucleon entails
\begin{equation}
\label{Sexp} 
{\cal S}(\ell;P) = s_1(\ell;P)\,I_{\rm D} + \left(i\gamma\cdot \hat\ell - \hat\ell \cdot \hat P\, I_{\rm D}\right)\,s_2(\ell;P)\,, 
\end{equation} 
where $(I_{\rm D})_{rs}= \delta_{rs}$, $\hat \ell^2=1$, $\hat P^2= - 1$.  In the nucleon rest frame, $s_{1,2}$ describe, respectively, the upper, lower component of the bound-state nucleon's spinor.  Placing the same constraint on the axial-vector component, one has
\begin{equation}
\label{Aexp}
 {\cal A}^i_\nu(\ell;P) = \sum_{n=1}^6 \, p_n^i(\ell;P)\,\gamma_5\,A^n_{\nu}(\ell;P)\,,\; i=+,0,-\,,
\end{equation}
where ($ \hat \ell^\perp_\nu = \hat \ell_\nu + \hat \ell\cdot\hat P\, \hat P_\nu$, $ \gamma^\perp_\nu = \gamma_\nu + \gamma\cdot\hat P\, \hat P_\nu$)
\begin{equation}
\label{Afunctions}
\begin{array}{lll}
A^1_\nu= \gamma\cdot \hat \ell^\perp\, \hat P_\nu \,,\; &
A^2_\nu= -i \hat P_\nu \,,\; &
A^3_\nu= \gamma\cdot\hat \ell^\perp\,\hat \ell^\perp\,,\\
A^4_\nu= i \,\hat \ell_\mu^\perp\,,\; &
A^5_\nu= \gamma^\perp_\nu - A^3_\nu \,,\; &
A^6_\nu= i \gamma^\perp_\nu \gamma\cdot\hat \ell^\perp - A^4_\nu\,.
\end{array}
\end{equation}

One can now write the Faddeev equation satisfied by $\Psi_3$ as
\begin{equation} 
 \left[ \begin{array}{r} 
{\cal S}(k;P)\, u(P)\\ 
{\cal A}^i_\mu(k;P)\, u(P) 
\end{array}\right]\\ 
 = -\,4\,\int\frac{d^4\ell}{(2\pi)^4}\,{\cal M}(k,\ell;P) 
\left[ 
\begin{array}{r} 
{\cal S}(\ell;P)\, u(P)\\ 
{\cal A}^j_\nu(\ell;P)\, u(P) 
\end{array}\right] .
\label{FEone} 
\end{equation} 
The kernel in Eq.~(\ref{FEone}) is 
\begin{equation} 
\label{calM} {\cal M}(k,\ell;P) = \left[\begin{array}{cc} 
{\cal M}_{00} & ({\cal M}_{01})^j_\nu \\ 
({\cal M}_{10})^i_\mu & ({\cal M}_{11})^{ij}_{\mu\nu}\rule{0mm}{3ex} 
\end{array} 
\right] ,
\end{equation} 
with 
\begin{equation} 
 {\cal M}_{00} = \Gamma^{0^+}\!(k_q-\ell_{qq}/2;\ell_{qq})\, 
S^{\rm T}(\ell_{qq}-k_q) \,\bar\Gamma^{0^+}\!(\ell_q-k_{qq}/2;-k_{qq})\, 
S(\ell_q)\,\Delta^{0^+}(\ell_{qq}) \,, 
\end{equation} 
where: $\ell_q=\ell+P/3$, $k_q=k+P/3$, $\ell_{qq}=-\ell+ 2P/3$, 
$k_{qq}=-k+2P/3$ and the superscript ``T'' denotes matrix transpose; and
\begin{eqnarray}
\nonumber
\lefteqn{({\cal M}_{01})^j_\nu= {\tt t}^j \,
\Gamma_\mu^{1^+}\!(k_q-\ell_{qq}/2;\ell_{qq})} \\
&& \times 
S^{\rm T}(\ell_{qq}-k_q)\,\bar\Gamma^{0^+}\!(\ell_q-k_{qq}/2;-k_{qq})\, 
S(\ell_q)\,\Delta^{1^+}_{\mu\nu}(\ell_{qq}) \,, \label{calM01} \\ 
\nonumber \lefteqn{({\cal M}_{10})^i_\mu = 
\Gamma^{0^+}\!(k_q-\ell_{qq}/2;\ell_{qq})\, 
}\\ 
&&\times S^{\rm T}(\ell_{qq}-k_q)\,{\tt t}^i\, \bar\Gamma_\mu^{1^+}\!(\ell_q-k_{qq}/2;-k_{qq})\, 
S(\ell_q)\,\Delta^{0^+}(\ell_{qq}) \,,\\ 
\nonumber \lefteqn{({\cal M}_{11})^{ij}_{\mu\nu} = {\tt t}^j\, 
\Gamma_\rho^{1^+}\!(k_q-\ell_{qq}/2;\ell_{qq})}\\ 
&&\times \, S^{\rm T}(\ell_{qq}-k_q)\,{\tt t}^i\, \bar\Gamma^{1^+}_\mu\!(\ell_q-k_{qq}/2;-k_{qq})\, 
S(\ell_q)\,\Delta^{1^+}_{\rho\nu}(\ell_{qq}) \,. \label{calM11} 
\end{eqnarray} 

\subsection{Kernel of the Faddeev equation}
\label{completing}
To complete the Faddeev equations, Eq.\,(\ref{FEone}), one must specify the dressed-quark propagator, the diquark Bethe-Salpeter amplitudes and the diquark propagators.

\subsubsection{Dressed-quark propagator} 
\label{subsubsec:S} 
The dressed-quark propagator has the general form 
\begin{eqnarray} 
S(p) & = & -i \gamma\cdot p\, \sigma_V(p^2) + \sigma_S(p^2) = 1/[i\gamma\cdot p\, A(p^2) + B(p^2)]\label{SpAB} 
\end{eqnarray}
and can be obtained from QCD's gap equation; namely, the Dyson-Schwinger equation for the dressed-quark self-energy \cite{Roberts:1994dr}.  It is a longstanding prediction of DSE studies in QCD that for light-quarks the wave function renormalisation and dressed-quark mass: 
\begin{equation} 
\label{ZMdef}
Z(p^2)=1/A(p^2)\,,\;M(p^2)=B(p^2)/A(p^2)\,, 
\end{equation} 
respectively, receive strong momentum-dependent corrections at infrared momenta \cite{Lane:1974he,Politzer:1976tv,Roberts:1994dr}: $Z(p^2)$ is suppressed and $M(p^2)$ enhanced.  These features are an expression of dynamical chiral symmetry breaking (DCSB) and, plausibly, of confinement \cite{Roberts:2007ji}.  The enhancement of $M(p^2)$ is central to the appearance of a constituent-quark mass-scale and an existential prerequisite for Goldstone modes.   These DSE predictions are confirmed in numerical simulations of lattice-regularised QCD \cite{bowman}, and the conditions have been explored under which pointwise agreement between DSE results and lattice simulations may be obtained \cite{Bhagwat:2003vw,Bhagwat:2006tu,Bhagwat:2004kj}.

The impact of this infrared dressing on hadron phenomena has long been emphasised \cite{Roberts:1994hh} and, while numerical solutions of the quark DSE are now readily obtained, the utility of an algebraic form for $S(p)$ when calculations require the evaluation of numerous multidimensional integrals is self-evident.  An efficacious parametrisation 
of $S(p)$, which exhibits the features described above, has been used 
extensively in hadron studies \cite{Roberts:2007jh}.  It is expressed via
\begin{eqnarray} 
\bar\sigma_S(x) & =&  2\,\bar m \,{\cal F}(2 (x+\bar m^2)) + {\cal
F}(b_1 x) \,{\cal F}(b_3 x) \,  
\left[b_0 + b_2 {\cal F}(\epsilon x)\right]\,,\label{ssm} \\ 
\label{svm} \bar\sigma_V(x) & = & \frac{1}{x+\bar m^2}\, \left[ 1 - {\cal F}(2 (x+\bar m^2))\right]\,, 
\end{eqnarray}
with $x=p^2/\lambda^2$, $\bar m$ = $m/\lambda$, 
\begin{equation}
\label{defcalF}
{\cal F}(x)= \frac{1-\mbox{\rm e}^{-x}}{x}  \,, 
\end{equation}
$\bar\sigma_S(x) = \lambda\,\sigma_S(p^2)$ and $\bar\sigma_V(x) =
\lambda^2\,\sigma_V(p^2)$.  The mass-scale, $\lambda=0.566\,$GeV, and
parameter values\footnote{$\epsilon=10^{-4}$ in Eq.\ (\ref{ssm}) acts only to
decouple the large- and intermediate-$p^2$ domains.}
\begin{equation} 
\label{tableA} 
\begin{array}{ccccc} 
   \bar m& b_0 & b_1 & b_2 & b_3 \\\hline 
   0.00897 & 0.131 & 2.90 & 0.603 & 0.185 
\end{array}\;, 
\end{equation} 
were fixed in a least-squares fit to light-meson observables \cite{mark,valencedistn}.  The dimensionless $u=d$ current-quark mass in Eq.~(\ref{tableA}) corresponds to
\begin{equation} 
\label{mcq}
m=5.08\,{\rm MeV} = :m^{\rm phys}\,. 
\end{equation} 
The parametrisation yields a Euclidean constituent-quark mass
\begin{equation} 
\label{MEq} M_{u,d}^E = 0.33\,{\rm GeV}, 
\end{equation} 
defined as the solution of $p^2=M^2(p^2)$.  

The ratio $M^E/m = 65$ is one expression of DCSB in the parametrisation of $S(p)$.  It emphasises the dramatic enhancement of the dressed-quark mass function at infrared momenta. Another is the chiral-limit vacuum quark condensate \cite{Roberts:1994hh}
\begin{equation}
\label{qbqparam}
-\langle \bar q q \rangle_\zeta^0 = \lambda^3 \, \frac{3}{4\pi^2}\, \frac{b_0}{b_1 b_3} \, \ln \frac{\zeta^2}{\Lambda_{\rm QCD}^2},
\end{equation}
which assumes the value ($\Lambda_{\rm QCD} = 0.2\,$GeV)
\begin{equation}
-\langle \bar q q \rangle_{\zeta=1\,{\rm GeV}}^0 = (0.221 \,{\rm GeV})^3.
\end{equation}
A detailed discussion of the vacuum quark condensate in QCD can be found in Ref.\,\cite{Langfeld:2003ye,Chang:2006bm}

\subsubsection{Diquark Bethe-Salpeter amplitudes}
\label{qqBSA}
The rainbow-ladder DSE truncation yields asymptotic diquark states in the strong interaction spectrum.  Such states are not observed and their appearance is an artefact of the truncation.  Higher-order terms in the quark-quark scattering kernel, whose analogue in the quark-antiquark channel do not much affect the properties of vector and flavour non-singlet pseudoscalar mesons, ensure that QCD's quark-quark scattering matrix does not exhibit singularities which correspond to asymptotic diquark states~\cite{Bhagwat:2004hn}.  Nevertheless, studies with kernels that don't generate diquark bound states, do support a physical interpretation of the masses, $m_{(qq)_{J^P}}$, obtained using the rainbow-ladder truncation: the quantity $l_{(qq)_{J^P}}=1/m_{(qq)_{J^P}}$ may be interpreted as a range over which the diquark correlation can propagate within a baryon.  These observations motivate an {\it Ansatz} for the quark-quark scattering matrix that is employed in deriving the Faddeev equation: \begin{equation} 
[M_{qq}(k,q;K)]_{rs}^{tu} = \sum_{J^P=0^+,1^+,\ldots} \bar\Gamma^{J^P}\!(k;-K)\, \Delta^{J^P}\!(K) \, \Gamma^{J^P}\!(q;K)\,. \label{AnsatzMqq} 
\end{equation}  

One manner of specifying the $\Gamma^{J^P}\!\!$ in Eq.\,(\ref{AnsatzMqq}) is to employ the solutions of a rainbow-ladder quark-quark Bethe-Salpeter equation (BSE), as e.g.\ in Refs.\ \cite{Eichmann:2008ef,Maris:2002yu,Maris:2004bp}.  Using the properties of the Gell-Mann matrices one finds easily that $\Gamma^{J^P}_C:= \Gamma^{J^P}C^\dagger$ satisfies exactly the same equation as the $J^{-P}$ colour-singlet meson {\it but} for a halving of the coupling strength \cite{Cahill:1987qr}.  This makes clear that the interaction in the ${\bar 3_c}$ $(qq)$ channel is strong and attractive.  

A solution of the BSE equation requires a simultaneous solution of the quark-DSE.  However, since we choose to simplify the calculations by parametrising $S(p)$, we also employ that expedient with $\Gamma^{J^P}\!$, using the following one-parameter forms: 
\begin{eqnarray} 
\label{Gamma0p} \Gamma^{0^+}(k;K) &=& \frac{1}{{\cal N}^{0^+}} \, 
H^a\,C i\gamma_5\, i\tau_2\, {\cal F}(k^2/\omega_{0^+}^2) \,, \\ 
\label{Gamma1p} {\tt t}^i \Gamma^{1^+}_\mu (k;K) &=& \frac{1}{{\cal N}^{1^+}}\, 
H^a\,i\gamma_\mu C\,{\tt t}^i\, {\cal F}(k^2/\omega_{1^+}^2)\,, 
\end{eqnarray} 
with the normalisation, ${\cal N}^{J^P}\!$, fixed by requiring 
\begin{eqnarray}
\label{BSEnorm} 
2 \,K_\mu & = & 
\left[ \frac{\partial}{\partial Q_\mu} \Pi(K,Q) \right]_{Q=K}^{{K^2=-m_{J^P}^2}},\\
\Pi(K,Q) & = & tr\!\! \int\!\! 
\frac{d^4 q}{(2\pi)^4}\, \bar\Gamma(q;-K) \, S(q+Q/2) \, \Gamma(q;K) \, S^{\rm T}(-q+Q/2) .
\end{eqnarray}

The {\it Ans\"atze} of Eqs.\,(\ref{Gamma0p}), (\ref{Gamma1p}) retain only that single Dirac-amplitude which would represent a point particle with the given quantum numbers in a local Lagrangian density.  They are usually the dominant amplitudes in a solution of the rainbow-ladder BSE for the lowest mass $J^P$ diquarks \cite{Burden:1996nh,Maris:2002yu} and mesons \cite{Maris:1997tm,Maris:1999nt,Maris:1999bh}.

\subsubsection{Diquark propagators}
\label{qqprop}
Solving for the quark-quark scattering matrix using the rainbow-ladder truncation yields free particle propagators for $\Delta^{J^P}$ in 
Eq.~(\ref{AnsatzMqq}).  As already noted, however, higher-order contributions 
remedy that defect, eliminating asymptotic diquark states from the spectrum.  The attendant modification of $\Delta^{J^P}$ can be modelled efficiently by simple functions that are free-particle-like at spacelike momenta but pole-free on the timelike axis \cite{Bhagwat:2004hn}; namely,\footnote{These forms satisfy a sufficient condition for confinement because of the associated violation of reflection positivity.  See Sect.~2 of Ref.\,\cite{Roberts:2007ji} for a brief discussion.}
\begin{eqnarray} 
\Delta^{0^+}(K) & = & \frac{1}{m_{0^+}^2}\,{\cal F}(K^2/\omega_{0^+}^2)\,,\\ 
\Delta^{1^+}_{\mu\nu}(K) & = & 
\left(\delta_{\mu\nu} + \frac{K_\mu K_\nu}{m_{1^+}^2}\right) \, \frac{1}{m_{1^+}^2}\, {\cal F}(K^2/\omega_{1^+}^2) \,,
\end{eqnarray} 
where the two parameters $m_{J^P}$ are diquark pseudoparticle masses and 
$\omega_{J^P}$ are widths characterising $\Gamma^{J^P}\!$.  Herein we require additionally that
\begin{equation}
\label{DQPropConstr}
\left. \frac{d}{d K^2}\,\left(\frac{1}{m_{J^P}^2}\,{\cal F}(K^2/\omega_{J^P}^2)\right)^{-1} \right|_{K^2=0}\! = 1 \; \Rightarrow \; \omega_{J^P}^2 = \sfrac{1}{2}\,m_{J^P}^2\,,
\end{equation} 
which is a normalisation that accentuates the free-particle-like propagation characteristics of the diquarks {\it within} the hadron. 

\section{Euclidean Conventions}
\label{App:EM} 
In our Euclidean formulation: 
\begin{equation} 
p\cdot q=\sum_{i=1}^4 p_i q_i\,; 
\end{equation} 
\begin{equation}
\{\gamma_\mu,\gamma_\nu\}=2\,\delta_{\mu\nu}\,;\; 
\gamma_\mu^\dagger = \gamma_\mu\,;\; 
\sigma_{\mu\nu}= \sfrac{i}{2}[\gamma_\mu,\gamma_\nu]\,; \;
{\rm tr}\,[\gamma_5\gamma_\mu\gamma_\nu\gamma_\rho\gamma_\sigma]= 
-4\,\epsilon_{\mu\nu\rho\sigma}\,, \epsilon_{1234}= 1\,.  
\end{equation}

A positive energy spinor satisfies 
\begin{equation} 
\bar u(P,s)\, (i \gamma\cdot P + M) = 0 = (i\gamma\cdot P + M)\, u(P,s)\,, 
\end{equation} 
where $s=\pm$ is the spin label.  It is normalised: 
\begin{equation} 
\bar u(P,s) \, u(P,s) = 2 M \,,
\end{equation} 
and may be expressed explicitly: 
\begin{equation} 
u(P,s) = \sqrt{M- i {\cal E}}\left( 
\begin{array}{l} 
\chi_s\\ 
\displaystyle \frac{\vec{\sigma}\cdot \vec{P}}{M - i {\cal E}} \chi_s 
\end{array} 
\right)\,, 
\end{equation} 
with ${\cal E} = i \sqrt{\vec{P}^2 + M^2}$, 
\begin{equation} 
\chi_+ = \left( \begin{array}{c} 1 \\ 0  \end{array}\right)\,,\; 
\chi_- = \left( \begin{array}{c} 0\\ 1  \end{array}\right)\,. 
\end{equation} 
For the free-particle spinor, $\bar u(P,s)= u(P,s)^\dagger \gamma_4$. 
 
The spinor can be used to construct a positive energy projection operator: 
\begin{equation} 
\label{Lplus} \Lambda_+(P):= \frac{1}{2 M}\,\sum_{s=\pm} \, u(P,s) \, \bar 
u(P,s) = \frac{1}{2M} \left( -i \gamma\cdot P + M\right). 
\end{equation} 
 
A negative energy spinor satisfies 
\begin{equation} 
\bar v(P,s)\,(i\gamma\cdot P - M) = 0 = (i\gamma\cdot P - M) \, v(P,s)\,, 
\end{equation} 
and possesses properties and satisfies constraints obtained via obvious analogy 
with $u(P,s)$. 
 
A charge-conjugated Bethe-Salpeter amplitude is obtained via 
\begin{equation} 
\label{chargec}
\bar\Gamma(k;P) = C^\dagger \, \Gamma(-k;P)^{\rm T}\,C\,, 
\end{equation} 
where ``T'' denotes a transposing of all matrix indices and 
$C=\gamma_2\gamma_4$ is the charge conjugation matrix, $C^\dagger=-C$. 

\section{Nucleon-Photon Vertex}
\label{NPVertex}
In order to explicate the vertex depicted in Fig.\,\ref{vertex} we write the scalar and axial-vector components of the nucleons' Faddeev amplitudes in the form [cf.\ Eq.\,(\ref{FEone})]
\begin{equation}
\label{NucWF}
\Psi(k;P) = \left[
\begin{array}{l}
\Psi^{s}(k;P)\\
\Psi_{\mu}^{i}(k;P)
\end{array}
\right]
= \left[ 
\begin{array}{l}
\mathcal{S}(k;P) u(P)\\
\mathcal{A}_{\mu}^{i}(k;P)u(P)
\end{array}
\right],
\qquad i=s,+,0,-\,.
\end{equation}
For explicit calculations, we work in the Breit frame: $P_\mu=P_\mu^{BF}-Q_\mu /2$, $P'_\mu=P_\mu^{BF}+Q_\mu /2$ and $P_\mu^{BF}=(0,0,0,i\sqrt{M_n^2+Q^2/4})$, and write the electromagnetic current matrix element as [cf.\ Eq.\,(\ref{Jnucleon})]
\begin{eqnarray}
\label{ABcurrent}
\left\langle P' | \hat{J}_{\mu}^{em} | P \right\rangle
&=& \Lambda^+(P')
\left[ \gamma_\mu G_E + M_n \frac{P_{\mu}^{BF}}{P_{BF}^{2}}
(G_E-G_M) \right] \Lambda^+(P)\,,
\\
&=& \int \frac{d^4 p}{(2\pi)^4}\,\frac{d^4 k}{(2\pi)^4}\,
\bar{\Psi}(-p,P') J_{\mu}^{em}(p,P';k,P) \Psi(k,P)\,.
\end{eqnarray}
In Fig.\,\ref{vertex} we have separated the current, $J_{\mu}^{em}(p,P';k,P)$, into a sum of six terms, each of which we subsequently make precise.  NB.\ Diagrams~1, 2 and 4 are one-loop integrals, which we evaluate by Gau{\ss}ian quadrature.  The remainder, Diagrams 3, 5 and 6, are two-loop integrals, for whose evaluation Monte-Carlo methods are employed.  A technical aspect concerning the computation is described in \ref{CHexpand}~\emph{Chebyshev expansion}.

\subsection{Diagram~1}
\label{Diag1}
This represents the photon coupling directly to the bystander quark. It is expressed as \begin{eqnarray}
\label{B1}
J_{\mu}^{qu} &=&  S(p_q) \hat{\Gamma}_{\mu}^{qu}(p_q;k_q) S(k_q) 
\left(\Delta^{0^+}(k_s) + \Delta^{1^+}(k_s) \right)
(2\pi)^4 \delta^4(p-k-\hat{\eta}Q)\,,
\end{eqnarray}
where $\hat\Gamma_{\mu}^{qu}(p_q;k_q)= Q_q \, \Gamma_{\mu}(p_q;k_q)$, with $Q_q={\rm diag}[2/3,-1/3]$ being the quark electric charge matrix, and $\Gamma_{\mu}(p_q;k_q)$ is the dressed-quark-photon vertex.  In Eq.\,(\ref{B1}) the momenta are
\begin{eqnarray}
\label{etavalue}
\begin{array}{lc@{\qquad}l}
k_q=\eta P+k\,, & & p_q=\eta P'+p\,, \\
k_d=\hat{\eta}P-k\,, & & p_d=\hat{\eta}P'-p\,,
\end{array}
\end{eqnarray}
with $\eta + \hat{\eta}=1$.  The results reported herein were obtained with $\eta=1/3$, which provides a single quark with one-third of the baryon's total momentum and is thus a natural choice.  Notably, as our approach is manifestly Poincar\'e covariant, the precise value is immaterial so long as the numerical methods preserve that covariance.  Calculations converge most quickly with the natural choice. 

It is a necessary condition for current conservation that the quark-photon vertex satisfy the Ward-Takahashi identity:
\begin{equation}
\label{vwti}
Q_\mu \, i\Gamma_\mu(\ell_1,\ell_2) = S^{-1}(\ell_1) - S^{-1}(\ell_2)\,,
\end{equation}
where $Q=\ell_1-\ell_2$ is the photon momentum flowing into the vertex.  Since the quark is dressed, Sec.\,\ref{subsubsec:S}, the vertex is not bare; i.e., $\Gamma_\mu(\ell_1,\ell_2) \neq \gamma_\mu$.  It can be obtained by solving an inhomogeneous Bethe-Salpeter equation, which was the procedure adopted in the DSE calculation that successfully predicted the electromagnetic pion form factor \cite{Maris:2000sk,Maris:1999bh}.  However, since we have parametrised $S(p)$, we follow Ref.~\cite{Roberts:1994hh} and write \cite{bc80}
\begin{equation}
\label{bcvtx}
i\Gamma_\mu(\ell_1,\ell_2)  =  
i\Sigma_A(\ell_1^2,\ell_2^2)\,\gamma_\mu +
2 k_\mu \left[i\gamma\cdot k_\mu \,
\Delta_A(\ell_1^2,\ell_2^2) + \Delta_B(\ell_1^2,\ell_2^2)\right] \!;
\end{equation}
with $k= (\ell_1+\ell_2)/2$, $Q=(\ell_1-\ell_2)$ and
\begin{equation}
\Sigma_F(\ell_1^2,\ell_2^2) = \sfrac{1}{2}\,[F(\ell_1^2)+F(\ell_2^2)]\,,\;
\Delta_F(\ell_1^2,\ell_2^2) =
\frac{F(\ell_1^2)-F(\ell_2^2)}{\ell_1^2-\ell_2^2}\,,
\label{DeltaF}
\end{equation}
where $F= A, B$; viz., the scalar functions in Eq.\,(\ref{SpAB}).  It is
critical that $\Gamma_\mu$ in Eq.\ (\ref{bcvtx}) satisfies Eq.\ (\ref{vwti})
and very useful that it is completely determined by the dressed-quark
propagator.  

\subsection{Diagram~2} 
\label{dqff}
This figure depicts the photon coupling directly to a diquark correlation.  It is expressed as 
\begin{eqnarray}
\label{B2}
J_{\mu}^{dq} &=& \Delta^i(p_{d}) 
\left[ \hat{\Gamma}_{\mu}^{dq}(p_{d};k_{d}) \right]^{i j} 
\Delta^{j}(k_{d}) S(k_q)
(2\pi)^4 \delta^4(p-k+\eta Q)\,,
\end{eqnarray}
with $ [\hat{\Gamma}_{\mu}^{dq}(p_{d};k_{d})]^{i j}={\rm diag}[Q_{0^+} \Gamma_\mu^{0^+},Q_{1^+}\Gamma_\mu^{1^+}] $, where $Q_{0^+}=1/3$ and $\Gamma_\mu^{0^+}$ is given in Eq.\,(\ref{Gamma0plus}), and $Q_{1^+}={\rm diag}[4/3,1/3,-2/3]$ with $\Gamma_\mu^{1^+}$ given in Eq.\,(\ref{AXDQGam}).  Naturally, the diquark propagators match the line to which they are attached. 

In the case of a scalar correlation, the general form of the diquark-photon vertex is
\begin{equation}
\Gamma_\mu^{0^+}(\ell_1,\ell_2) = 2\, k_\mu\, f_+(k^2,k\cdot Q,Q^2) + Q_\mu  \, f_-(k^2,k\cdot Q,Q^2)\,.
\end{equation}
If one is dealing with an elementary scalar correlation, then the Ward-Takahashi identity reads: 
\begin{equation}
\label{VWTI0}
Q_\mu \,\Gamma_\mu^{0^+}(\ell_1,\ell_2) = \Pi^{0^+}(\ell_1^2)  - \Pi^{0^+}(\ell_2^2)\,,\; \Pi^{J^P}(\ell^2) = \{\Delta^{J^P}(\ell^2)\}^{-1}.
\end{equation} 
However, for a composite system of the type we consider this identity is modified; viz. \cite{Frank:1993ye}, 
\begin{equation}
\label{VWTI0Comp}
Q_\mu \,\Gamma_\mu^{0^+}(\ell_1,\ell_2) = \left[\Pi^{0^+}(\ell_1^2)  - \Pi^{0^+}(\ell_2^2)\right] F_{qq}(Q^2)\,,
\end{equation}
where 
\begin{equation}
\label{Fqqform}
F_{qq}(Q^2) = \frac{1}{1 + \frac{1}{6} r_{qq}^2 Q^2}
\end{equation}
is a form factor describing the distribution of charge within the correlation.  

The evaluation of scalar diquark elastic electromagnetic form factors in Ref.\,\cite{Maris:2004bp} is a first step toward calculating $\Gamma_\mu^{0^+}(\ell_1,\ell_2)$.  However, in providing only an on-shell component, it is insufficient for our requirements.  We choose to adapt Eq.\,(\ref{bcvtx}) to our needs and employ
\begin{equation}
\label{Gamma0plus}
\Gamma_\mu^{0^+}(\ell_1,\ell_2) =  k_\mu\,
\Delta_{\Pi^{0^+}}(\ell_1^2,\ell_2^2)\,F_{qq}(Q^2)\,,
%
\end{equation}  
with the definition of $\Delta_{\Pi^{0^+}}(\ell_1^2,\ell_2^2)$ apparent from Eq.\,(\ref{DeltaF}) and the value of $r_{qq}$ given in Eq.\,(\ref{rqqvalue}).

Equation~(\ref{Gamma0plus}) is an \textit{Ansatz} that satisfies Eq.\,(\ref{VWTI0Comp}), is completely determined by quantities introduced already and is free of kinematic singularities on the relevant domain.  It implements $f_- \equiv 0$, which is a requirement for elastic form factors, and guarantees a valid normalisation of electric charge; viz., 
\begin{equation}
\lim_{\ell^\prime\to \ell} \Gamma_\mu^{0^+}(\ell^\prime,\ell) = 2 \, \ell_{\mu} \, \frac{d}{d\ell^2}\, \Pi^{0^+}(\ell^2) \stackrel{\ell^2\sim 0}{=} 2 \, \ell_{\mu}\,,
\end{equation}
owing to Eq.\,(\ref{DQPropConstr}).  NB.\ We have factored the fractional diquark charge, which therefore appears subsequently in our calculations as a simple multiplicative factor. 

For the case in which the struck diquark correlation is axial-vector and the scattering is elastic, the vertex assumes the form \cite{HawesPichowsky99}:\,\footnote{If the scattering is inelastic the general form of the vertex involves eight scalar functions \protect\cite{Salam64}.  For simplicity, we ignore the additional structure in this \textit{Ansatz}.  
}
\begin{equation}
\label{AXDQGam}
\Gamma^{1^+}_{\mu\alpha\beta}(\ell_1,\ell_2) 
= -\sum_{i=1}^{3} \Gamma^{\rm [i]}_{\mu\alpha\beta}(\ell_1,\ell_2)\,,
\end{equation}
with ($T_{\alpha\beta}(\ell) = \delta_{\alpha\beta} - \ell_\alpha \ell_\beta/\ell^2$)
\begin{eqnarray}
\label{AXDQGam1}
\Gamma^{\rm [1]}_{\mu\alpha\beta}(\ell_1,\ell_2) 
&=& (\ell_1+\ell_2)_\mu \, T_{\alpha\lambda}(\ell_1) \, T_{\lambda\beta}(\ell_2)\; F_1(\ell_1^2,\ell_2^2)\,,
\\
\label{AXDQGam2}
\Gamma^{\rm [2]}_{\mu\alpha\beta}(\ell_1,\ell_2)
&=& \left[ T_{\mu\alpha}(\ell_1)\, T_{\beta\rho}(\ell_2) \, \ell_{1 \rho}
+ T_{\mu\beta}(\ell_2) \, T_{\alpha\rho}(\ell_1) \, \ell_{2\rho} \right] F_{2}(\ell_1^2,\ell_2^2) \,,
\\ 
\label{AXDQGam3}
\Gamma^{\rm [3]}_{\mu\alpha\beta}(\ell_1,\ell_2)
&=& -\frac{1}{2 m_{1^+}^2}\, (\ell_1+\ell_2)_\mu\, T_{\alpha\rho}(\ell_1)\, \ell_{2 \rho}
\, T_{\beta\lambda}(\ell_2)\, \ell_{1 \lambda}\; F_{3}(\ell_1^2,\ell_2^2) \,.
\end{eqnarray}
This vertex satisfies:
\begin{equation}
\ell_{1\alpha} \, \Gamma^{1^+}_{\mu\alpha\beta}(\ell_1,\ell_2) = 0 = 
\Gamma^{1^+}_{\mu\alpha\beta}(\ell_1,\ell_2) \, \ell_{2\beta} \,,
\end{equation}
which is a general requirement of the elastic electromagnetic vertex of axial-vector bound states and guarantees that the interaction does not induce a pseudoscalar component in the axial-vector correlation.  We note that the electric, magnetic and quadrupole form factors of an axial-vector bound state are expressed \cite{HawesPichowsky99}
\begin{eqnarray}
\label{GEDQ}
& &
G_{\cal E}^{1^+}(Q^2) = F_1 + \sfrac{2}{3}\, \tau_{1^+}\, 
G_{\cal Q}^{1^+}(Q^2) \,, \; \tau_{1^+} = \frac{Q^2}{ 5 \,m_{1^{+}}^{2}}\,,
\\
\label{GMDQ}
& &
G_{\cal M}^{1^+}(Q^2) = - F_2(Q^2) ~,
\\
& &
\label{GQDQ}
G_{\cal Q}^{1^+}(Q^2) = F_1(Q^2) + F_2(Q^2) + \left( 1 + \tau_{1^+}\right) F_3(Q^2) \,.
\end{eqnarray}

Owing to the fact that $\Gamma^{J^P}_C:= \Gamma^{J^P}C^\dagger$ satisfies exactly the same Bethe-Salpeter equation as the $J^{-P}$ colour-singlet meson {\it but} for a halving of the coupling strength, the vector meson form factor calculation in Ref.\,\cite{Bhagwat:2006pu} might become useful as a guide in understanding the form factors in Eqs.\,(\ref{AXDQGam})--(\ref{AXDQGam3}).  However, in providing only an on-shell component, that information is insufficient for our requirements.  Hence we employ the following \textit{Ans\"atze}:
\begin{eqnarray}
\label{AnsatzF1}
F_{1}(\ell_1^2,\ell_2^2) &=& \Delta_{\Pi^{1^+}}(\ell_1^2,\ell_2^2)\,F_{qq}(Q^2)\,, \\
\label{AnsatzF2}
F_{2}(\ell_1^2,\ell_2^2) &=& -\, F_{1}  + 
(1-\tau_{1^+}) \,( \tau_{1^+} F_{1}+1 - \mu_{1^+})\, d(\tau_{1^+})\,, \\
\label{AnsatzF3}
F_{3}(\ell_1^2,\ell_2^2) &=& -\,(\chi_{1^+}\,(1- \tau_{1^+})\,d(\tau_{1^+})+F_1 + F_2)\, d(\tau_{1^+})\,,
\end{eqnarray}
with $d(x)=1/(1+x)^2$.  This construction ensures a valid electric charge normalisation for the axial-vector correlation; viz., 
\begin{equation}
\lim_{\ell^\prime \to\ell} \, \Gamma^{1^+}_{\mu\alpha\beta}(\ell^\prime,\ell) = T_{\alpha\beta}(\ell) \,\frac{d}{d\ell^2}\, \Pi^{1^+}(\ell^2) 
\stackrel{\ell^2\sim 0}{=}  T_{\alpha\beta}(\ell) \,2 \,\ell_{ \mu}\,,
\end{equation}
owing to Eq.\,(\ref{DQPropConstr}), and current conservation 
\begin{equation}
\lim_{\ell_2\to\ell_1} \, Q_\mu \Gamma^{1^+}_{\mu\alpha\beta}(\ell_1,\ell_2) = 0\,.
\end{equation}
The diquark's static electromagnetic properties follow: 
\begin{equation}
\label{pointp}
G_{\cal E}^{1^+}(0) = 1\,,\;
G_{\cal M}^{1^+}(0) = \mu_{1^+}\,,\;
G_{\cal Q}^{1^+}(0) = -\chi_{1^+}\,.
\end{equation}
For an on-shell or pointlike axial-vector: $\mu_{1^+}=2$; and $\chi_{1^+}=1$.  In addition, Eqs.\,(\ref{AXDQGam})--(\ref{AXDQGam3}) with Eqs.\,(\ref{AnsatzF1})--(\ref{AnsatzF3}) realise the constraints of Ref.\,\cite{brodskyhiller92}; namely, independent of the values of $\mu_{1^+}$ \& $\chi_{1^+}$, the form factors assume the ratios
\begin{equation}
\label{pQCDavdq}
G_{\cal E}^{1^+}(Q^2): G_{\cal M}^{1^+}(Q^2): G_{\cal Q}^{1^+}(Q^2)
\stackrel{Q^2\to \infty}{=} (1 - \sfrac{2}{3} \tau_{1^+}) : 2 : - 1 \,.
\end{equation}

It is noteworthy that within a nucleon the diquark correlation is not on-shell.  Hence, in contrast with Ref.\,\cite{Alkofer:2004yf}, we do not assume herein that a point-particle value for the magnetic moment in Eq.\,(\ref{pointp}) serves as a good reference point.  Instead we employ the value determined in Ref.\,\cite{Cloet:2008wg}:
\begin{equation}
\mu_{1^+}=0.37\,,
\end{equation}
which is in accord with that obtained following the approach in Ref.\,\cite{Eichmann:2008ef}.  While equally one need not employ the point-particle value for $\chi_{1^+}$, changing to $\chi_{1^+}=0$ has little impact on the results \protect\cite{Alkofer:2004yf}.  We therefore stay with  $\chi_{1^+}=1$.

\subsection{Diagram~3}
This image depicts a photon coupling to the quark that is exchanged as one diquark breaks up and another is formed.  It is expressed as
\begin{equation}
\label{B3}
J_{\mu}^{ex} = -\frac{1}{2} S(k_{q}) \Delta^{i}(k_{d})
\Gamma^{i}(p_1,k_{d})
S^T(q) \hat{\Gamma}_{\mu}^{quT}(q',q) S^T(q')
\bar{\Gamma}^{jT}(p'_2,p_{d})
\Delta^j(p_{d}) S(p_{q})\,,
\end{equation}
wherein the vertex $\hat{\Gamma}_{\mu}^{qu}$ appeared in Eq.\,(\ref{B1}).  While this is the first two-loop diagram we have described, no new elements appear in its specification: the dressed-quark-photon vertex was discussed in Sec.~\ref{Diag1}.  In Eq.\,(\ref{B3}) the momenta are 
\begin{eqnarray}
\begin{array}{lc@{\qquad}l}
q = \hat{\eta}P-\eta P'-p-k\,, & & q' = \hat{\eta}P'-\eta P-p-k \,,\\
p_1 = (p_q-q)/2\,,& & p'_2 = (-k_q+q')/2 \,,\\
p'_1 = (p_q-q')/2\,, & & p_2 = (-k_q+q)/2 \,.
\end{array}
\end{eqnarray}

It is noteworthy that the process of quark exchange provides the attraction necessary in the Faddeev equation to bind the nucleon.  It also guarantees that the Faddeev amplitude has the correct antisymmetry under the exchange of any two dressed-quarks.  This key feature is absent in models with elementary (noncomposite) diquarks.  The full contribution is obtained by summing over the superscripts $i,j$, which can each take the values $0^+$, $1^+$.  

\subsection{Diagram 4}
\label{kTsec}
This differs from Diagram~2 in expressing the contribution to the nucleons' form factors owing to an electromagnetically induced transition between scalar and axial-vector diquarks.  The explicit expression is given by Eq.\,(\ref{B2}) with $[\hat{\Gamma}_{\mu}^{dq}(p_{d};k_{d})]^{i= j}=0$, and $[\hat{\Gamma}_{\mu}^{dq}(p_{d};k_{d})]^{1,2}=\Gamma_{SA}$ and $[\hat{\Gamma}_{\mu}^{dq}(p_{d};k_{d})]^{2,1}=\Gamma_{AS}$.  This transition vertex is a rank-2 pseudotensor, kindred to the matrix element describing the $\rho\, \gamma^\ast \pi^0$  transition \cite{maristransition}, and can therefore be expressed 
\begin{equation}
\label{SAPhotVertex}
\Gamma_{SA}^{\gamma\alpha}(\ell_1,\ell_2) = -\Gamma_{AS}^{\gamma\alpha}(\ell_1,\ell_2) 
= \frac{i}{M_N} \, {\cal T}(\ell_1,\ell_2) \, \varepsilon_{\gamma\alpha\rho\lambda}\ell_{1\rho} \ell_{2 \lambda}\,,
\end{equation}
where $\gamma$, $\alpha$ are, respectively, the vector indices of the photon and axial-vector diquark.  For simplicity we proceed under the assumption that
\begin{equation}
\label{calTvalue}
{\cal T}(\ell_1,\ell_2) = \kappa_{\cal T}\,;
\end{equation}
A typical on-shell value for the dimensionless normalisation is
$\kappa_{\cal T} \sim 2$ \cite{Oettel:2000jj}.  
However, as with $\mu_{1^+}$, we recognise herein that this value is not a useful reference point because, for the processes described by Fig.\,\ref{vertex}, $\kappa_{\cal T}$ can be much smaller in magnitude.  We use the value determined in Ref.\,\cite{Cloet:2008wg}:
\begin{equation}
\label{kappaTfit}
\kappa_{\cal T} = 0.12\,.
\end{equation}
This diagram impacts upon the nucleons' magnetic form factors \cite{Cloet:2007pi,Alkofer:2004yf,Cloet:2008wg}.

\subsection{Diagrams~5 \& 6}
\label{X56}
These two-loop diagrams are the so-called ``seagull'' terms, which appear as partners to Diagram~3 and arise because binding in the nucleons' Faddeev equations is effected by the exchange of a dressed-quark between \textit{nonpointlike} diquark correlations \cite{Oettel:1999gc}.  The explicit expression for their contribution to the nucleons' form factors is 
\begin{eqnarray}
\label{B5}
J_{\mu}^{sg} &=& \frac{1}{2} S(k_{q}) \Delta^{i}(k_{d}) 
\left( X_{\mu}^{i}(p_q,q',k_d) S^T(q')
\bar{\Gamma}^{jT}(p'_2,p_{d})
\right.
\nonumber\\
& & -
\left. 
\Gamma^{i}(p_1,k_{d}) S^T(q) 
\bar{X}_{\mu}^{j}(-k_q,-q,p_d)
\right) \Delta^{j}(p_{d}) S(p_{q})\,,
\end{eqnarray}
where, again, the superscripts are summed. 
  
The new elements in these diagrams are the couplings of a photon to two dressed-quarks as they either separate from (Diagram~5) or combine to form (Diagram~6) a diquark correlation.  As such they are components of the five point Schwinger function which describes the coupling of a photon to the quark-quark scattering kernel.  This Schwinger function could be calculated, as is evident from the computation of analogous Schwinger functions relevant to meson observables \cite{marisgppp}.  However, such a calculation provides relevant input only when a uniform truncation of the DSEs has been employed to calculate each of the elements described hitherto.  We must instead employ an algebraic parametrisation \cite{Oettel:1999gc}, which for Diagram~5 reads
\begin{eqnarray}
\nonumber
X^{J^P}_\mu(k,Q) & =&  e_{\rm by}\,\frac{4 k_\mu- Q_\mu}{4 k\cdot Q - Q^2}\,\left[\Gamma^{J^P}\!(k-Q/2)-\Gamma^{J^P}\!(k)\right]\\
& +& e_{\rm ex}\,\frac{4 k_\mu+ Q_\mu}{4 k\cdot Q + Q^2}\,\left[\Gamma^{J^P}\!(k+Q/2)-\Gamma^{J^P}\!(k)\right], \label{X5}
\end{eqnarray}
with $k$ the relative momentum between the quarks in the initial diquark, $e_{\rm by}$ the electric charge of the quark which becomes the bystander, and $e_{\rm ex}$ the charge of the quark that is reabsorbed into the final diquark.  Diagram~6 has
\begin{eqnarray}
\nonumber
\bar X^{J^P}_\mu(k,Q) & =&  e_{\rm by}\,\frac{4 k_\mu+ Q_\mu}{4 k\cdot Q + Q^2}\,\left[\bar\Gamma^{J^P}\!(k+Q/2)-\bar\Gamma^{J^P}\!(k)\right]\\
& +& e_{\rm ex}\,\frac{4 k_\mu-Q_\mu}{4 k\cdot Q - Q^2}\,\left[\bar\Gamma^{J^P}\!(k-Q/2)-\bar\Gamma^{J^P}\!(k)\right], \label{X6}
\end{eqnarray}
where $\bar\Gamma^{J^P}\!(\ell)$ is the charge-conjugated amplitude, Eq.\,(\ref{chargec}).  Plainly, these terms vanish if the diquark correlation is represented by a momentum-independent Bethe-Salpeter-like amplitude; i.e., the diquark is pointlike.

It is naturally possible to use more complicated \textit{Ans\"atze} \cite{Eichmann:2008ef}.  However, like Eq.\,(\ref{Gamma0plus}), Eqs.\,(\ref{X5}) \& (\ref{X6}) are simple forms, free of kinematic singularities and sufficient to ensure the nucleon-photon vertex satisfies the Ward-Takahashi identity when the composite nucleon is obtained from the Faddeev equation.

\section{Chebyshev expansion}
\label{CHexpand}
In solving the Faddeev equation we employ a Chebyshev expansion of the scalar functions appearing in the Faddeev amplitude and wave function in order to restrain the use of computer memory.  (See, e.g., Ref.\,\cite{Maris:1997tm}.)  The results herein were obtained with twelve terms in both.  The Chebyshev-expanded functions then define the Faddeev amplitude that appears and is evaluated in the expressions for the form factors.  Without due care, this can lead to a problem; namely, with increasing $Q^2$ a function can be computed outside the expansion's domain of convergence.  

Consider a function $F(k^2,k\cdot P; P^2)$, which represents a term in the Faddeev amplitude.  It is a function of only two variables: $k^2$ and $k\cdot P$, where $k$ is the relative quark-diquark momentum, because the total momentum always satisfies $P^2=-M^2$, where $M$ is the bound-state's mass.  In the bound-state's rest frame one can define an angle $\alpha$ through
\begin{equation}
\label{cosalpha}
i |k| M \cos \alpha := k\cdot P\,.
\end{equation}
Then, with $\{U_i(x),\,j=1\,\ldots\infty\}$ being Chebyshev polynomials of the second kind, 
\begin{equation}
\label{cheb}
F(k^2,k\cdot P; -M^2) = \lim_{N_m\to \infty} \sum_{j=0}^{N_m} \,^j\! F(|k|,i M;-M^2) \,  U_j(\cos\beta)\,.
\end{equation}
For any finite $N_m$ the expansion in Eq.\,(\ref{cheb}) is a true approximation to the $k\cdot P$-dependence of the function $F$ in the sense that, with increasing $N_m$, the right-hand-side (rhs) is uniformly pointwise an increasingly accurate representation of the function.  The lhs of Eq.\,(\ref{cheb}) is Poincar\'e invariant.  Hence, in the limit $N_m \to \infty$, so is the rhs.  These statements are true so long as $\cos\alpha$ defined in Eq.\,(\ref{cosalpha}) satisfies $-1\leq \cos\alpha \leq 1$.  

In calculating a form factor one must compute the Faddeev amplitude of a bound-state that is not at rest.  In the Breit frame, e.g., the total momentum can be written as $P= (0,0,\pm Q/2,i E(Q/2))$, where $E^2(Q/2) = M^2 + Q^2/4$, the bound-state is moving with three momentum $\pm Q/2$ and 
\begin{equation}
\label{kcdotPo}
k\cdot P = \pm \frac{1}{2} |k| Q \cos\theta \sin\beta + i |k| E(Q) \cos\beta\,,
\end{equation}
with $k$ expressed using the standard definition of hyperspherical coordinates. In principle, as demonstrated in Ref.\,\cite{Bhagwat:2006pu}, this is not a problem in a Poincar\'e covariant framework.  However, it can consume large amounts of computer memory and time.  We therefore proceed by writing
\begin{equation}
\label{kcdotP}
k\cdot P = i |k| E(Q) \left[\mp  \frac{iQ}{2E(Q)} \cos\theta \sin\beta + \cos\beta\right] =: i |k| E(Q) z\,,
\end{equation}
in which case the real and imaginary parts of $z$ are bounded in magnitude by one, and then define
\begin{equation}
F(k^2,k\cdot (P\pm Q/2); -M^2) =  \sum_{j=0}^{N_m} \,^j\! F(|k|,i E(Q);-M^2)\, U_j(z)\,.
\end{equation}

\section{Form Factor Notation}
\label{FFN}
We represent all form factors by their usual symbols.  Hence, the notation can be introduced via an exemplar; viz., the proton's Pauli form factor, $F_1^p$.
\begin{itemize}
\item $F_1^{p,q}$ -- \label{nota}{Sum} of all contributions to $F_1^p$ that can be represented by Diagram~1 in Fig.\,\ref{vertex}; i.e., in which the photon interacts with a bystander quark, either $u$ or $d$.  $P_1^{p,q}=F_1^{p,q}(Q^2=0)$ gauges the probability that the photon interacts with a bystander quark.

\item $F_1^{p,c}$ -- Sum of all contributions to $F_1^p$ that can be represented by either Diagram~2 or 4; i.e., in which the photon interacts with a diquark correlation, either scalar or axial-vector, or excites a transition between them.  $P_1^{p,c}=F_1^{p,c}(Q^2=0)$ gauges the probability that the photon interacts with a diquark.

\item $F_1^{p,e}$ -- Sum of all contributions to $F_1^p$ that can be represented by one of Diagrams~3, 5 or 6; i.e., in which the photon interacts with a diquark in association with its breakup. $P_1^{p,e}=F_1^{p,e}(Q^2=0)$ gauges the probability that the photon acts in association with diquark breakup.
    
    NB.\ $F_1^{p,q} + F_1^{p,c} + F_1^{p,e} = F_1^p$.

\item $F_1^{p,u}$ -- Sum of all contributions to $F_1^p$ in Fig.\,\ref{vertex} that are proportional to the charge of a $u$-quark, $e_u$; i.e., the total $u$-quark contribution $F_1^p$.  

\item $F_1^{p,q,u}$ -- Sum of all contributions to $F_1^{p,u}$ that can be represented by Diagram~1 in Fig.\,\ref{vertex}; i.e., in which the photon interacts with a bystander $u$-quark.

\item $F_1^{p,c,u}$ -- Sum of all contributions to $F_1^p$ that can be represented by either Diagram~2 or 4 and are proportional to $e_u$; i.e., in which the photon resolves a $u$-quark within a diquark correlation.

\item $F_1^{p,e,u}$ -- Sum of all contributions to $F_1^{p,u}$ that can be represented by one of Diagrams~3, 5 or 6 and are proportional to $e_u$; i.e., in which the photon interacts with a $u$-quark in association with the breakup of a diquark.
    
    NB.\ $F_1^{p,q,u} + F_1^{p,c,u} + F_1^{p,e,u} = F_1^{p,u}$; $F_1^{p,u}(0) = 2 e_u$; $2 e_u P_1^{p,\alpha,u}:=F_1^{p,\alpha,u}(Q^2=0)$, $\alpha=q,d,e$.
    
\item $F_1^{p,d}$ and related functions are defined in direct analogy with those connected to $F_1^{p,u}$.
    
    NB.\ $F_1^{p,q,d} + F_1^{p,c,d} + F_1^{p,e,d} = F_1^{p,d}$; $F_1^{p,d}(0) =  e_d$; $ e_d P_1^{p,\alpha,d}:=F_1^{p,\alpha,d}(Q^2=0)$, $\alpha=q,d,e$.
    
\end{itemize}

\begin{itemize}
\item $F_1^{p,s}$ -- Sum of all contributions to $F_1^p$ in Fig.\,\ref{vertex} that involve a scalar diquark component in both $\Psi_i$ and $\Psi_f$.  $P_1^{p,s}=F_1^{p,s}(Q^2=0)$ gauges the probability that the photon interacts with a scalar diquark component of the nucleon.

\item $F_1^{p,a}$ -- Sum of all contributions to $F_1^p$ that involve an axial-vector diquark component in both $\Psi_i$ and $\Psi_f$.  $P_1^{p,a}=F_1^{p,a}(Q^2=0)$ gauges the probability that the photon interacts with an axial-vector diquark component of the nucleon.

\item $F_1^{p,m}$ -- Sum of all contributions to $F_1^p$ in which the diquark component of $\Psi_i$ is different to that in $\Psi_f$. $P_1^{p,m}=F_1^{p,m}(Q^2=0)$ gauges the probability that the photon induces a transition between diquark components of the incoming and outgoing nucleon.
    
    NB.\ $F_1^{p,s} + F_1^{p,a} + F_1^{p,m} = F_1^p$.
    
\item $F_1^{p,s,u}$ -- Sum of all contributions to $F_1^p$ in Fig.\,\ref{vertex} that involve a scalar diquark component in both $\Psi_i$ and $\Psi_f$, and are proportional to $e_u$; i.e., in which a $u$-quark is resolved in the presence of a scalar diquark.

\item $F_1^{p,a,u}$ -- Sum of all contributions to $F_1^p$ that involve an axial-vector diquark component in both $\Psi_i$ and $\Psi_f$, and are proportional to $e_u$.

\item $F_1^{p,m,u}$ -- Sum of all contributions to $F_1^p$ that are proportional to $e_u$ and in which the diquark component of $\Psi_i$ is different to that in $\Psi_f$.
    
    NB.\ $F_1^{p,s,u} + F_1^{p,a,u} + F_1^{p,m,u} = F_1^{p,u}$; 
    $2 e_u P_1^{p,\alpha,u}:=F_1^{p,\alpha,u}(Q^2=0)$, $\alpha=s,a,m$.

\item $F_1^{p,s,d}$ \label{notb}{and} similar functions are defined in direct analogy with those connected to $F_1^{p,s,u}$.
    
    NB.\ $F_1^{p,s,d} + F_1^{p,a,d} + F_1^{p,m,d} = F_1^{p,d}$; 
    $e_d P_1^{p,\alpha,d}:=F_1^{p,\alpha,d}(Q^2=0)$, $\alpha=s,a,m$.
    
\end{itemize}


\end{document}